\documentclass{aa}

\usepackage{txfonts}
\usepackage{subcaption}
\usepackage{graphicx}
\usepackage{multirow, makecell}
\begin{document}
%\raggedbottom
\authorrunning{E. Samara, et al.,}

   %\title{The influence of the shape of coronal holes on the solar wind speed at Earth}
   \title{Influence of coronal hole morphology on the solar wind speed at Earth}

   \author{Evangelia Samara
          \inst{1,2},
          Jasmina Magdaleni\'{c}\inst{1,2},
          Luciano Rodriguez\inst{1},
          Stephan G. Heinemann\inst{3},
          Manolis K. Georgoulis\inst{4},
          Stefan J. Hofmeister\inst{5},
          Stefaan Poedts\inst{2,6}
          }

   \institute{1. Solar-Terrestrial Centre of Excellence -- SIDC, Royal Observatory of Belgium, Ringlaan 3, 1180 Brussels, Belgium\\
             2. Centre for mathematical Plasma Astrophysics (CmPA), KU Leuven, Celestijnenlaan 200B, 3001 Leuven, Belgium\\
             3. Max-Planck-Institut for Solar System Research, Justus-von-Liebig-Weg 3, 37077, Göttingen, Germany \\
             4. Research Center for Astronomy and Applied Mathematics, Academy of Athens, 4 Soranou Efesiou Street, 11527 Athens, Greece
             5. Leibniz-Institut for Astrophysics Potsdam (AIP), An der Sternwarte 16, 14482, Potsdam, Germany  \\
             6. Institute of Physics, University of Maria Curie-Sk{\l}odowska, plac Marii Curie-Sk{\l}odowskiej 5, 20-400 Lublin, Poland\\}

%   \date{Received September 15, 1996; accepted March 16, 1997}

% \abstract{}{}{}{}{} 
% 5 {} token are mandatory

    \abstract
  % context heading (optional)
  % {} leave it empty if necessary  
   %{}
  % aims heading (mandatory)
   {It has long been known that the high-speed stream (HSS) peak velocity at Earth directly depends on the area of the coronal hole (CH) on the Sun. Different degrees of association between the two parameters have been shown by many authors. In this study, we revisit this association in greater detail for a sample of 45 nonpolar CHs during the minimum phase of solar cycle 24. The aim is to understand how CHs of different properties influence the HSS peak speeds observed at Earth and draw from this to improve solar wind modeling. }
  % methods heading (mandatory)
   {The CHs were extracted based on the Collection of Analysis Tools for Coronal Holes (CATCH) which employs an intensity threshold technique applied to extreme-ultraviolet (EUV) filtergrams. We first examined all the correlations between the geometric characteristics of the CHs and the HSS peak speed at Earth for the entire sample. The CHs were then categorized in two different groups based on morphological criteria, such as the aspect ratio and the orientation angle. We also defined the geometric complexity of the CHs, a parameter which is often neglected when the formation of the fast solar wind at Earth is studied. The quantification of complexity was done in two ways. First, we considered the ratio of the maximum inscribed rectangle over the convex hull area of the CH.
   The maximum inscribed rectangle provides an estimate of the area from which the maximum speed of the stream originates. The convex hull area is an estimate of how irregular the CH boundary is. The second way of quantifying the CH complexity was carried out by calculating the CH's fractal dimension which characterizes the raggedness of the CH boundary and internal structure. }
  % results heading (mandatory)
   {When treating the entire sample, the best correlations were achieved between the HSS peak speed observed in situ, and the CH longitudinal extent. When the data set was split into different subsets, based on the CH aspect ratio and orientation angle, the correlations between the HSS maximum velocity and the CH geometric characteristics significantly improved in comparison to the ones estimated for the whole sample. By further dividing CHs into subsets based on their fractal dimension, we found that the Pearson's correlation coefficient in the HSS peak speed -- CH area plot decreases when going from the least complex toward the most complex structures. Similar results were obtained when we considered categories of CHs based on the ratio of the maximum inscribed rectangle over the convex hull area of the CH. To verify the robustness of these results, we applied the bootstrapping technique. The method confirmed our findings for the entire CH sample. It also confirmed the improved correlations, compared to the ones found for the whole sample, between the HSS peak speed and the CH geometric characteristics when we divided the CHs into groups based on their aspect ratio and orientation angle. Bootstrapping results for the CH complexity categorizations are, nonetheless, more ambiguous. }
  % conclusions heading (optional), leave it empty if necessary 
   {Our results show that the morphological parameters of CHs such as the aspect ratio, orientation angle, and complexity play a major role in determining the HSS peak speed at 1 AU. Therefore, they need to be taken into consideration for empirical models that aim to forecast the fast solar wind at Earth based on the observed CH solar sources.}

   \keywords{Sun: coronal holes --
                Sun: high speed streams --
                Sun: solar wind
               }
               
\maketitle

%________________________________________________________________
\section{Introduction} \label{Sec:Introduction}

The solar atmosphere is a dynamic and complex system in which the solar wind, a continuous outflow of plasma filling the heliosphere, is generated. It has long been known that the solar wind is bimodal, i.e., it has a slow and a fast component \citep[see e.g.,][and references therein]{McComas98, Schwenn2006, cranmer17}. The sources of both wind components are a subject of intense scientific research. More specifically, it is considered that the slow solar wind is mostly associated with the streamer belt region \citep[e.g.,][]{schwenn90, Habbal97}, while the fast solar wind is associated with the coronal holes \citep[CHs; e.g.,][]{Nolte76, Cranmer2009}. 

Coronal holes are regions with open magnetic field topology in the solar corona. They are characterized by lower densities and temperatures compared to the ambient solar plasma \citep[][]{DelZannaBromage99, Flundra99, Bromage2000,Heinemann2021a} and, therefore, they appear as dark structures in extreme-ultraviolet (EUV) and X-ray images \citep[see review by][]{Cranmer2002}. Already in the 1960s, an association between coronal conditions and periodicities of 27 to 28 days in the solar wind bulk speed, as well as in the interplanetary magnetic sector, was suggested \citep[][]{Snyder&Neugebauer66, Wilcox1968}. During that time, it was also considered that high speed solar wind streams might originate from magnetically open regions in the corona, rather than from regions with elevated coronal temperatures \citep{BillingsRoberts64, Davis65}. In the 1970s, the term ``coronal hole'' was introduced based on X-ray observations in which CHs appeared as regions with reduced X-ray emission compared to the ambient environment. They also appeared to have a diverging magnetic field configuration, so they were directly connected as the sources of fast solar wind streams \citep[][]{Krieger1972}, completing the puzzle of their theoretical prediction. Evidence for this association was provided by tracing the peak speed of a high speed stream (HSS) back to the Sun and comparing it with the position of the associated CH \citep[][]{Krieger1972}. Other works directly compared the CH characteristics with properties of the in situ solar wind detected at Earth \citep[see e.g.,][]{Nolte76}. 

Up to this day, significant progress has been made toward understanding the association between the properties of the observed HSSs and their source regions on the Sun. For example, \citet{WangSheeley90} studied the long-term variation in the solar wind bulk speed with the total CH area occupied by all CHs observed at the Sun and found a good correlation. The authors also showed that the wind speed at $1\;$AU and the rate of the flux-tube divergence near the Sun are inversely correlated. Analyzing the relationship between the CH area and position as well as physical characteristics of the associated corotating HSS at $1\;$AU, \citet{vrsnak07} showed that an accurate prediction can be obtained for the solar wind velocity, but not for the solar wind density, temperature, or magnetic field. \citet{Rotter2012} analyzed the relationship between the CH area, position, and intensity levels in EUV with the solar wind speed, proton temperature, proton density, and magnetic field strength measured in situ at $1\;$AU. They also confirmed a good linear correlation (cc = 0.78) between the peaks of the observed and the predicted solar wind velocities, the latter being estimated based on the CH areas. A mediocre correlation (cc = 0.41) was found between the peaks of the observed and predicted solar wind temperature, as well as between the observed and predicted solar wind magnetic field. No correlation was obtained between observed and predicted proton densities. Moreover, \citet{Hofmeister2018} studied a number of CH topological properties with respect to the in situ solar wind speed at $1\;$AU for a sample of 115 CHs. They show that the peak velocities of the HSSs are linearly dependent on the areas and the latitudinal separation angle between the CH and the satellite taking the in situ measurements. 

%Other authors have also tried to go deeper into the study of the magnetic characteristics of CHs and their association with the solar wind in situ bulk speed. More specifically, \citet{Abramenko09} tried to explain the lack of correlation between the density net flux and the solar wind speed measured at $1\;$AU. They claim that the reason for that should stem from the structural organization of the magnetic flux and its multi-fractal nature. 

In this paper, we focus on the relations of the morphological and geometric CH characteristics with the in situ HSS properties at Earth for a sample of 45 CHs. The CHs are characterized by different sizes, shapes, and complexities. We analyze how quantities such as the CH area, longitudinal, and latitudinal CH extent are associated with the maximum fast solar wind speed and HSS duration at $1\;$AU. The sample was categorized in two different groups, based on the aspect ratio and the orientation angle. The aim is to understand how important the role of these parameters is for the formation of HSS peak velocity at Earth. Particular emphasis is given to the definition of the geometric complexity of CHs, a concept that is usually neglected. The geometric complexity is relevant to the hollowness of the internal part of the CH as well as to the irregularity of the CH boundary. Thus, we investigate how this parameter is associated with the CH area and HSS peak speed in situ. The ensemble of different criteria will help us identify important parameters and correlations that need to be addressed for better understanding and forecasting the solar wind observed at Earth. 

The structure of the paper is as follows: in Section~2 we present the data set, the selection of the events, and the employed models. %In Section~3, we present the extra motivation of this work and its connection with the solar wind modeling with EUHFORIA. 
Section~3 contains the statistical analysis of the CH geometric characteristics with the in situ HSS peak speed and duration. The categorization of our sample based on the CH aspect ratio and orientation angle is also discussed there, as well as results on how these parameters influence the HSS peak speed at Earth. In Section~4 we describe how the geometric complexity of the CHs can be evaluated based on two different techniques and how this parameter is related to the CH area and HSS peak velocity at $1\;$AU. In Section~5, we summarize our results and main conclusions.

%From the modelling point of view, \citet{Hofmeister2020} showed that for a CH of an ideal rectangular shape, the slope of the linear relationship between the CH area and peak HSS velocity at Earth decrease with increasing latitudes of the coronal holes, and the peak velocities saturate at a value of about 730km/s, similar to observations (see \citet{Heinemann2020, Garton2018} who showed that for a random CH sample, the well-known CH area vs HSS peak velocity does not indicate a linear trend but converges to a speed limit of 700 km/s and 710 km/s, respectively, for large CHs\footnote{Large CHs refer to an area of larger than 10 $\times$ 10$^_{10}$ km$^_{2}$}. 

\section{Data sets, selection of events, and models} 

We analyzed 1-minute in situ solar wind observations during the recent solar minimum interval from November 2017 to September 2019. The data were obtained from OMNIweb and were smoothed based on a moving average of 1h. The solar wind in situ signatures (bulk speed $v_{b}$; proton density $n$; temperature $T$; magnitude of the interplanetary magnetic field (IMF) $B$; the IMF latitudinal component $B_{z}$; and the IMF longitudinal $\phi$-angle) observed at the Lagrangian point 1 (L1) by the \text{Wind} spacecraft \citep{WIND} were examined in order to identify clear HSS cases. In order to focus only on clearly undisturbed HSSs, we excluded all the intervals for which an interplanetary coronal mass ejection (ICME) was recorded based on the Richardson-Cane list \citep{RClist2003, RClist2010}.

In the considered time window of 25 Carrington rotations, we isolated 63 events that contained all the signatures of a HSS following the common definition presented by \citet[][]{Jian2006}. A HSS was identified when a simultaneous increase in the solar wind speed and the temperature was recorded, preceded by a density increase. The density increase indicated the arrival of the compression region that was formed from the interaction of the upcoming fast solar wind with the preceding slow solar wind. An increase in the total magnetic field strength was also observed due to that compression. Additionally, the longitudinal angle of the IMF vector, the so-called $\phi$-angle, indicated if the field was pointing toward or away from the Sun. This angle should be predominantly in one of those two directions and match the polarity of the source CH in order to confirm the arrival of the HSS. It was calculated by taking into account the arctan of the Bx, By components of the IMF in the Geocentric Solar Ecliptic (GSE) frame. These are the criteria we followed for the detection of HSSs, for velocities exceeding $400\;$km/s \citep{Schwenn2006, vrsnak07}. 

After isolating the HSSs in the in situ observations, we searched for the CHs from which these fast streams originated. For that purpose, we inspected images of the Sun two to five days (which corresponds to the usual travel time of HSSs from the Sun to Earth) prior to the beginning of each HSS. The inspection was done by employing EUV observations at $193\;\r{A}$ which observe the corona at about 1.6 MK, obtained by the Atmospheric Imaging Assembly \citep[AIA;][]{Lemen2012} on board the Solar Dynamics Observatory \citep[SDO;][]{SDO}. Additionally, we examined the NOAA full-Sun drawings\footnote{https://www.ngdc.noaa.gov/stp/space-weather/solar-data/solar-imagery/composites/full-sun-drawings/boulder/} in order to confirm the position and polarity of the CH. The NOAA drawings are based on H-alpha observations, sunspots, magnetogams, as well as on He I 10830$\AA$ ground-based observations which help define the CH boundaries. In order to confirm the association between a HSS and a particular CH, we required that the polarity of the $\phi$-angle detected in the in situ observations was the same as the polarity of the CH in the NOAA drawings.

We extracted the CH characteristics using the Collection of Analysis Tools for Coronal Holes \citep[CATCH;][]{Heinemann2019}. CATCH enables the user to download data, perform guided manual CH extraction, as well as analyze the photospheric magnetic field characteristics of CHs. The basic principle of the CH extraction relies on an intensity-based threshold technique applied to EUV filtergrams of sufficient contrast, a preliminary form of which was developed by \citet{rotter12}. 

We note that, even if all CHs are individually processed and extracted by CATCH, the procedure is not sufficiently accurate for polar CHs due to the strong projection effects and the lack of reliable information from the Sun's polar regions. Therefore, we excluded  all the extensions of the polar CHs (18 out of the total 63 CHs) from our set and only focused on the nonpolar ones (45 out of 63 in total). The EUV images of the nonpolar CHs and their extraction boundaries are presented in Appendix A, in  Fig.~\ref{Fig:EUV_images}, \ref{Fig:EUV_images2}, \ref{Fig:EUV_images3}. Our data set includes all nonpolar CHs found in the interval of interest (i.e., November 2017 - September 2019), which, first, are associated with HSSs observed at Earth and, second, their center of mass (CoM) is found on (or very close to) the central meridian since the connectivity of the CH with Earth, through the produced HSS, is optimal in this case. No other preselection criteria were applied, i.e., our sample contains CHs of different sizes and shapes.

\section{The geometric characteristics of CHs and the HSSs' properties at $1\;$AU}

In this section, we address the association between the geometric characteristics of the CHs with the maximum HSS velocity and duration at Earth, as observed in situ by the Wind satellite. We discuss the following parameters: the~CH area,~CH longitudinal and latitudinal extent,~CH CoM,~CH aspect ratio (i.e., ratio of the longitudinal to the latitudinal extent), and CH orientation angle (the angle of the major axis with respect to the solar equator, once the CH is fitted by an ellipse). All parameters, except the orientation angle, were obtained with CATCH. Each CH was extracted separately by using an individual intensity threshold that permitted the optimal identification of the CH structure from its surrounding environment. 

\begin{figure}[h!]
\centering

\begin{subfigure}[]{0.99\linewidth}
\includegraphics[width=\linewidth]{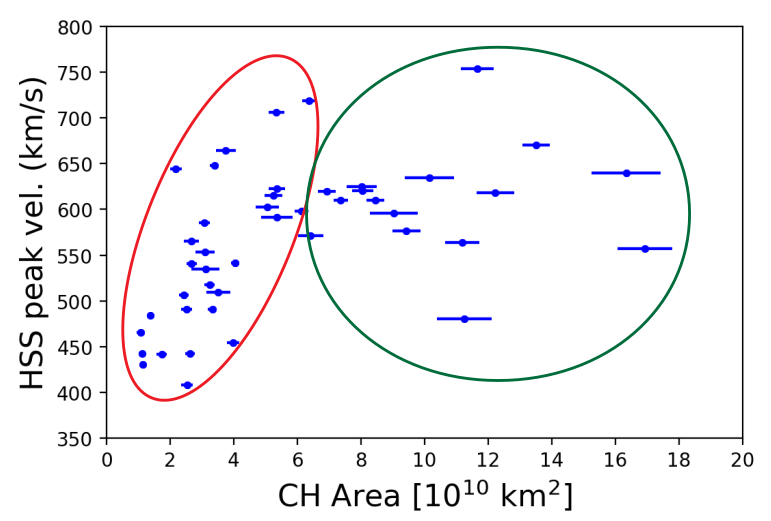}
\caption{}
\end{subfigure}%

\begin{subfigure}[]{0.95\linewidth}
\includegraphics[width=\linewidth]{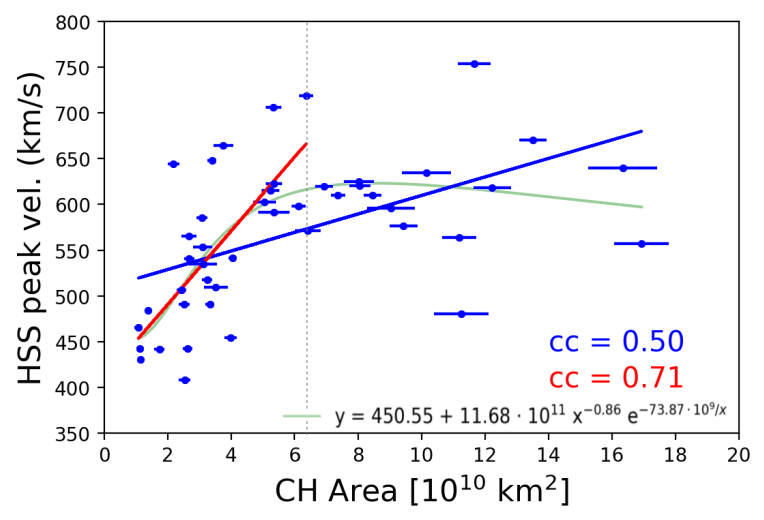}
\caption{}
\end{subfigure}%
\caption{HSS peak velocity as a function of the CH area. Panel (a): Red and green ovals divide our CH sample into two different groups. Panel (b): Fits for the different CH groups. The blue line shows the linear fit applied to the whole sample, while the red line is the linear fit applied for the subset that provided the highest cc. The gray dotted line separates the two groups and it sets the threshold beyond which the HSS peak velocity does not increase with an increasing CH area. The light green line corresponds to a power-exponential fit for all points and confirms the physical threshold in velocities.}
\label{Fig:CHarea_vs_speed}
\end{figure}

\subsection{CH area versus peak HSS speed}
\label{subsection:CHarea}

The relation between the CH area and the HSS peak speed at $1\;$AU (Fig.~\ref{Fig:CHarea_vs_speed}) is one of the most studied relationships used to predict the solar wind observed in situ \citep[][]{Nolte76, vrsnak07, Abramenko09, karachik11, rotter12, Hofmeister2018, Heinemann2018, Heinemann2020}. The linear associations presented in previous studies are found within a wide range of Pearson's correlation coefficients (cc) and depend, not only on the considered data set, but also on the selection criteria, as we show in the following subsections. The CH areas were not modified by employing the colatitude correction suggested by \citet{Hofmeister2018}. This is because the latitude of the CoM for the majority of CHs was relatively close to the equator, so such a correction was not necessary as it would not induce a significant difference in the obtained results. However, for completeness, the corrected and uncorrected statistics of our sample are presented in Appendix~\ref{appendix:colatitude_corr}.

In Fig.~\ref{Fig:CHarea_vs_speed}a we observe two potential groups of CHs, marked by the red and green ovals, respectively. The first group extends up to the CH area of $\approx 6.4 \times 10^{10}\;$km$^{2}$ and includes points that indicate a good linear correlation between the HSS peak velocity and the CH area (cc = 0.71, see Fig.~\ref{Fig:CHarea_vs_speed}b for more details). Beyond that point, the linear correlation gets worse. This is also where the second group of CHs begins and extends up to $\approx 17 \times 10^{10}\;$km$^{2}$. We found that 6.4 $\times 10^{10}$\;km$^{2}$ is the threshold beyond which, regardless of the CH growth, there is no increase in the HSS peak speed observed at Earth. It was identified by finding the point at which the cc of the fitted linear regression line was maximized. The best-fit upper threshold in speeds is found at $\approx 630\;$km/s for our sample. This is confirmed by fitting a power-exponential function of the form $f(x) = A + Bx^{\gamma}exp(C/x) $ (light green line in Fig.~\ref{Fig:CHarea_vs_speed}b). For $\gamma \longrightarrow  1$ and $|C| \longrightarrow 0$, this equation becomes linear. We used such a fit because, of all fitting curves, the power-exponential one has the physical meaning that the HSS peak velocity does not increase without restrictions. Our observational results are qualitatively similar to a recent study by \citet{Hofmeister2020} who showed, via modeling, that the HSS peak velocities saturate at a value of about $730\;$km/s. From the observational point of view, \citet{Heinemann2020} identified a threshold at $700\;$km/s for CHs larger than $\approx 10 \times 10^{10}\;$km$^{2}$, while \citet{Garton2018} showed that an upper threshold of $710\;$km/s is imposed when they correlated the HSS peak velocity with the CH longitudinal extent. In the latter study, the threshold in the CH longitudinal extent, above which the solar wind velocity saturates ($\approx$ 67 deg), was identified by the intersection of two lines: one fitted for CHs lower than $\approx$ 70 deg and one for CHs bigger than $\approx$70 deg. The differences in the results among the studies mostly arise from the different CH samples that are used, as well as from different methods of extracting CH characteristics. In our case, the extraction tool was CATCH and we took into account all nonpolar CHs observed in the considered time interval, regardless of the geometric complexity, size, or latitudinal position.

\begin{figure*}[t]
\centering
\begin{subfigure}[]{0.48\linewidth}
\includegraphics[width=\linewidth]{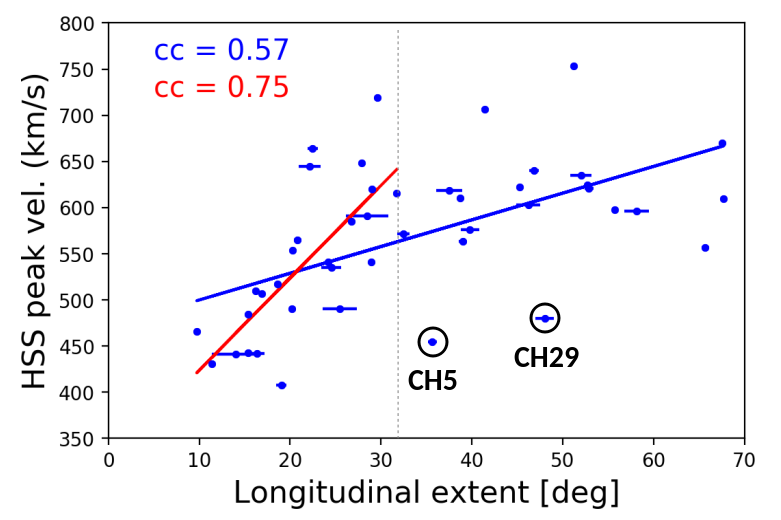}
\caption{}
\end{subfigure}
%\hspace*{\fill}
\begin{subfigure}[]{0.48\linewidth}
\includegraphics[width=\linewidth]{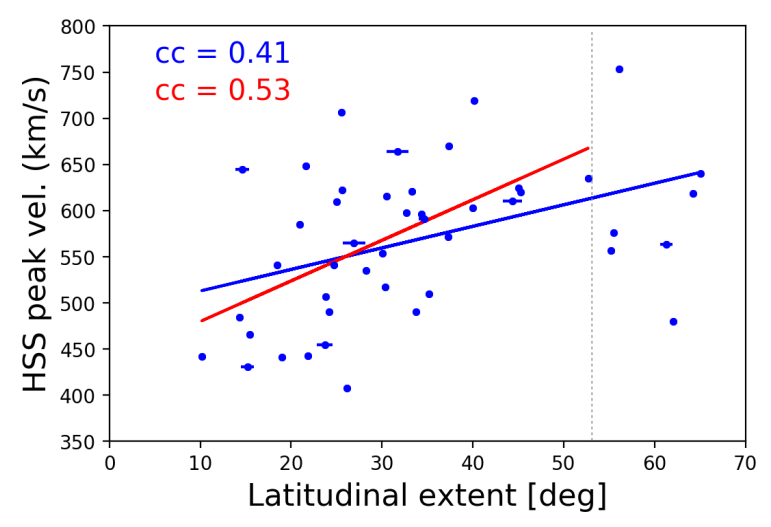}
\caption{}
\end{subfigure}%

\caption{HSS peak velocity as a function of the CH longitudinal extent (a) and latitudinal extent (b). A linear regression line (red color) was fitted up until the point that the highest cc was achieved. A second linear regression line (blue color) was fitted to the whole sample. The gray dotted line separates the two clusters and sets the threshold beyond which, regardless of the CH longitude or latitude growth, the HSS peak velocity does not increase further. In panel (a), we marked CH5 and CH29 as outliers with high longitudinal widths, but low HSS peak speeds at Earth.}
\label{Fig:long_lat_extent_vs_speed}
\end{figure*}

\begin{table*}[ht]
\centering
\caption{Comparison of the Pearson's cc for the entire CH sample and the different subgroups. }
\begin{center}
 \begin{tabular}{|l | c| c | c  |} 
 \hline
  Groups &  \makecell{Pearson's cc \\ (HSS V$_{b}$ - CH area)} & \makecell{Pearson's cc \\ (HSS V$_{b}$ - CH long. extent)}  & \makecell{Pearson's cc \\ (HSS V$_{b}$ - CH lat. extent)} \\
 \hline\hline

Total sample        &   0.50 &  0.57 & 0.41 \\
\hline
Aspect ratio $>$ 1      &       0.52 &  0.49 & 0.49 \\
\hline
Aspect ratio $<$ 1      &       0.69 &  0.77 & 0.70     \\
\hline
Orientation angle $>$ 0 & 0.84 & 0.77 & 0.56 \\
\hline
Orientation angle $<$ 0 & 0.93 & 0.88 & 0.89 \\
\hline

\end{tabular}
\end{center}
\label{Table:Summarized_cc}
\end{table*}

\begin{table*}[ht]
\centering
\caption{Comparison of the Pearson's cc median value as well as of the first (Q1) and third (Q3) quartiles shown within brackets as [Q1, Q3] for the entire CH sample and the different subgroups, after applying the pair bootstrapping method.} 
\begin{center}
 \begin{tabular}{|l | c| c | c  |} 
 \hline
  Groups &  \makecell{Pearson's cc \\ (HSS V$_{b}$ - CH area)} & \makecell{Pearson's cc \\ (HSS V$_{b}$ - CH long. extent)}  & \makecell{Pearson's cc \\ (HSS V$_{b}$ - CH lat. extent)} \\
 \hline\hline

Total sample        & 0.51 [0.43, 0.58] & 0.57 [0.50, 0.63] & 0.42 [0.33, 0.50] \\
\hline
Aspect ratio $>$ 1      & 0.53 [0.41, 0.63] & 0.50 [0.37, 0.61] & 0.49 [0.36, 0.61] \\
\hline
Aspect ratio $<$ 1      & 0.70 [0.64, 0.76] & 0.77 [0.71, 0.83] & 0.72 [0.66, 0.77]   \\
\hline

Orientation angle $>$ 0 & 0.85 [0.79, 0.89] & 0.78 [0.71, 0.85] & 0.57 [0.39, 0.71] \\
\hline
Orientation angle $<$ 0 & 0.94 [0.91, 0.96] & 0.89 [0.87, 0.92] & 0.90 [0.87, 0.93]  \\
\hline

\end{tabular}
\end{center}
\label{Table:bootstrapping_aspectratio}
\end{table*}

Many authors have shown that the correlation between the HSS peak velocity at Earth and the CH area on the Sun range between $ 0.40 \leq$ cc $\leq 0.96$ \citep[see e.g.,][]{Nolte76, vrsnak07, karachik11, rotter12, Hofmeister2018, Heinemann2020}. Our results, with or without splitting the data into two groups, fall within the range of coefficients obtained by previous studies. The linear fit to the full data set in Fig.~\ref{Fig:CHarea_vs_speed}b (blue line) results in a cc of 0.50, while it increases to 0.71 (red line) for the first group of CHs, with areas smaller than 6.4 $\times 10^{10}$\;km$^{2}$. The cc then seems to lie closer to the higher limits of that range, while in the former case the correlation is poorer, lying closer to the lower limits. This means that differentiating between small- to medium-size CHs, and larger CHs, does make a difference in the association between the CH areas and the HSS peak velocity recorded at Earth. Nevertheless, it is clear that a linear fit is not sufficient to approximate the behavior of our sample. In order to confirm this, we performed a chi-square goodness-of-fit analysis for the linear (blue) and the power-exponential (light green) line. The power exponential line gave better, but still not optimal results due to scatter.

%Considering now both of the discussed possibilities (namely, splitting our sample into two data sets, or not), we see why such a broad spread on the correlation coefficients was obtained in the previous studies. 
% When the sample is consisting of CHs with a diverse characteristics i.e. large spread of the CH area sizes, the resulting correlation coefficient reflects the point that at least two different data sets are consider as a single data set. 
%The correlation coefficient between the HSS peak velocity and the CH area highly depends on the considered CH sample and the size of the structures. 

\subsection{Longitudinal and latitudinal extent of CHs versus peak HSSs' speed}
\label{subsection:CHlong_lat}

In this section we show how the HSS peak velocity at Earth depends on the longitudinal and latitudinal extent of the CH
(Fig.~\ref{Fig:long_lat_extent_vs_speed}). %The latitudinal extents were not modified by using the co-latitude correction from \citet{Hofmeister2018}, likewise for the CH areas in the previous section. %A linear regression line was fitted for the CH longitudinal and latitudinal extent (Fig.~\ref{Fig:long_lat_extent_vs_speed}). \sout{we find a Pearson's correlation coefficient of 0.56 and 0.39, respectively.} 
Figure~\ref{Fig:long_lat_extent_vs_speed}a shows the dependence of the HSS peak velocity on the CH longitudinal width. The existence of two separate groups is also noted here, similar to Fig.~\ref{Fig:CHarea_vs_speed}. Therefore, we aim to identify a threshold regarding the longitudinal extent by finding the point in which the cc is maximized when we fit a linear regression line.
%but somewhat less distinct than in the case when the CH area and the HSS peak velocity are considered (see Fig.~\ref{Fig:CHarea_vs_speed}). 
The first group extends up to $32\;$deg longitudinal width and the second covers CHs larger than that. Our results are qualitatively in agreement with the ones by \citet{Garton2018}, who show, for a sample of 47 CHs, that the HSS peak velocities are well correlated with the longitudinal extent of the associated CHs. They also found that, at the longitudinal extent of 67 deg, a saturation limit in the velocities of $\approx 710\;$km/s is imposed. Our study shows a similar pattern, but the saturation limit is obtained for CHs with a lower longitudinal extent (up to 32 deg) and a velocity threshold at $\approx 600\;$km/s. The difference in the results may be explained by the different CH samples used in each study. Our sample includes both relatively small and big CHs and, in general, the peak HSS speeds recorded do not reach very high values. The highest speed is recorded at $\approx$ 750 km/s, while peaks in the range 700 - 750 km/s are only found in three cases. 

Overall, the HSS peak velocity and longitudinal extent relationship is qualitatively similar to the HSS peak velocity and CH area relationship (Fig.~\ref{Fig:CHarea_vs_speed}). Namely, the correlation is lower when we consider the whole data set (cc = 0.57, see blue line in Fig.~\ref{Fig:long_lat_extent_vs_speed}a) and higher when the first subset is considered (cc = 0.75, see red line in Fig.~\ref{Fig:long_lat_extent_vs_speed}a). We distinguish two outlier points (CHs 5 and 29, see Fig.~\ref{Fig:EUV_images}, \ref{Fig:EUV_images2}, respectively), which have very low HSS speeds but large longitudinal widths. CH 5 has a CoM at 32.70 deg, and it is the most distant one from the equator in our data sample. Its longitudinal width is quite extended, but the peak velocity recorded at Earth is very small. Our initial guess was that this happens exclusively due to the higher CoM CH latitude which induces a larger latitudinal angle with L1, resulting in the satellite to only be affected by the HSS’s flanks \citep[see][for more details]{Hofmeister2018}. Nevertheless, for CH 3 (see Fig.~\ref{Fig:EUV_images}), which is not an outlier and has similar characteristics to CH 5 (e.g., CoM at 32.05 deg and similar longitudinal and latitudinal extents), we recorded a HSS peak speed $\approx$ 170 km/s higher than in the CH 5 case. The discrepancy in speeds between the two cases could arise from the different signs in the $B_{0}$ angle (the heliographic latitude of the central point of the solar disk at the time the CHs were observed). In the case of CH 3, $B_{0}$ is 0.66 deg and in the case of CH5, $B_{0}$ is -2.76 deg in the Heliocentric Earth EQuatorial (HEEQ) frame. Therefore, the solar equator was in opposite sides of the ecliptic for these two cases. More specifically, during the passage of CH 5 from the central meridian, the solar southern hemisphere was facing Earth more than the northern hemisphere, that is the solar equator was above the ecliptic, causing CH 5 to be more distant from L1. Additionally, the latitudinal position of the satellite at L1 was at -3.14 deg from the solar equator (HEEQ frame), $\approx$ 0.39 lower than the ecliptic. The cumulative effect of the abovementioned two points probably impeded the direct impact of that HSS at L1 position. The second outlier is CH 29, with a very irregular structure (see Fig.~ \ref{Fig:EUV_images2}) which seems to be composed of two parts. These two CHs correspond to 10$\%$ of the population in the second group.

The situation is somewhat different for the association of the latitudinal extent of CHs with the peak velocity of the HSS. Figure~\ref{Fig:long_lat_extent_vs_speed}b shows a high spread of the data points. Two data groups can hardy be defined, with one extending up to 53 deg and the other going up to 70 deg in latitudinal extent. Correlations for the considered groups are low and not very different from each other, indicating that there is no reason to divide the initial data set.

Our results show better correlations for the CH longitudinal extent compared to the CH area, and significantly better correlations compared to the CH latitudinal extent (see first row of summarized Table \ref{Table:Summarized_cc}). To verify the robustness of these results, we employed the pair bootstrapping technique \citep[see][for an extensive description on bootstrapping]{bootstrap2}. 
This technique is particularly efficient when the number of points in a study is relatively small, as is our case here. The method allows one to redraw numerous data sets of the same length as the original data set with replacement. This means that a pair of data points from the original sample can be selected multiple times. For each of our cases, we redrew 100.000 data sets generating a distribution of 100.000 Pearson's cc\footnote{For bootstrapping applications with python, see http://bebi103.caltech.edu.s3-website-us-east-1.amazonaws.com/2019a/content/lessons/lesson$\_$07/index.html}. In Table \ref{Table:bootstrapping_aspectratio} (first row), we present the median value, the first (Q1) and third (Q3) quartiles of those distributions after applying bootstrapping to the relationships of HSS peak velocity and CH area, HSS peak velocity and CH longitudinal extent, and HSS peak velocity and CH latitudinal extent. The Q1, median, and Q3 parameters represent the values below which 25$\%$, 50$\%$, and 75$\%$ of the data are contained, respectively. These three parameters describe our results best due to the skewed nature of the Pearson's cc distributions generated from bootstrapping. From Table \ref{Table:bootstrapping_aspectratio} (first row), we see that the median Pearson's cc are very similar to the ones estimated for our original sample (Table \ref{Table:Summarized_cc}, first row).

The fact that the CH longitudinal extent is better correlated to the HSS peak velocity at Earth, compared to the CH area and latitudinal extent (confirmed by bootstrapping, as well), could be explained by the propagation effects which the HSSs undergo on the way from Sun to Earth. More specifically, the solar rotation combined with the mostly radial propagation of the solar wind make the fastest, western plasma parcels of the HSS impinge into the preceding stream interaction region (SIR), drive the stream interface to the slow solar wind, and  thereby get decelerated. The following fast parcels subsequently interact with the SIR less and less, better keeping their fast velocities. These phenomena take place on the azimuthal plane which is directly connected to the CH longitudinal widths, rather than the latitudinal ones \citep[see][for more details]{Hofmeister2021}.

We have also inspected the possible association between the CH area and the HSS peak velocity by splitting the CHs into three groups based on their CoM latitudes: |CoM| < 10 deg, |CoM| = 10 - 20 deg, and |CoM| = 20 - 35 deg. We found that only in the former case is a correlation of cc = 0.66 observed. For the second and third group, a cc of $\approx$ 0 and 0.48 is found, respectively. This shows that this specific parameter does not necessarily influence the HSS peak speed. Furthermore, since all CHs are at the central meridian at the moment of estimating their characteristics, we did not investigate the influence of the CoM longitude.

%of the CHs will more influence the type of the impact the associated HSS will have at Earth than the HSS velocity. 

\subsection{CH geometric characteristics and HSS duration}
\label{subsection:HSSduration}

\begin{figure}[h!]
\centering
\begin{subfigure}[]{0.95\linewidth}
\includegraphics[width=\linewidth]{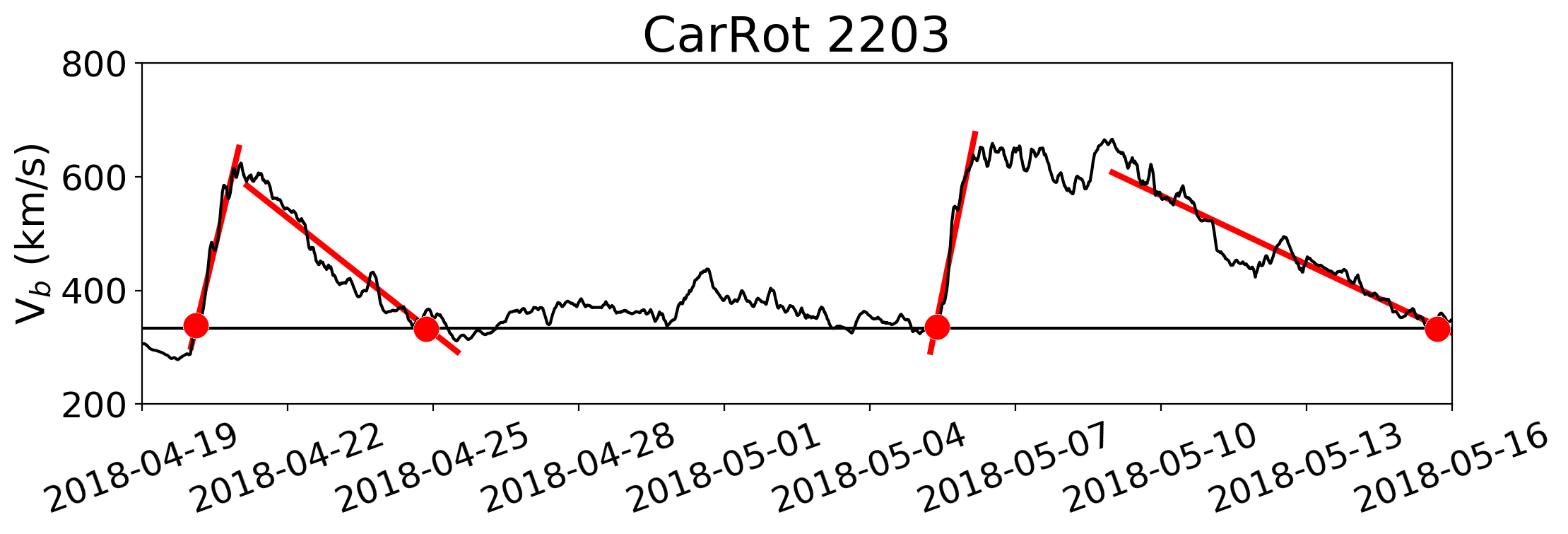}
\caption{}
\end{subfigure}
\caption{Example of how the HSS duration at Earth was estimated for two HSSs of our sample recorded during Carrington Rotation 2203. The ascending and descending phase of each HSS were linearly fitted (red lines). The intersection of these red lines with the mean background solar wind (red circles) indicate the arrival and end times of each HSS, based on which the HSS durations were calculated. }
\label{FittingHSSs}
\end{figure}

After studying the geometric characteristics of the CHs and their association with the maximum velocity of the HSS detected at Earth, we inspected the association between the CH characteristics and the HSS duration at the same point. To estimate the duration of the HSS at Earth, we employed the method by \citet{Garton2018}. More specifically, we found the intersection between the slope of the ascending and descending HSS phase (red fits on Fig.~\ref{FittingHSSs}) with the mean background solar wind. These intersections roughly correspond to the arrival\footnote{In general, the arrival time of the HSS is given by the arrival time of the stream interface between the HSS and the preceding slow solar wind plasma. As the flow speed also starts to increase across the stream interface significantly, it is an acceptable proxy for the arrival time of HSSs for our purposes.} and end times of each HSS, based on which their duration was calculated. For the estimation of the mean background solar wind speed, we took into account the average speed of the intervals in which no HSS was detected (see definition of HSS in section 2). This estimation was done for each Carrington rotation separately, resulting in mean background solar wind speeds in the range [329, 376] km/s.

\begin{figure*}[ht]
\centering
\begin{subfigure}[]{0.3\linewidth}
\includegraphics[width=\linewidth]{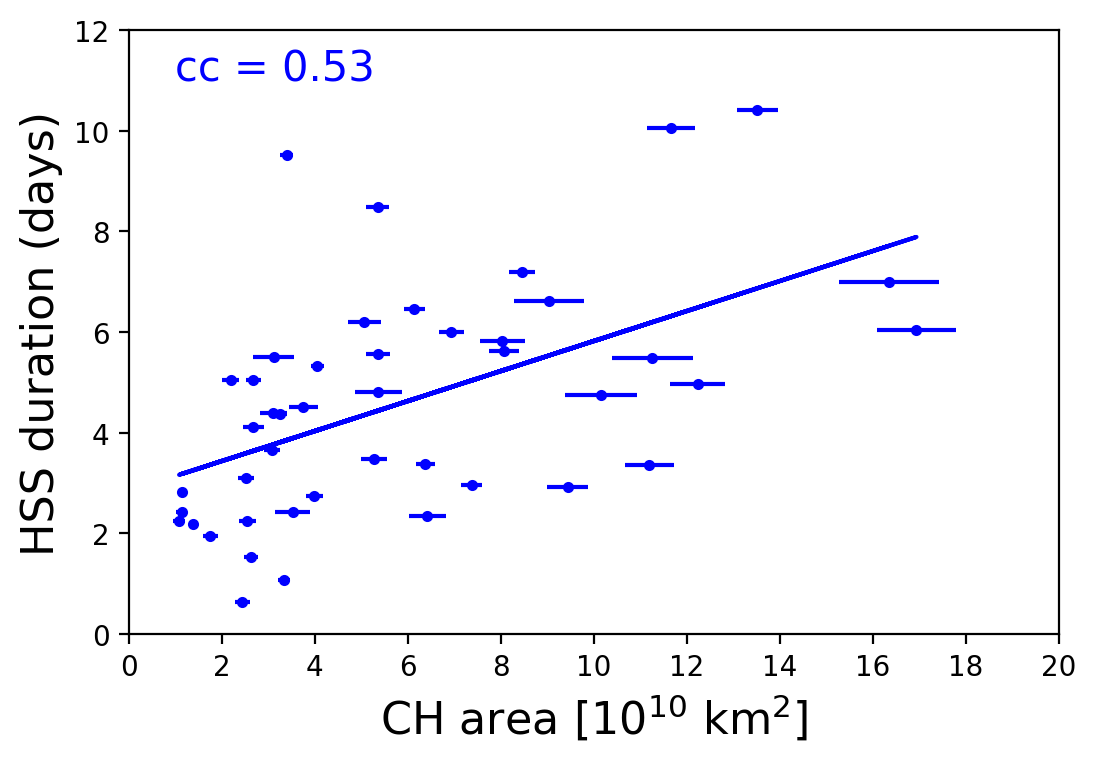}
\caption{}
\end{subfigure}
\hspace*{\fill}
\begin{subfigure}[]{0.3\linewidth}
\includegraphics[width=\linewidth]{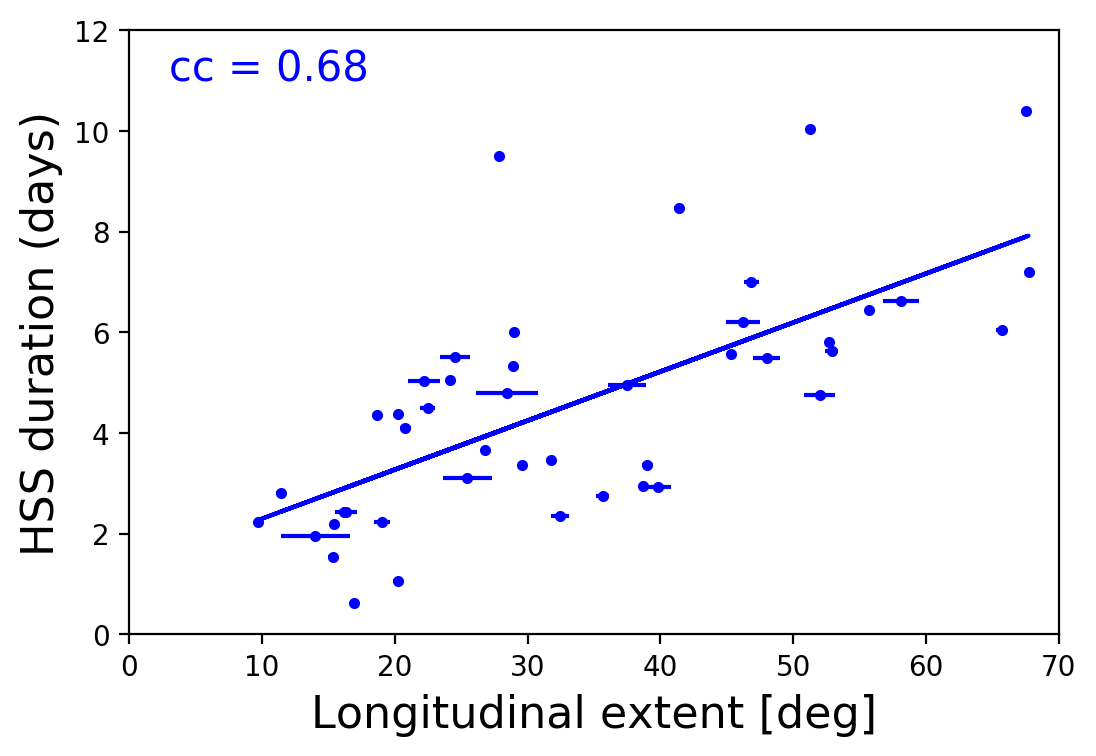}
\caption{}
\end{subfigure}%
\hspace*{\fill}
\begin{subfigure}[]{0.3\linewidth}
\includegraphics[width=\linewidth]{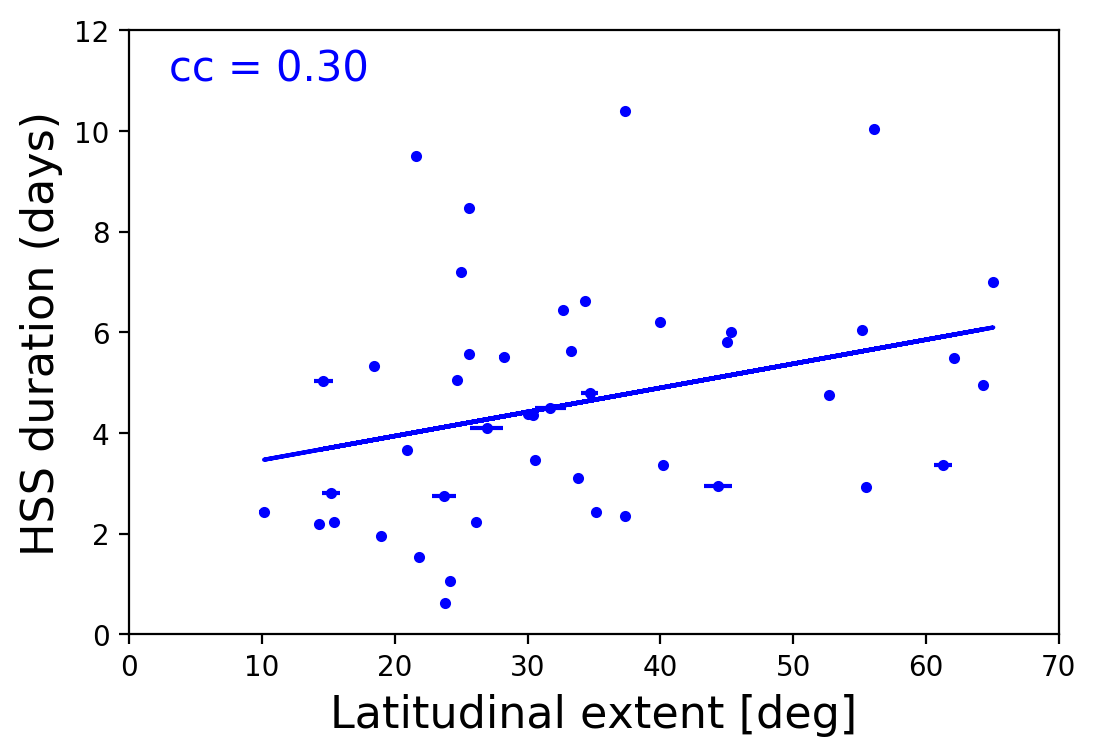}
\caption{}
\end{subfigure}%
\caption{Duration of the HSS recorded at Earth as a function of the CH area (a), CH longitudinal extent (b), and CH latitudinal extent (c).}
\label{Fig:Duration}
\end{figure*}

The associations between the HSS duration and the CH parameters are presented in Fig. \ref{Fig:Duration}. As expected, the HSS duration shows a good correlation with the CH longitudinal extent (cc = 0.68), that is CHs of larger longitudinal extents are the sources of longer lasting HSSs. Figure~\ref{Fig:Duration}c shows that there is a weak dependence of the HSS duration on the latitudinal extent (cc = 0.30). We also found a mediocre correlation (cc = 0.53) between the HSS duration and the CH area. This last correlation, in between that of the longitudinal and latitudinal extents, simply reflects the differences between them in correlating with the HSS peak speed.

\begin{figure*}[h!]
\centering
\begin{subfigure}[]{0.30\linewidth}
\includegraphics[width=\linewidth]{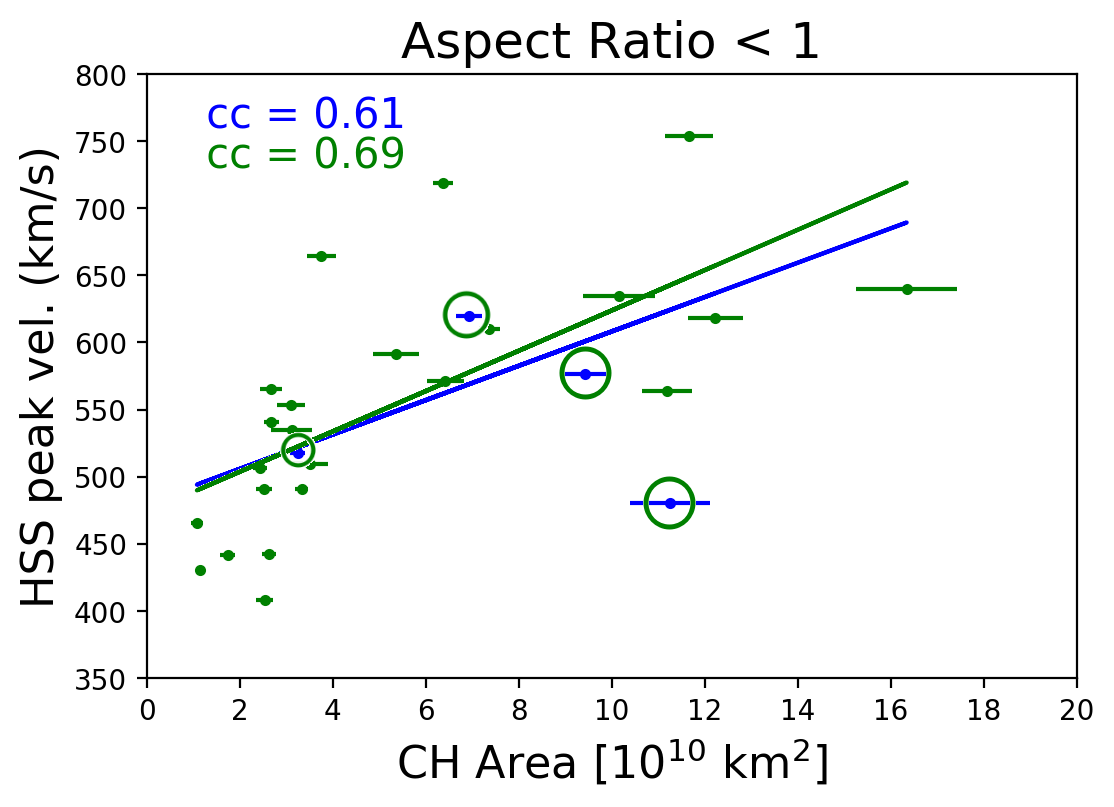}
\caption{}
\end{subfigure}
 \hspace*{\fill}
\begin{subfigure}[]{0.30\linewidth}
\includegraphics[width=\linewidth]{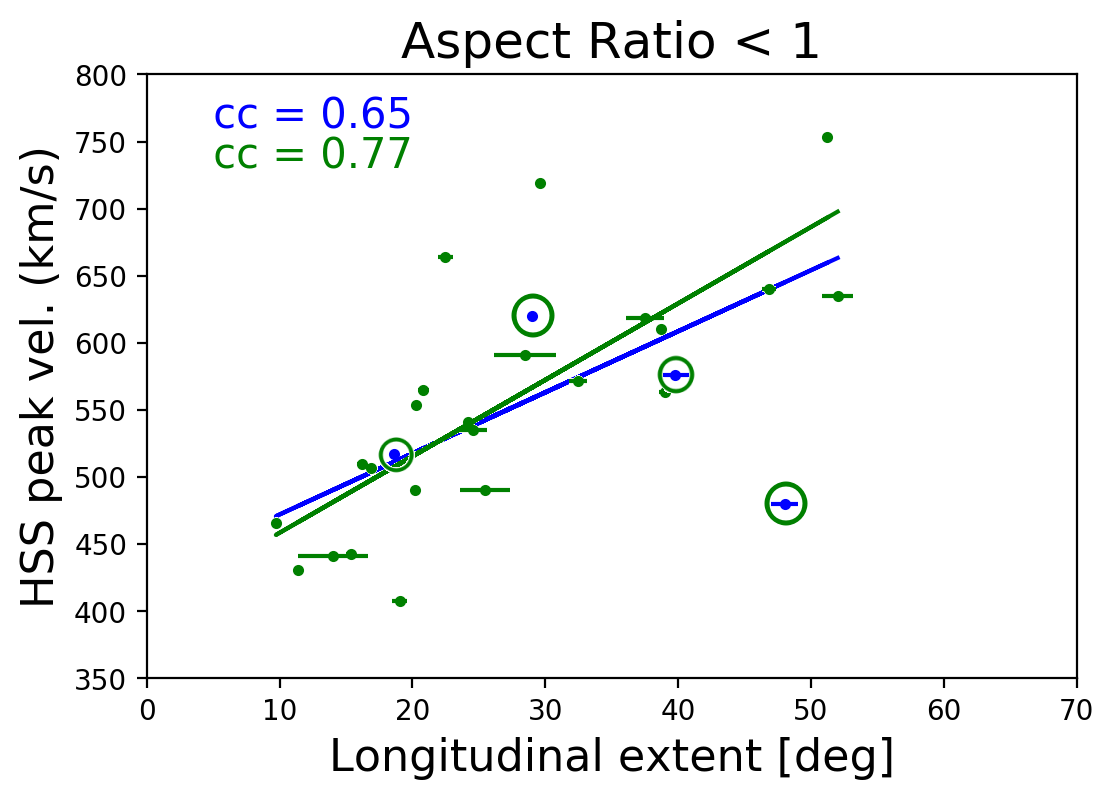}
\caption{}
\end{subfigure}%
 \hspace*{\fill}
\begin{subfigure}[]{0.30\linewidth}
\includegraphics[width=\linewidth]{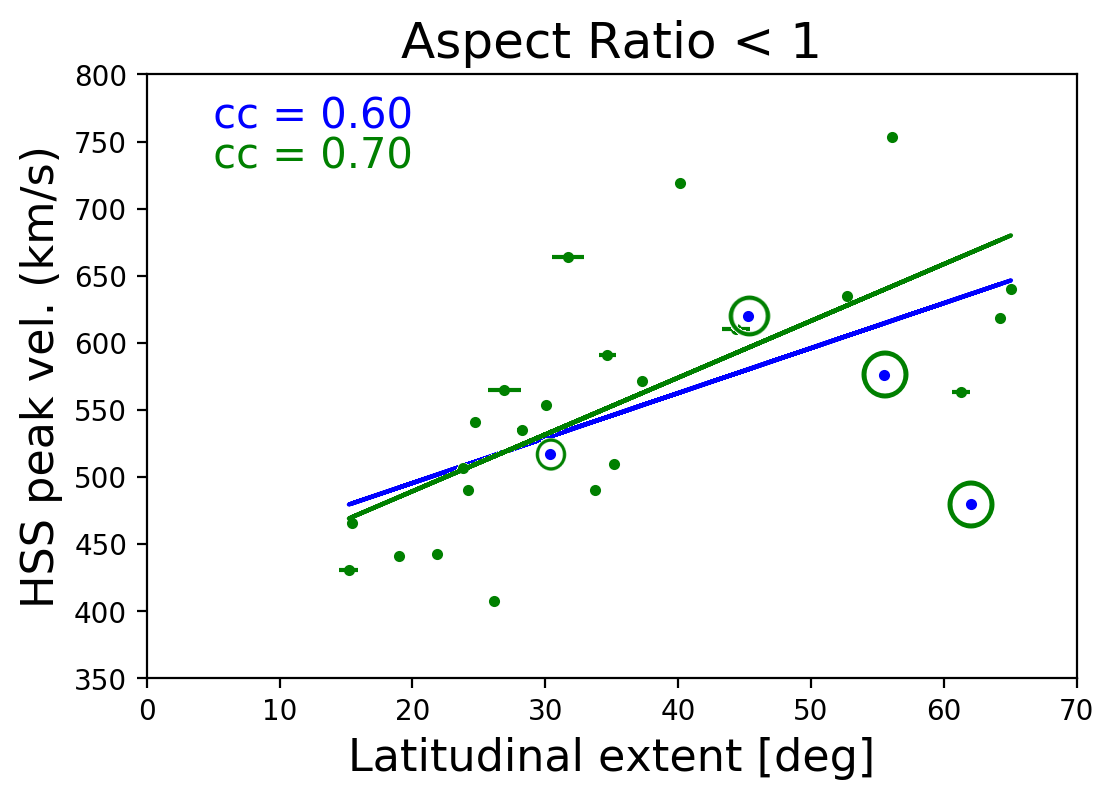}
\caption{}
\end{subfigure}%
\caption{HSS peak velocity as a function of the CH area (a), CH longitudinal extent (b), and CH latitudinal extent (c) for CHs with an aspect ratio $<$ 1. The blue line shows the linear fit applied for all CHs in the considered group. The green line shows the linear fit applied after we excluded the V-shaped and very irregular CHs, marked by green circles.}
\label{Fig:AspectRatio<1}
\end{figure*}

\begin{figure*}[h!]
\centering
\begin{subfigure}[]{0.30\linewidth}
\includegraphics[width=\linewidth]{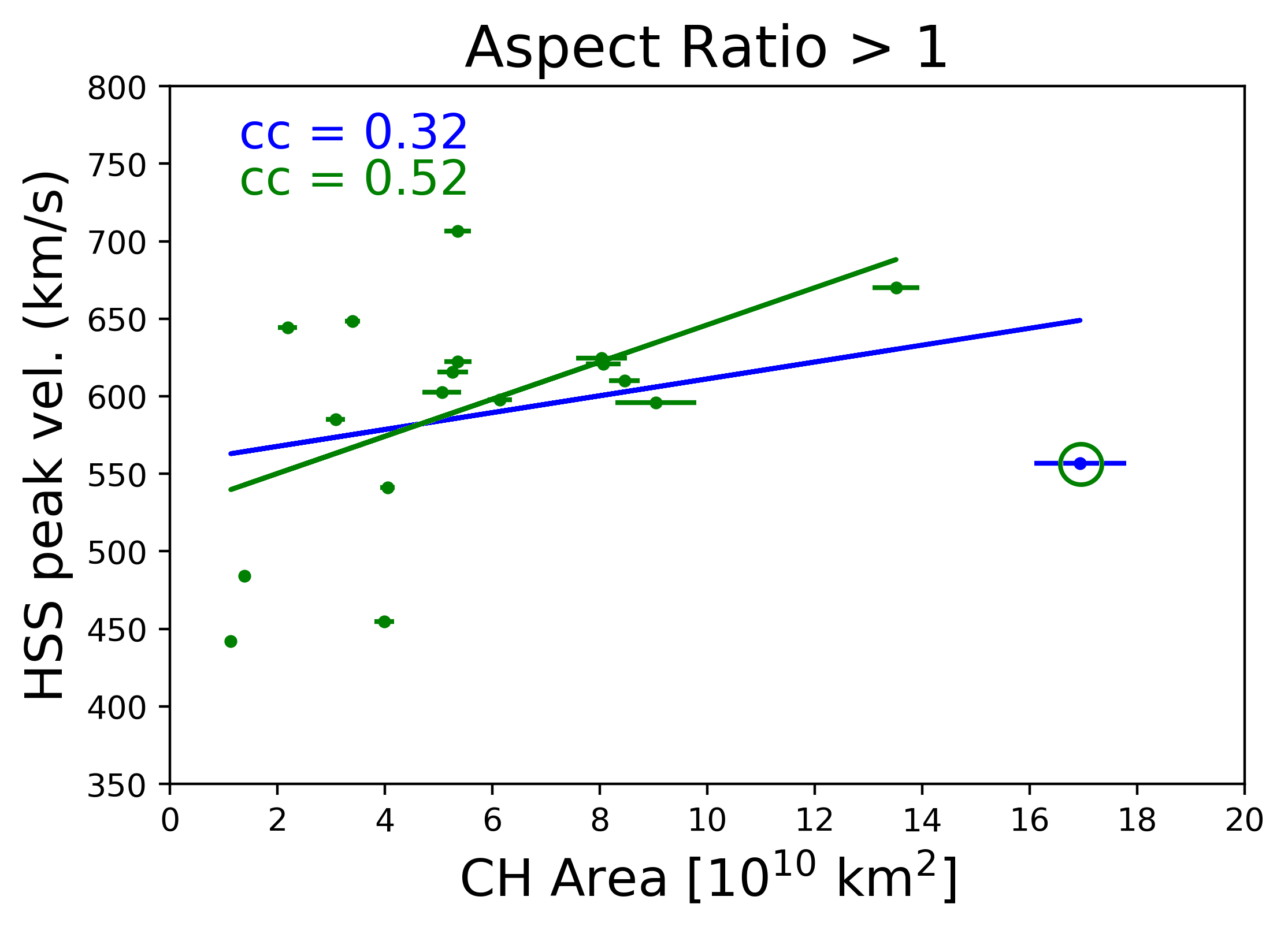}
\caption{}
\end{subfigure}
 \hspace*{\fill}
\begin{subfigure}[]{0.30\linewidth}
\includegraphics[width=\linewidth]{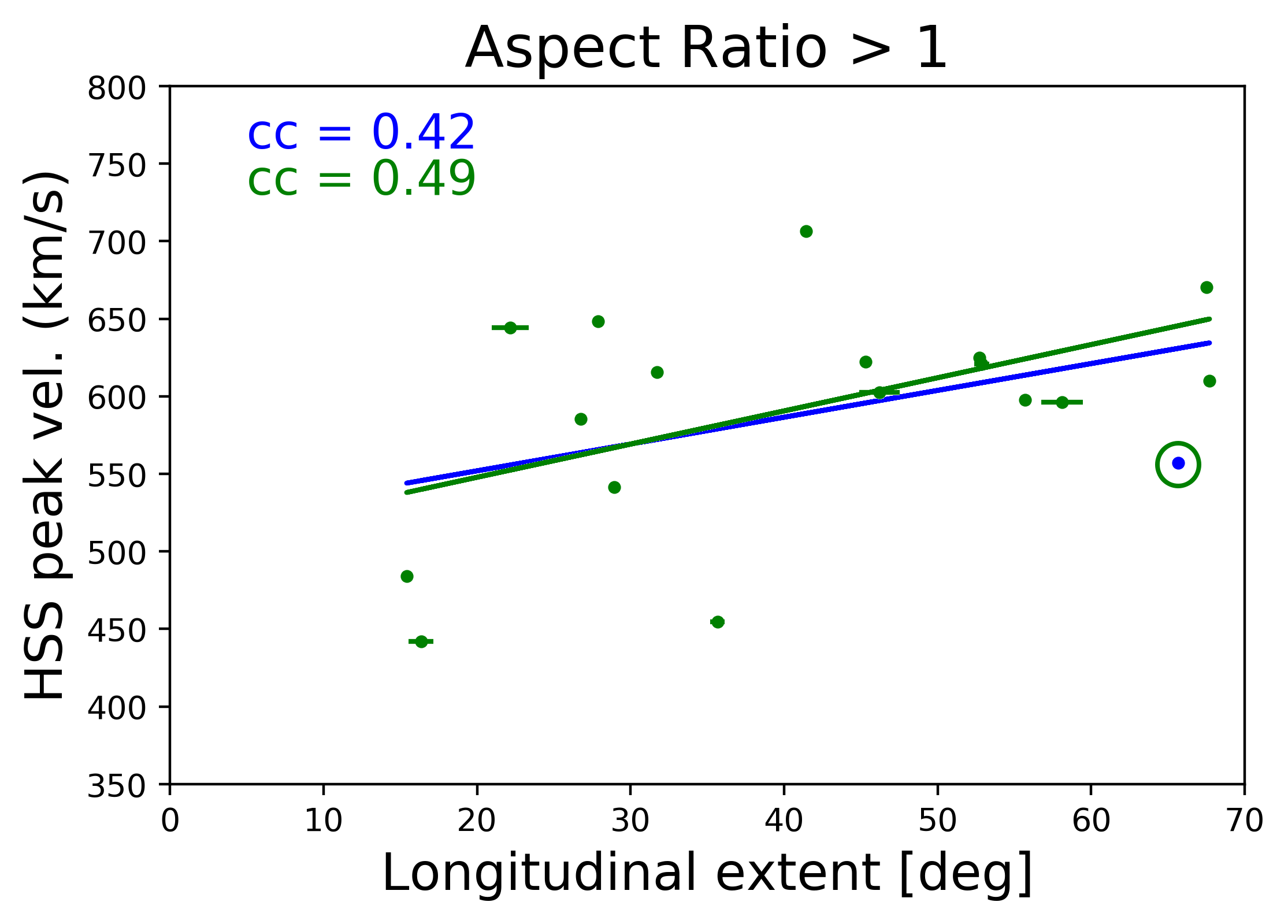}
\caption{}
\end{subfigure}
 \hspace*{\fill}
\begin{subfigure}[]{0.30\linewidth}
\includegraphics[width=\linewidth]{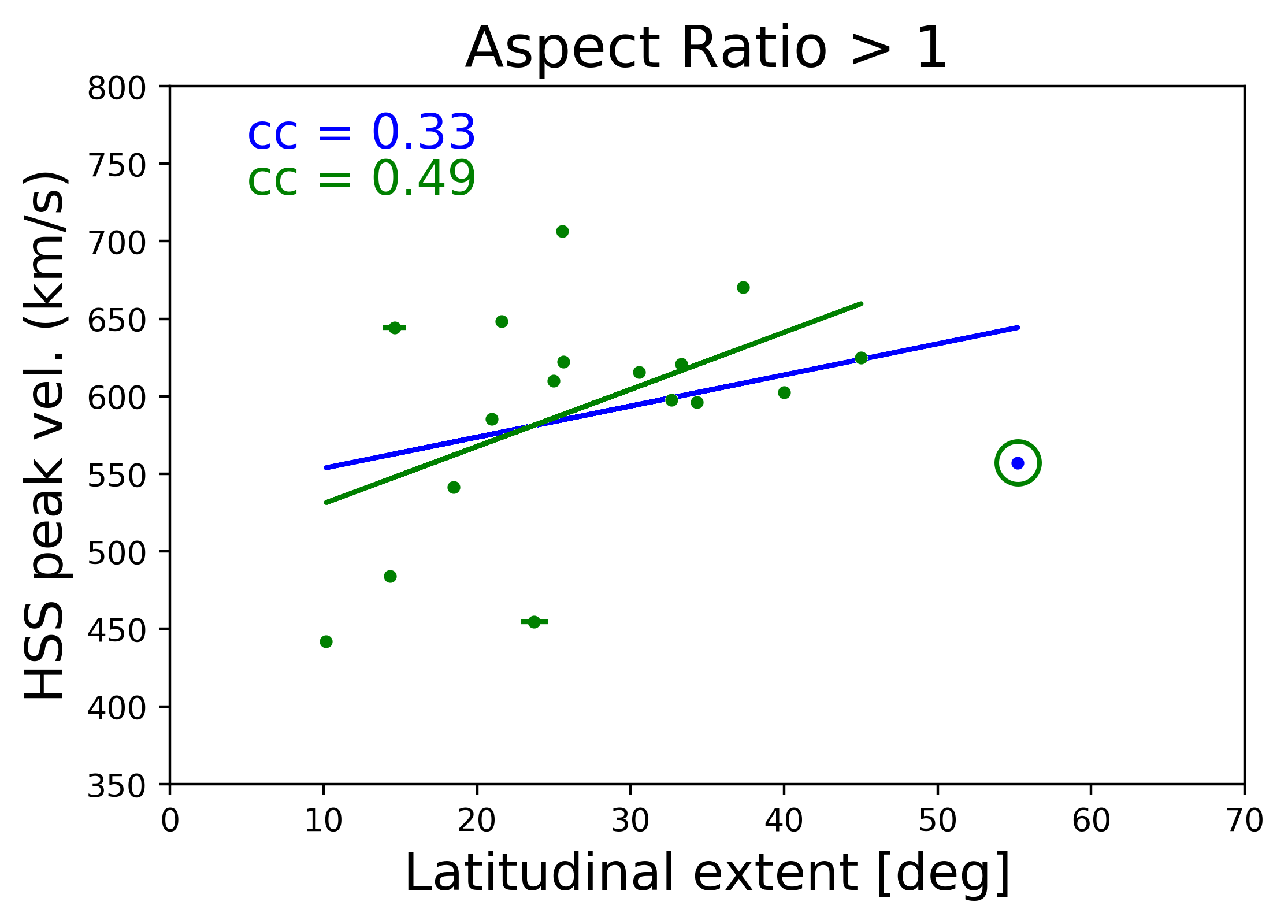}
\caption{}
\end{subfigure}%
\caption{HSS peak velocity as a function of the CH area (a), CH longitudinal extent (b), and CH latitudinal extent (c) for CHs with an aspect ratio $> 1$. The blue line shows the linear fit applied for all CHs in the considered group. The green line shows the linear fit applied after we excluded the V-shaped and very irregular CHs, marked by green circles.}
\label{Fig:AspectRatio>1}
\end{figure*}

\subsection{Categorizing the CHs in different groups}

\subsubsection{Aspect ratio}

The aim of this subsection is to categorize the CHs of our sample in different groups based on their geometric characteristics and to examine the associations with the HSS peak velocity. First, we categorize the CHs based on their aspect ratio, defined as the ratio of their longitudinal to their latitudinal extent. Two groups are formed. The first includes CHs with an aspect ratio lower than one (aspect ratio $< 1$) which are elongated more in latitude than in longitude. We nonetheless notice that, for the majority of the CHs in this group, the aspect ratio is close to one, meaning they have roughly circular or square shapes (see e.g., CHs 19, 20, 23 in Fig.~\ref{Fig:EUV_images}, \ref{Fig:EUV_images2}) or they are elongated and inclined with similar longitudinal and latitudinal extents (see e.g., CHs 26, 35 in Fig.~\ref{Fig:EUV_images2}). The second group contains all CHs with an aspect ratio greater than one (aspect ratio $> 1$). It includes CHs that have a larger longitudinal than latitudinal extent, and in fact we find aspect ratios significantly larger than one for the majority of cases.

In Fig.~\ref{Fig:AspectRatio<1} and \ref{Fig:AspectRatio>1}, the HSS peak velocity is shown as a function of the CH area, the CH longitudinal and CH latitudinal extent for each of the two groups. Five CHs from our sample (CH18, CH21, CH24, CH29, and CH31, marked by green circles in Fig.~\ref{Fig:AspectRatio<1} and Fig.~\ref{Fig:AspectRatio>1}) have very irregular shapes and were eventually excluded from the statistics. This is because four out of five of them have a triangular V-shape probably resulting from the differential solar rotation and they cannot be described by the aspect ratio rationale. The same is valid for the fifth CH (CH31) which is very complex and circular in shape. Excluding these structures eventually leads to a higher cc for both considered groups (see fittings in green in Fig.~\ref{Fig:AspectRatio<1} and \ref{Fig:AspectRatio>1}).

By comparing the results from the two groups separately, we conclude that the subset with an aspect ratio $< 1$ shows a better association between the maximum HSS velocity at Earth and the CH geometric parameters, compared to the subset with an aspect ratio $> 1$ (see Table \ref{Table:Summarized_cc}). Among the geometric parameters of the CH area, the longitudinal and latitudinal extent (Fig.~\ref{Fig:AspectRatio<1}a, b, c), the second is best correlated with the HSS peak velocity having a cc of 0.77. This correlation is better than the one obtained for the entire sample (cc = 0.57, see Fig.~\ref{Fig:long_lat_extent_vs_speed}a). The same is valid for the CH area and latitudinal extent (see Table~\ref{Table:Summarized_cc} for a direct comparison). Our results are confirmed by the bootstrapping technique. Table \ref{Table:bootstrapping_aspectratio} shows the median, as well as Q1 and Q3 values of the Pearson's cc distributions for each subgroup in Fig.~\ref{Fig:AspectRatio<1} and Fig.~\ref{Fig:AspectRatio>1} after the removal of the outliers and the redrawing of 100.000 data sets with replacement. The subset with an aspect ratio $< 1$ has higher Pearson's cc median values compared to the subset with an aspect ratio $> 1$ and the entire sample. Also, the CH longitude continues to be the best correlated quantity for that group, compared to the CH area and latitudinal extent.

Coronal holes with large latitudinal extents produce latitudinally extended HSSs, as well. Then, small variations in the latitudinal position of the measuring satellite within the HSS, during the lifetime of the CH, are not important compared to HSS cases that are latitudinally restricted (e.g., HSSs originating from CHs with an aspect ratio > 1, see Fig.~\ref{Fig:AspectRatio>1}). Therefore, the larger the latitudinal extent of a CH, the higher the probability is of the associated HSS being detected by the measuring satellite. In the case of an aspect ratio > 1, variations in the satellite's latitudinal position can have a significant effect on how close it is to the flanks of the HSS. These variations result in a larger scattering of the points and, therefore, a worse cc (comparison between Fig.~\ref{Fig:AspectRatio<1} and Fig.~\ref{Fig:AspectRatio>1}). Drawing from the aforementioned information and the discussion in section \ref{subsection:CHlong_lat}, we distinguish between two cases: CHs with large latitudinal extents and very restricted longitudes (aspect ratio $\ll$ 1) and CHs with large latitudinal extents and comparable longitudes (aspect ratio $\lessapprox$ 1), both of which being relevant to the former CH group with an aspect ratio $<$ 1. In the former case, an observer likely does not see the maximum velocity of the HSS because all fast plasma parcels would have impinged into the SIR (see discussion in Section \ref{subsection:CHlong_lat}), regardless of the latitudinal position of the measuring satellite. In the latter case, an observer is able to observe velocities close to the maximum velocity of the HSS because of both the adequately extended longitudinal width of the CH and also the high probability of the measuring satellite detecting the part that has originated from the CH core region (fastest parcels). This explains why we obtain better correlations in the first group (aspect ratio $\lessapprox$ 1), rather than in the second group (aspect ratio $>$ 1).

\subsubsection{Orientation}
\label{section:Orientation}

\begin{figure}[h!]
\centering
\begin{subfigure}[]{0.95\linewidth}
\includegraphics[width=\linewidth]{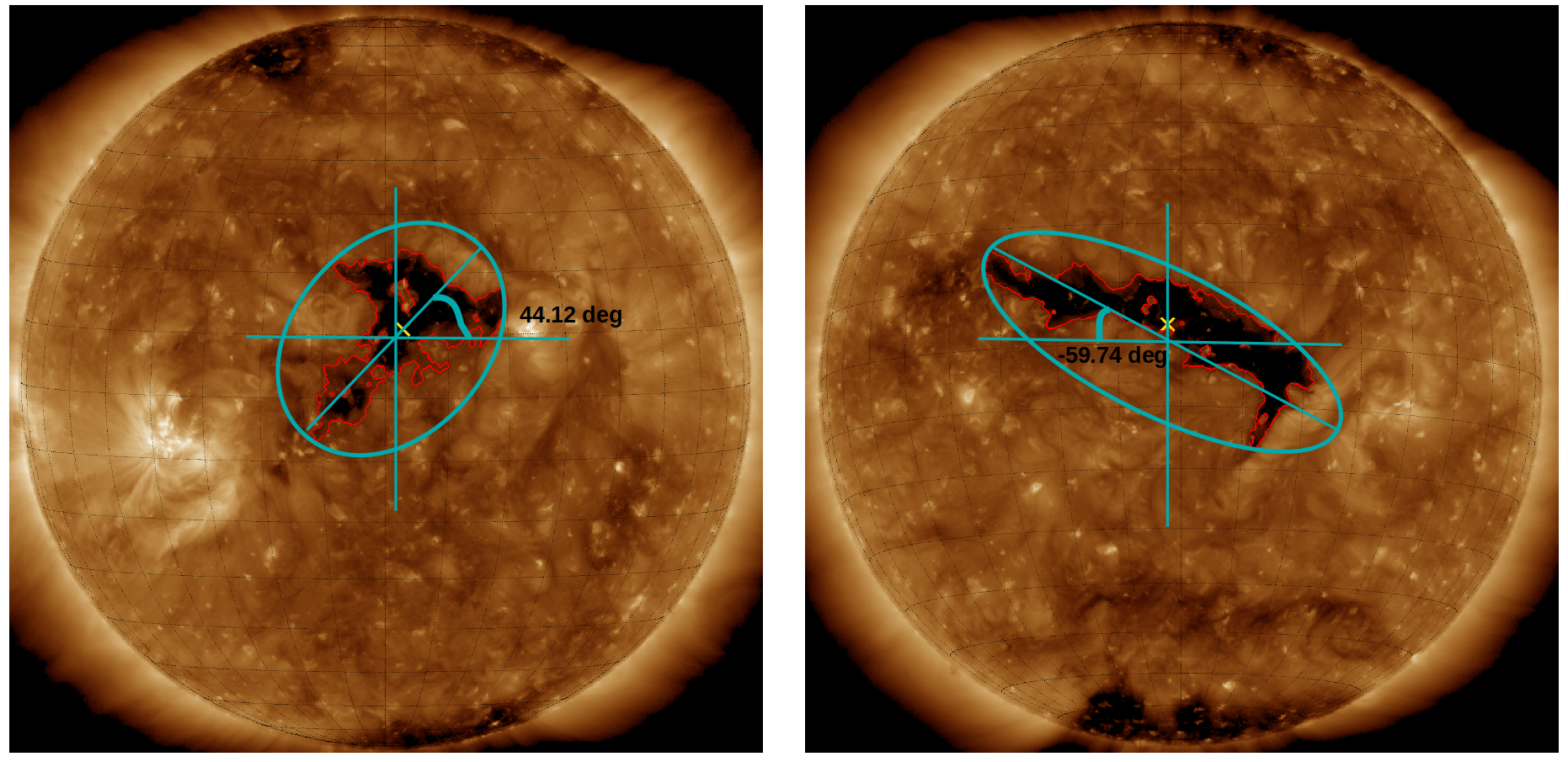}
\end{subfigure}
\caption{Definition of the CH orientation angle with respect to a $x$-axis parallel to the solar equator. Left: Positive orientation. Right: Negative orientation.}
\label{Fig:OrientationAngle}
\end{figure}

The majority of the CHs in our sample have a pronounced orientation with respect to the solar equator. Therefore, we fit an ellipse to each of them and estimated the angle of its major axis with respect to the solar equator (see Fig.~\ref{Fig:OrientationAngle}). Due to its simplicity, an ellipse is arguably the best shape to approximate the very variable shapes of studied CHs. In this way, we obtained two different CH groups. The first, is composed of CHs with a positive angle, namely a northward-oriented leading edge (see Fig.~\ref{Fig:OrientationAngle}a). The second includes CHs with negative angles with respect to the equator, namely a southward-oriented leading edge (see Fig.~\ref{Fig:OrientationAngle}b). We did not use the location at different solar hemispheres as a criterion for this categorization. Nevertheless, we noticed a posteriori that the majority of CHs with positive orientation angles were situated in the southern hemisphere with their leading edge oriented toward the equator and their trailing edge toward the south pole. Similarly, CHs with negative orientation angles with respect to the equator are mostly located in the northern hemisphere with the leading edge toward the equator and the trailing edge toward the north pole. This is likely an effect of the solar differential rotation \citep[][]{heinemann18}.

\begin{figure*}[h!]
\centering
\begin{subfigure}[]{0.30\linewidth}
\includegraphics[width=\linewidth]{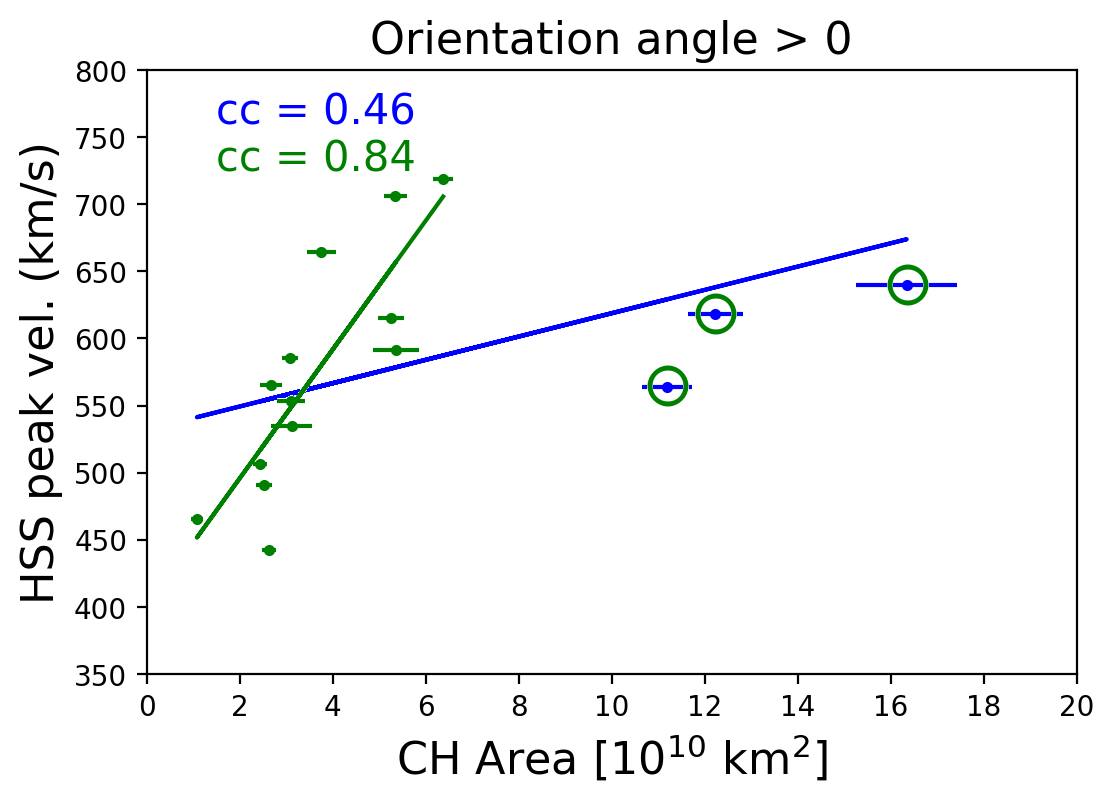}
\caption{}
\end{subfigure}
 \hspace*{\fill}
\begin{subfigure}[]{0.30\linewidth}
\includegraphics[width=\linewidth]{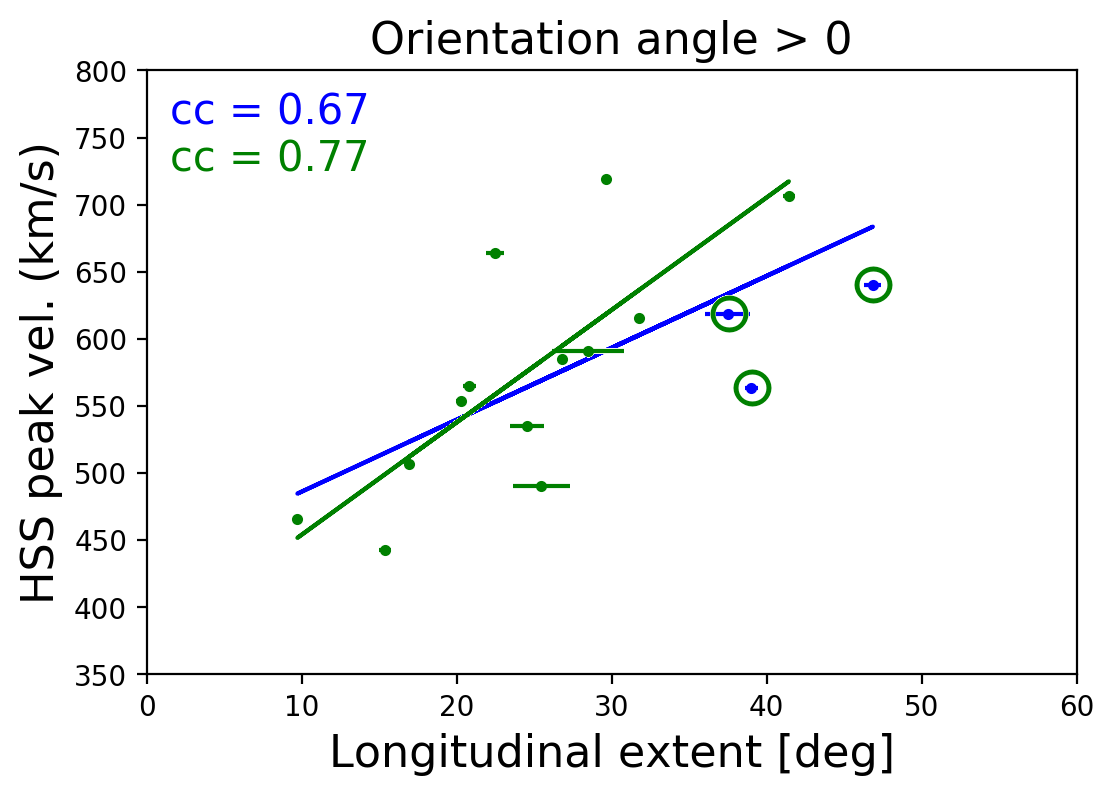}
\caption{}
\end{subfigure}%
 \hspace*{\fill}
\begin{subfigure}[]{0.30\linewidth}
\includegraphics[width=\linewidth]{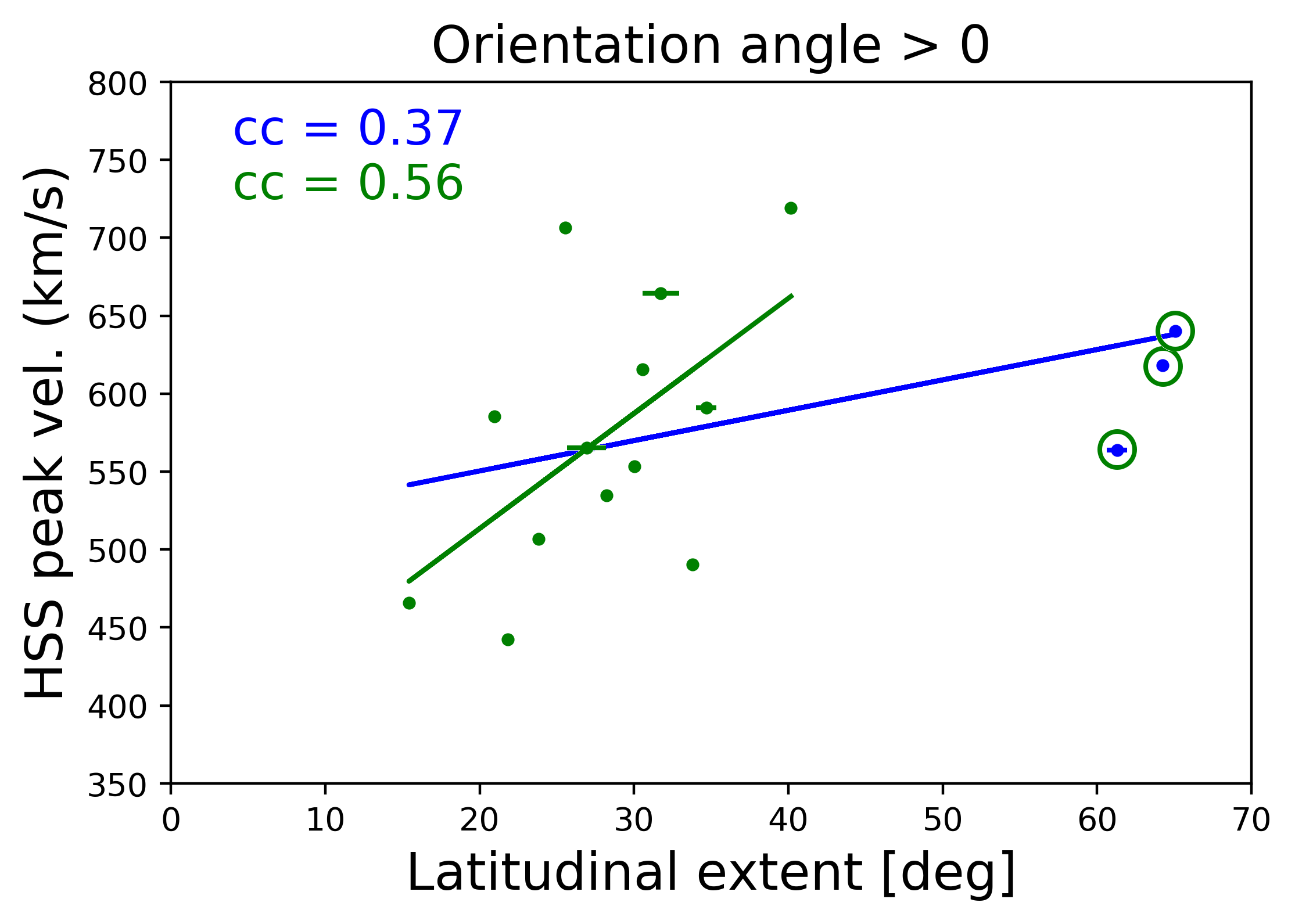}
\caption{}
\end{subfigure}%
\caption{HSS peak velocity as a function of the CH area (a), CH longitudinal extent (b), and CH latitudinal extent (c) for CHs with a positive orientation angle. The blue line shows the linear fit applied for all CHs in the considered group. The green line shows the linear fit applied after we excluded the three very irregular CHs marked in green.}
\label{Fig:Plots_Angle>0}
\end{figure*}

\begin{figure*}[h!]
\centering
\begin{subfigure}[]{0.30\linewidth}
\includegraphics[width=\linewidth]{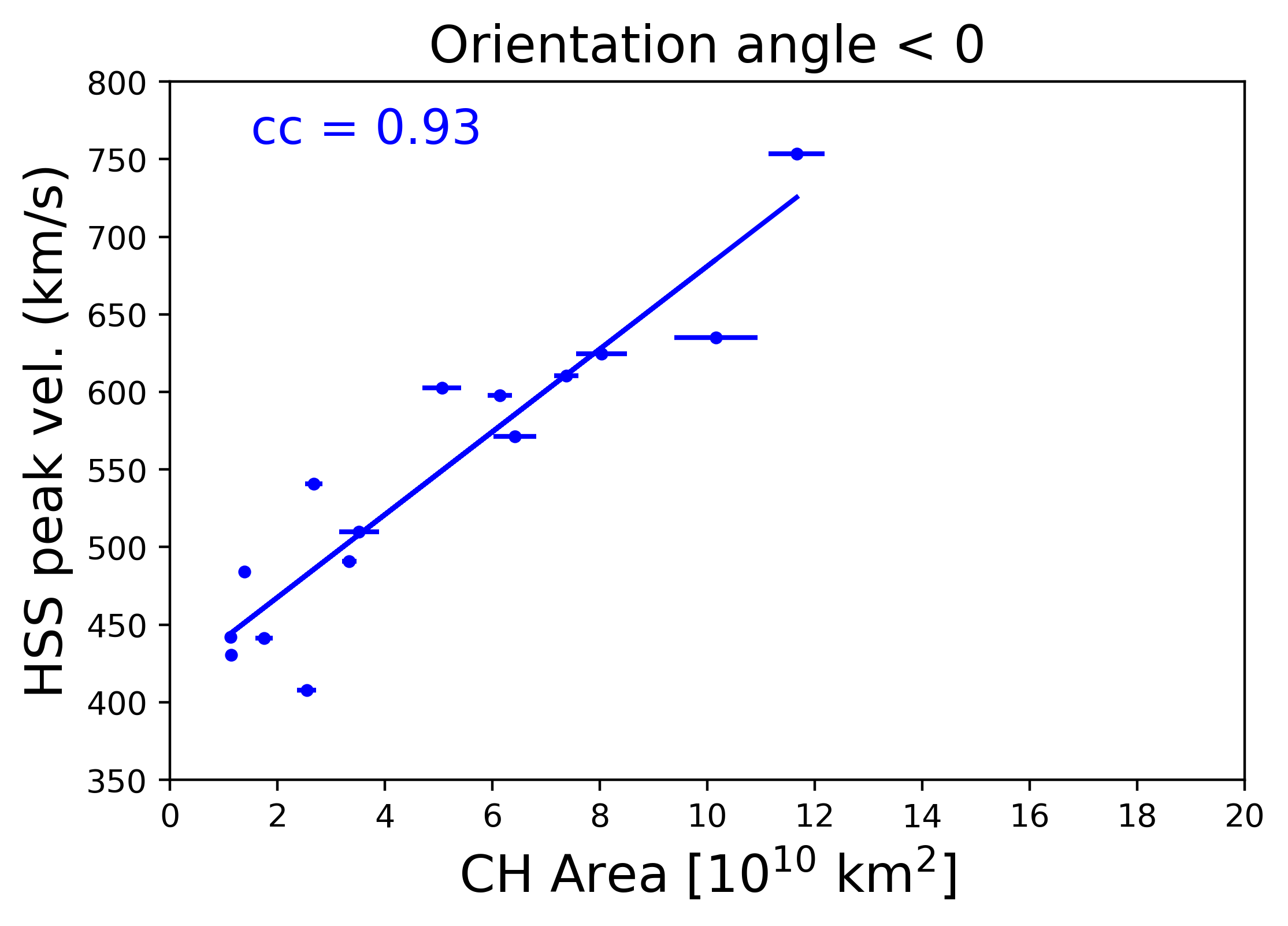}
\caption{}
\end{subfigure}
 \hspace*{\fill}
\begin{subfigure}[]{0.30\linewidth}
\includegraphics[width=\linewidth]{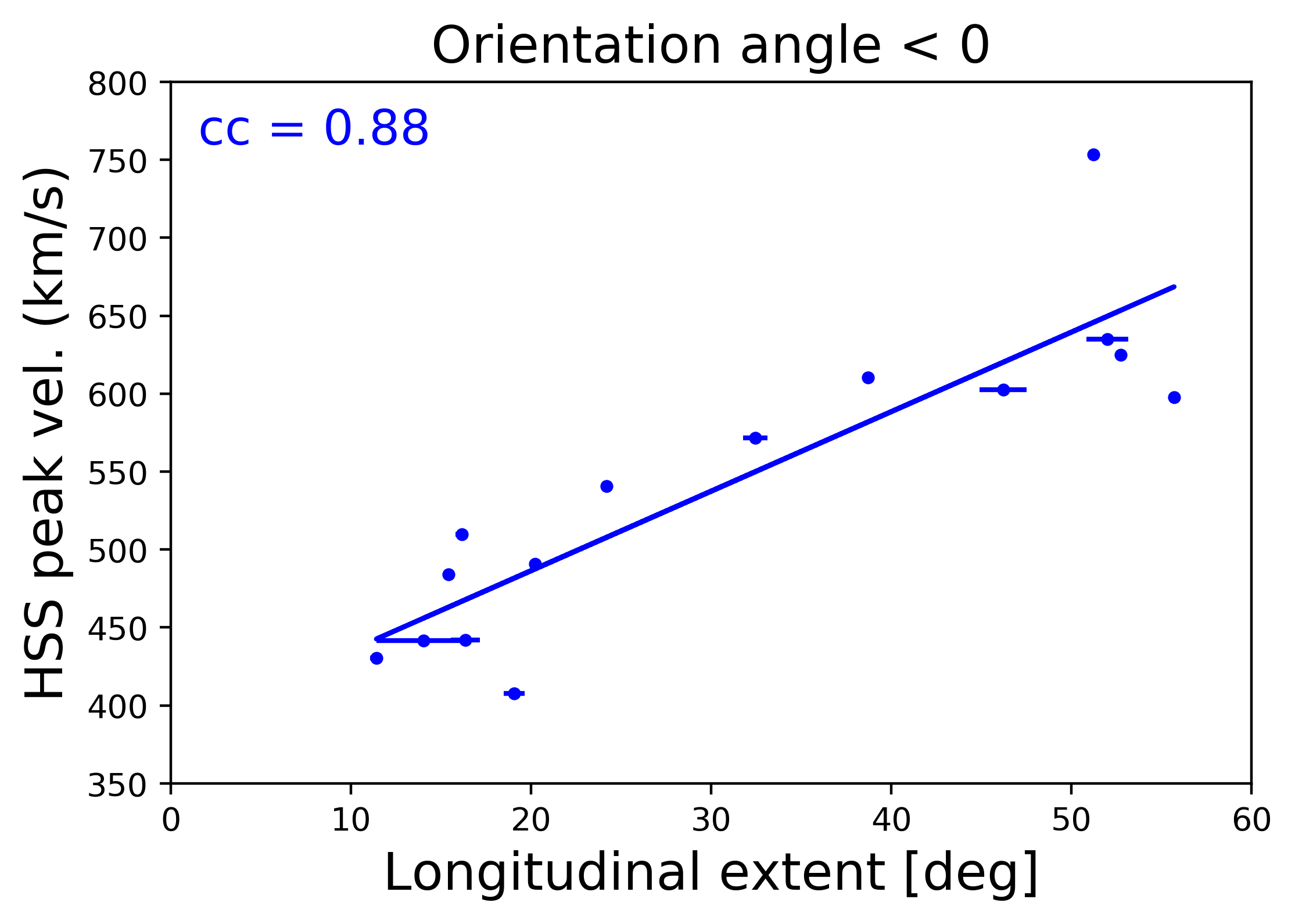}
\caption{}
\end{subfigure}%
 \hspace*{\fill}
\begin{subfigure}[]{0.30\linewidth}
\includegraphics[width=\linewidth]{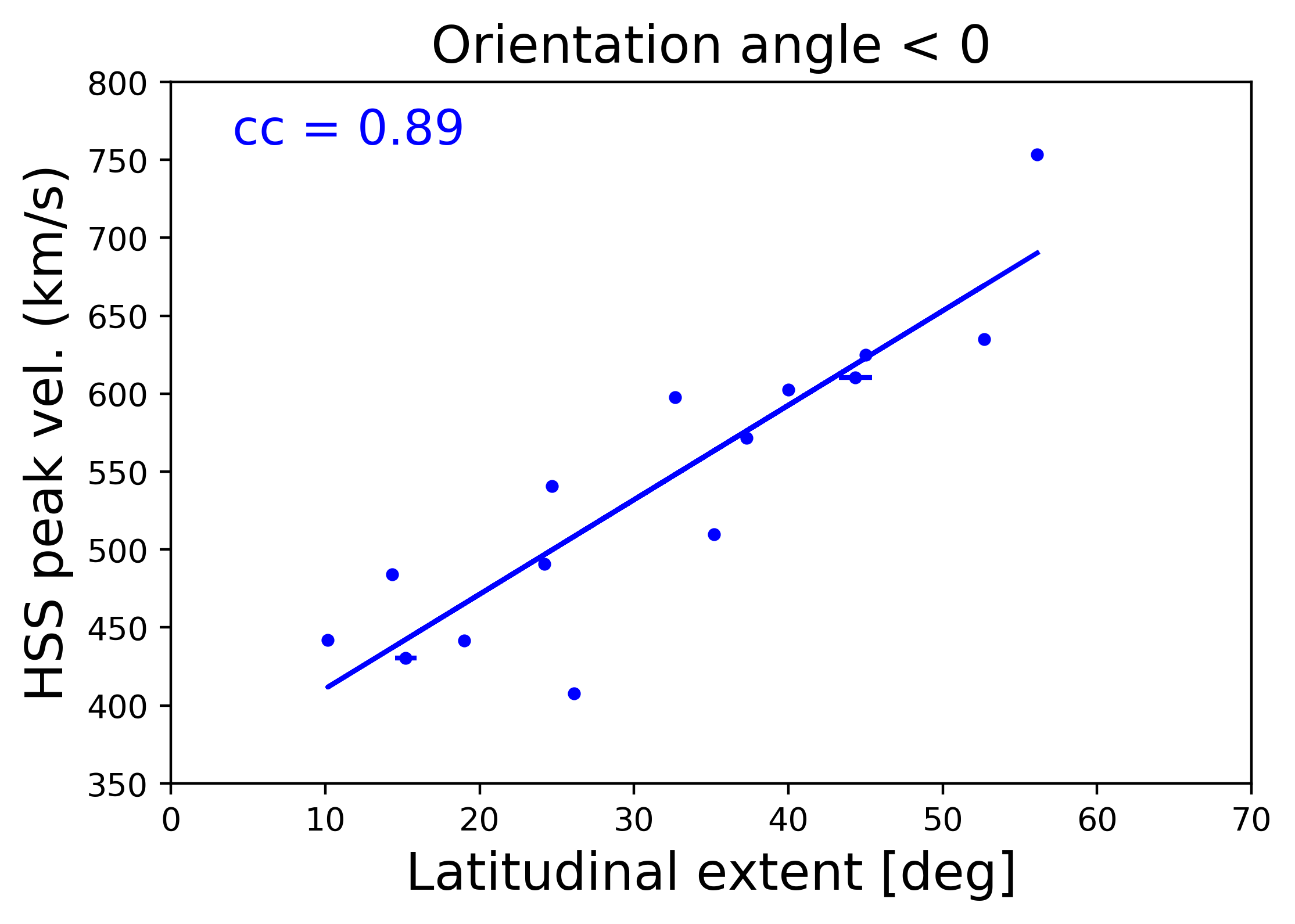}
\caption{}
\end{subfigure}%
\caption{HSS peak velocity as a function of the CH area (a), the CH longitudinal extent (b), and the CH latitudinal extent (c) for CHs with a negative orientation angle. }
\label{Fig:Plots_Angle<0}
\end{figure*}

For the five irregular CH cases mentioned in the previous subsection (CH18, CH21, CH24, CH29, and CH31), estimating their orientation angle was not possible due to their very peculiar V- or roundish shape which did not allow the fitting of a single ellipse with clearly defined major and minor axes. We also did not consider nine additional CHs with an approximately parallel axis with respect to the equator (e.g., CHs with very small orientation angles, either ranging from -20 to 0 deg or from 0 to 20 deg). These are CH3, CH4, CH5, CH7, CH8, CH9, CH12, CH34, and CH41\footnote{If these CHs stay in both subgroups, they change the cc in the HSS peak speed and CH area relationship to 0.84 (in the case of CHs with an orientation angle $ <$ 0) and 0.40 (in the case of CHs with an orientation angle $>$ 0). See Fig.~\ref{Fig:Plots_Angle>0} and Fig.~\ref{Fig:Plots_Angle<0} for more details.}. On the contrary, the vertically oriented CHs were taken into account (eight CHs, with angles between -90 to -70 deg and 70 to 90 deg). The latter structures form the biggest angles with respect to the solar equator, and they correspond to structures with an aspect ratio $<$ 1. We know from the previous subsection that their extended latitudes permit the elimination of errors in the position of an observer that encounters the HSS and, as a result, a good correlation is expected from them, as long as their longitudes are sufficiently large.

The correlations for the groups with positive and negative orientation angles are presented in Fig.~\ref{Fig:Plots_Angle>0} and \ref{Fig:Plots_Angle<0}. For structures with a positive orientation angle, the correlations between the HSS maximum velocity at Earth and the CH area, longitudinal, and latitudinal extent are cc = 0.46, cc = 0.67, and cc = 0.37, respectively. These associations are much poorer than in the case of CHs with a negative orientation angle for which we estimated a cc = 0.93, cc = 0.88, and cc = 0.89, respectively. This happens because of the three outliers in the former group, marked by green circles in Fig.~\ref{Fig:Plots_Angle>0}, which have CH areas higher than 10 $\cdot$ 10$^{10}$ km$^{2}$. These points correspond to CHs 22, 25, and 27 which are complex CHs with an irregular shape and they are composed of pronounced ``branches'' with different orientation directions (see Fig.~\ref{Fig:EUV_images2}). The estimation of their orientation angle when fitting an ellipse could not provide a structure with clearly defined major and minor axes, thus, the three CHs were excluded from our sample. Excluding these points substantially improved the correlations, with a cc of 0.84, 0.77, and 0.56 for the CH area, the CH longitudinal and latitudinal extent, respectively (see Fig.~\ref{Fig:Plots_Angle>0}). 

By comparing the correlations between the entire sample and the different orientation subgroups (see summarized Table~\ref{Table:Summarized_cc}), we conclude that the orientation angle is a significant geometric factor that should influence the way a HSS encounters Earth. We believe that this parameter should be taken into account for empirical relationships and models that aim to predict the fast solar wind based on the CH properties on the Sun. Our claim is further supported from the results of the bootstrapping technique after redrawing 100.000 data sets with replacement, and removing the outliers presented in Fig.~\ref{Fig:Plots_Angle>0} and Fig.~\ref{Fig:Plots_Angle<0}. In Table \ref{Table:bootstrapping_aspectratio} (fourth and fifth rows), we see that the median values of Pearson's cc for both subgroups of the orientation angle show much higher correlations compared to what we obtain for the entire sample (first row of the same table).

\section{Geometric complexity}

In the process of modeling and forecasting the HSS arrival, one of the frequently ignored parameters is the CH geometric complexity which had not been systematically discussed in the literature before. In this section, we quantify the geometric complexity of the CHs in the following two different ways: by taking the ratio of the maximum rectangle that can be inscribed within the CH, over the convex hull area of the CH, and by calculating the fractal dimension (FD) of the CHs. We note that the first way is simpler and more approximate. The second way is more precise since fractality estimates the self-similarity of a structure over different scales. To our knowledge, \citet[][]{Abramenko09} is the only other work that elaborates on the fractality of CHs. Nevertheless, their work is focused on the complexity of the CH magnetic fields, not their morphological structure. They claimed that the absence of the necessary complexity of the CH magnetic field seems to explain the noncorrelation between the solar wind speed and the net magnetic flux density, as derived from observations.

\citet{reiss16} were the first to mathematically introduce the parameter of ``compactness'' as a measure to quantify how irregular the shape of a CH is and distinguish among CHs and filament channels. Based on shape analysis, compactness ($C$) is defined as follows: 

\begin{equation}
C = U^{2}/4 \pi A,
\label{compactness_Reiss}
\end{equation}

\noindent where $U$ denotes the perimeter of the structure and $A$ is the CH area. The definition by \citet{reiss16} is not applicable to this study because it does not take into account how hollow the internal part of the CH is. The meaning of CH complexity should not be restricted to the irregularity of the outer boundary of the structure, but further include the internal part of it. This area is not always compact and it may well influence the maximum HSS velocity. Therefore, we sought a more encompassing measure to quantify the geometric complexity of the CHs by considering both the fragmentation in the interior and the raggedness along the boundary.

\subsection{Solidity: The ratio of the maximum inscribed rectangle over the convex hull area}
\label{section:Ratio}

\begin{figure}[h!]
\centering
\begin{subfigure}[]{0.415\linewidth}
\includegraphics[width=\linewidth]{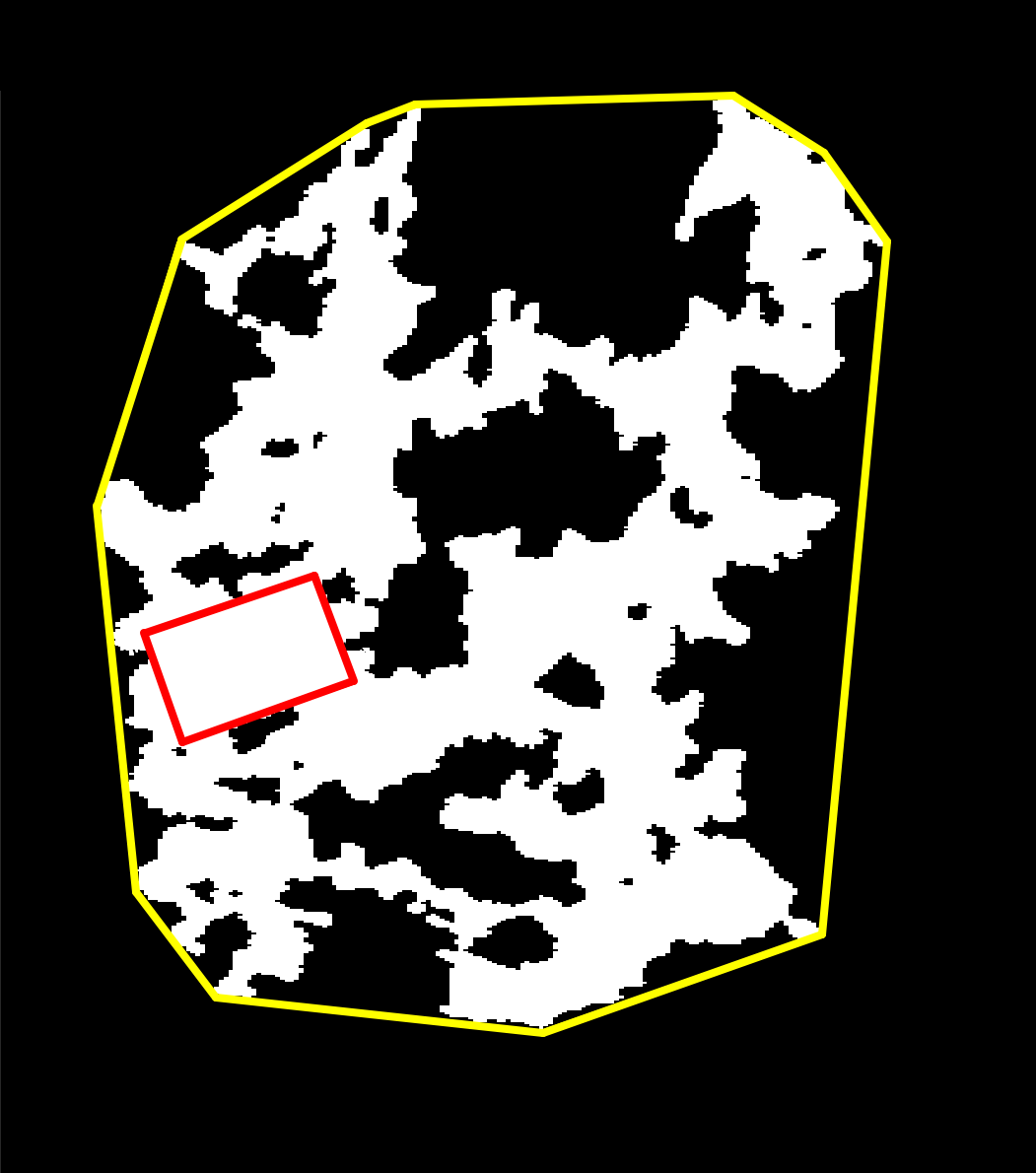}
\caption{}
\end{subfigure}
 %\hspace*{\fill}
\begin{subfigure}[]{0.435\linewidth}
\includegraphics[width=\linewidth]{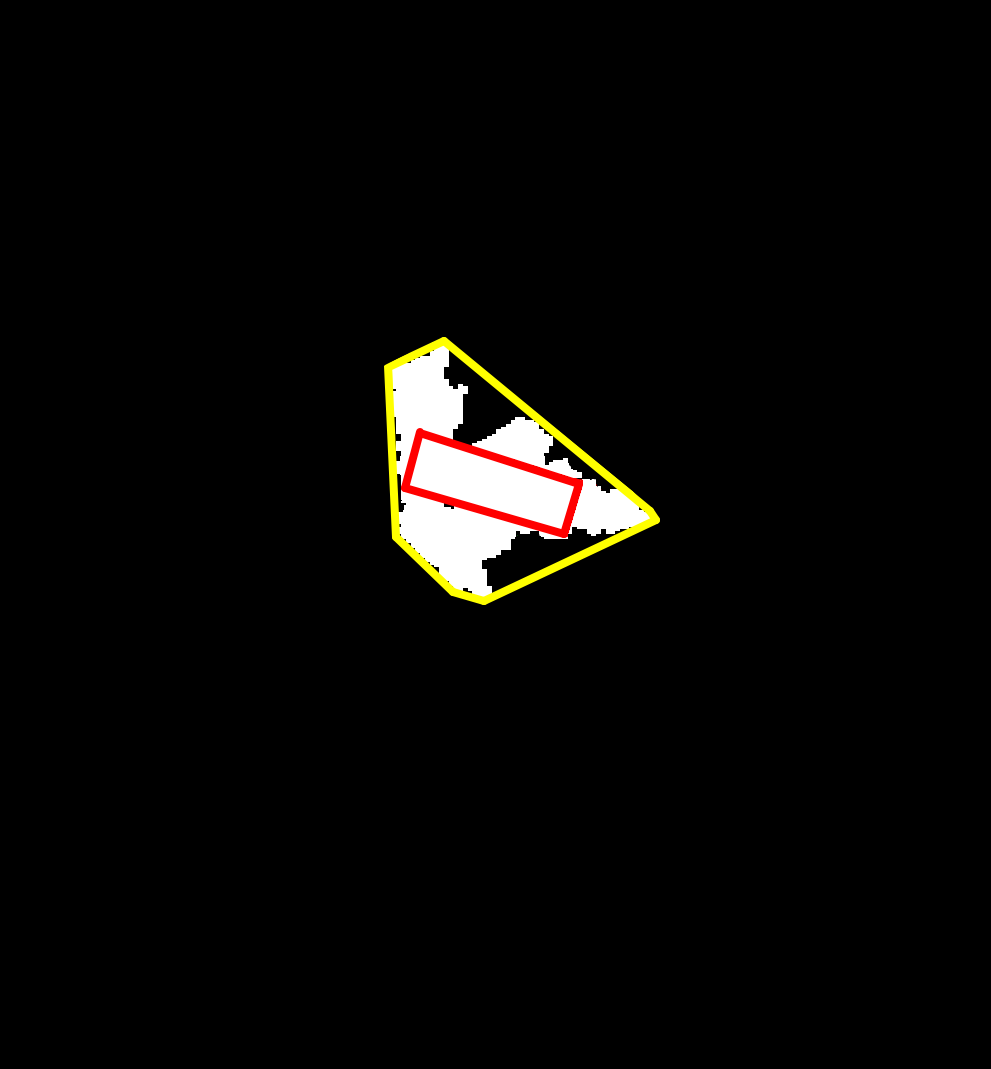}
\caption{}
\end{subfigure}%
\caption{Examples of the maximum inscribed rectangle in the interior part of CHs (red box) and the convex hull area (the area included inside the yellow lines of each structure). Panel (a): Very fragmented and patchy CH with a maximum inscribed rectangle = 1.08 $\times$ 10$^{10}$ km$^{2}$, a convex hull area = 31.69 $\times$ 10$^{10}$ km$^{2}$, and Solidity = 0.03. Panel (b): More compact CH with a maximum inscribed rectangle = 0.49 $\times$ 10$^{10}$ km$^{2}$, a convex hull area = 1.87 $\times$ 10$^{10}$ km$^{2}$, and Solidity = 0.26.}
\label{Fig:Ratio}
\end{figure}

The first, more approximate way to quantify the CH complexity is by estimating the ratio of the maximum rectangle that can be inscribed within the interior of a CH, over the convex hull area of that CH. The convex hull area envelops all line segments that join any two points of the edge of the structure. We refer to this parameter as ``Solidity''. Figure~\ref{Fig:Ratio} shows a small (Fig.~\ref{Fig:Ratio}a) and large (Fig.~\ref{Fig:Ratio}b) value of the Solidity, corresponding to a fragmented and compact CH, respectively. The maximum inscribed rectangle (red rectangle in Fig.~\ref{Fig:Ratio}) is an approximation of the biggest, most compact area found inside a CH. As a result, this quantity should be a good approximation of the area from which the fastest part of the solar wind originates. To calculate the maximum inscribed rectangle within the CH, we used the MATLAB function from \citet{MaxInscribedRect_MATLAB}. We chose a rectangular shape instead of a circular or square one, as it accounts for the maximum area of the compact part of the CH 
better. On the other hand, the convex hull area (the area included inside the yellow lines shown in Fig.~\ref{Fig:Ratio}) provides a quantification of the smallest convex set that contains all points of our structure. The more irregular the boundary of a CH is, the larger the convex hull area becomes. 
Therefore, for a Solidity close to 1, the CH is compact enough to have its edges close to its main body (Fig.~\ref{Fig:Ratio}b), whereas a Solidity much smaller than 1 implies a very fragmented, inherently hollow, structure (Fig.~\ref{Fig:Ratio}a).
%Therefore, for a Solidity close to 1, we have a maximum inscribed rectangle approximately equal to the convex hull area. That means that the CH of interest is compact and has its edges close to its main body (see Fig.~\ref{Fig:Ratio}b). On the other hand, a Solidity close to 0 means that the maximum inscribed rectangle fitted in the interior of the CH is much smaller than the convex hull area of the CH. In that case, the studied CH is a very fragmented structure (see Fig.~\ref{Fig:Ratio}a). 

\subsection{CHs as fractal objects}
\label{section:Fractals}

Another way of quantifying the complexity of a CH area is by calculating the FD of its contours (both internal and external). The fractal analysis is a mathematical method for measuring complexity in nature \citep{Rajkovic2017}, and it has been long used in different scientific domains \citep[see e.g.,][]{Smith1996}. The complexity and FD of an object increases as the object's border is more ragged; the branching pattern is more irregular and lines get more twisted \citep{Milosevic2013}. Fractality characterizes the self-similarity of a structure over different scales \citep{Falconer1990, Konatar2020}. The landmark book of \citet[][and references therein]{Mandelbrot1983} has shown that the boundary length of a fractal object can be mathematically expressed as a power law of the  following form: 

\begin{equation}
N(r) = \hbox{\rm const} \cdot r^{-D},
\label{fractal_equation}
\end{equation}

\noindent where $N(r)$ presents a measure of the fractal structure at scale $r$. The parameter $D$ is called ``fractal'' because it is not an integer number and is called a ``dimension'' because it provides a measure of how completely an object fills the space \citep{Fernandez2001}. If $D$ is an integer number, then it is equal to the Euclidean dimension in which an ideal point has dimension 0, a line has dimension 1, a plane has dimension 2, and an ideal solid volume has dimension 3. Applying this to our CH maps, which are not straight lines and do not cover the whole two-dimensional space, the FD is in the range $ 1 < D < 2$ \citep{Fernandez2001}.

A number of methods have been developed to estimate the FD of an object such as the Hausdorff dimension, the Minkowski-Bouligand, or the box-counting method \cite[see][and references therein]{Fernandez2001}. In the framework of our analysis, we followed the dimensionless box counting procedure by \citet{Georgoulis2005, Georgoulis2012}. Based on this method, the binary image of interest is covered by ($L/\lambda$)$^{2}$ boxes, where $L$ corresponds to the linear dimension of the image and $\lambda$ is the size of the box's edge. By varying $\lambda$, we counted the number of boxes $N(\lambda)$ that contain part of the structure of interest. The complexity results for our CHs, which are based on both the fractal dimension and Solidity measures, are listed in Table~\ref{Table:FDvalues} and ranked from the most to the least complex structures. The first two columns show the ranking based on the Solidity, while the next two columns show the results based on the fractal dimension.

\begin{table*}[ht]
\centering
\caption{CHs ranked by their complexity, from the most to the least complex structures. Ranking was achieved in two ways, namely, by using the Solidity (first two columns) and the fractal dimension (last two columns).}
 \begin{tabular}{||c | c || c | c || c| c || c | c ||} 
 \hline
 CH no. & Solidity & CH no. & FD \\ [0.5ex] 
 \hline\hline

31      &       0.0319  &       31      &       1.3949  $\pm$   0.0492  \\
\hline
27      &       0.0340  &       27      &       1.3809  $\pm$   0.0432  \\
\hline
29      &       0.0436  &   34  &       1.3602  $\pm$   0.0273  \\
\hline
25      &       0.0461  &       38      &       1.3333  $\pm$   0.0439  \\
\hline
11      &       0.0486  &       25      &       1.3329  $\pm$   0.0388  \\
\hline
45      &       0.0566  &       37      &       1.3317  $\pm$   0.0378  \\
\hline
26      &       0.0665  &   29  &       1.3197  $\pm$   0.0291  \\
\hline
30      &       0.0678  &       36      &       1.3176  $\pm$   0.0490  \\
\hline
32      &       0.0767  &       11      &       1.3010  $\pm$   0.0284  \\
\hline
12      &       0.0779  &       14      &       1.2994  $\pm$   0.0516  \\
\hline
34      &       0.0858  &       16      &       1.2976  $\pm$   0.0265  \\
\hline
16      &       0.0866  &       32      &       1.2971  $\pm$   0.0416  \\
\hline
10      &       0.0875  &       26      &       1.2920  $\pm$   0.0304  \\
\hline
21      &       0.0894  &       22      &       1.2907  $\pm$   0.0377  \\
\hline
38      &       0.0907  &       44      &       1.2822  $\pm$   0.0588  \\
\hline
22      &       0.0909  &       8       &       1.2722  $\pm$   0.0356  \\
\hline
43      &       0.0935  &       3       &       1.2701  $\pm$   0.0261  \\
\hline
44      &       0.0964  &       45      &       1.2680  $\pm$   0.0307  \\
\hline
14      &       0.0985  &  12   &       1.2664  $\pm$   0.0393  \\
\hline
36      &       0.1005  &       19      &       1.2564  $\pm$   0.0426  \\
\hline
3       &       0.1039  &       20      &       1.2519  $\pm$   0.0475  \\
\hline
37      &       0.1042  &       1       &       1.2510  $\pm$   0.0200  \\
\hline
33      &       0.1058  &       5       &       1.2417  $\pm$   0.0248  \\
\hline
9       &       0.1066  &       24      &       1.2373  $\pm$   0.0274  \\
\hline
8       &       0.1085  &   23  &       1.2350  $\pm$   0.0251  \\
\hline
24      &       0.1146  &       7       &       1.2272  $\pm$   0.0462  \\
\hline
1       &       0.1231  &   42  &       1.2234  $\pm$   0.0417  \\
\hline
18      &       0.1356  &       10      &       1.2149  $\pm$   0.0321  \\
\hline
35      &       0.1371  &   21  &       1.2120  $\pm$   0.0289  \\
\hline
42      &       0.1504  &   13  &       1.2108  $\pm$   0.0307  \\
\hline
20      &       0.1567  &       30      &       1.2107  $\pm$   0.0278  \\
\hline
6       &       0.1669  &       9       &       1.2021  $\pm$   0.0190  \\
\hline
41      &       0.1695  &       6       &       1.1927  $\pm$   0.0439  \\
\hline
19      &       0.1725  &       28      &       1.1926  $\pm$   0.0618  \\
\hline
23      &       0.1795  &       18      &       1.1849  $\pm$   0.0333  \\
\hline
13      &       0.1912  &       17      &       1.1831  $\pm$   0.0524  \\
\hline
5       &       0.2246  &       43      &       1.1810  $\pm$   0.0294  \\
\hline
7       &       0.2357  &       33      &       1.1808  $\pm$   0.0492  \\
\hline
2       &       0.2641  &       39      &       1.1804  $\pm$   0.0476  \\
\hline
4       &       0.2688  &       41      &       1.1787  $\pm$   0.0269  \\
\hline
39      &       0.2865  &       4       &       1.1775  $\pm$   0.0237  \\
\hline
17      &       0.3032  &       35      &       1.1666  $\pm$   0.0440  \\
\hline
28      &       0.3372  &       40      &       1.1631  $\pm$   0.0569  \\
\hline
40      &       0.3723  &   15  &       1.1556  $\pm$   0.0477  \\
\hline
15      &       0.4755  &       2       &       1.1455  $\pm$   0.0491  \\
\hline

\end{tabular}

\label{Table:FDvalues}
\end{table*}

\begin{figure}[h!]
\centering
\begin{subfigure}[]{0.99\linewidth}
\includegraphics[width=\linewidth]{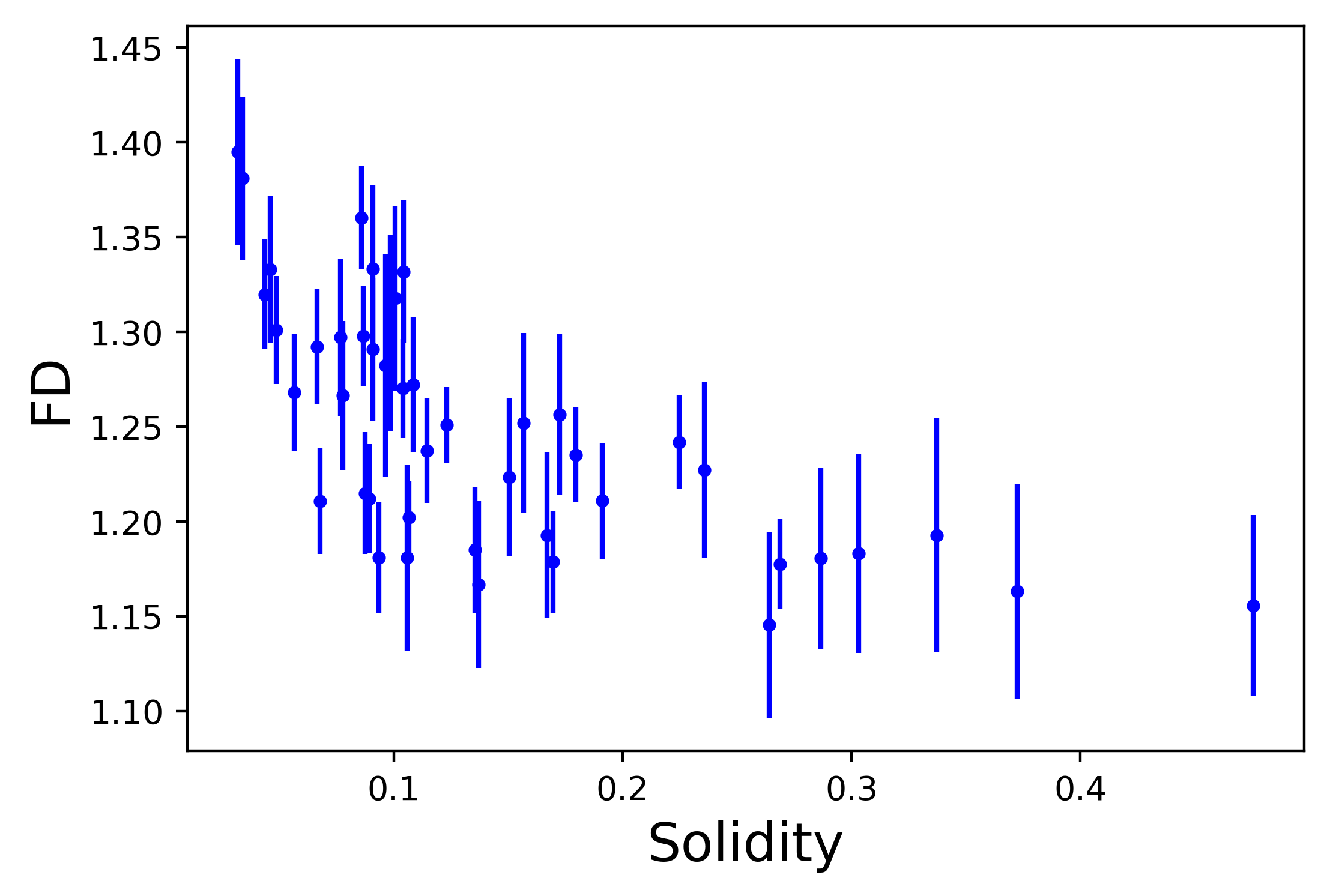}
\end{subfigure}%
\caption{FD of each CH structure, including applicable uncertainties, as a function of the Solidity.}
\label{Fig:FD_vs_Ratio}
\end{figure}

\begin{figure}[ht]
\centering
\begin{subfigure}[]{0.99\linewidth}
\includegraphics[width=\linewidth]{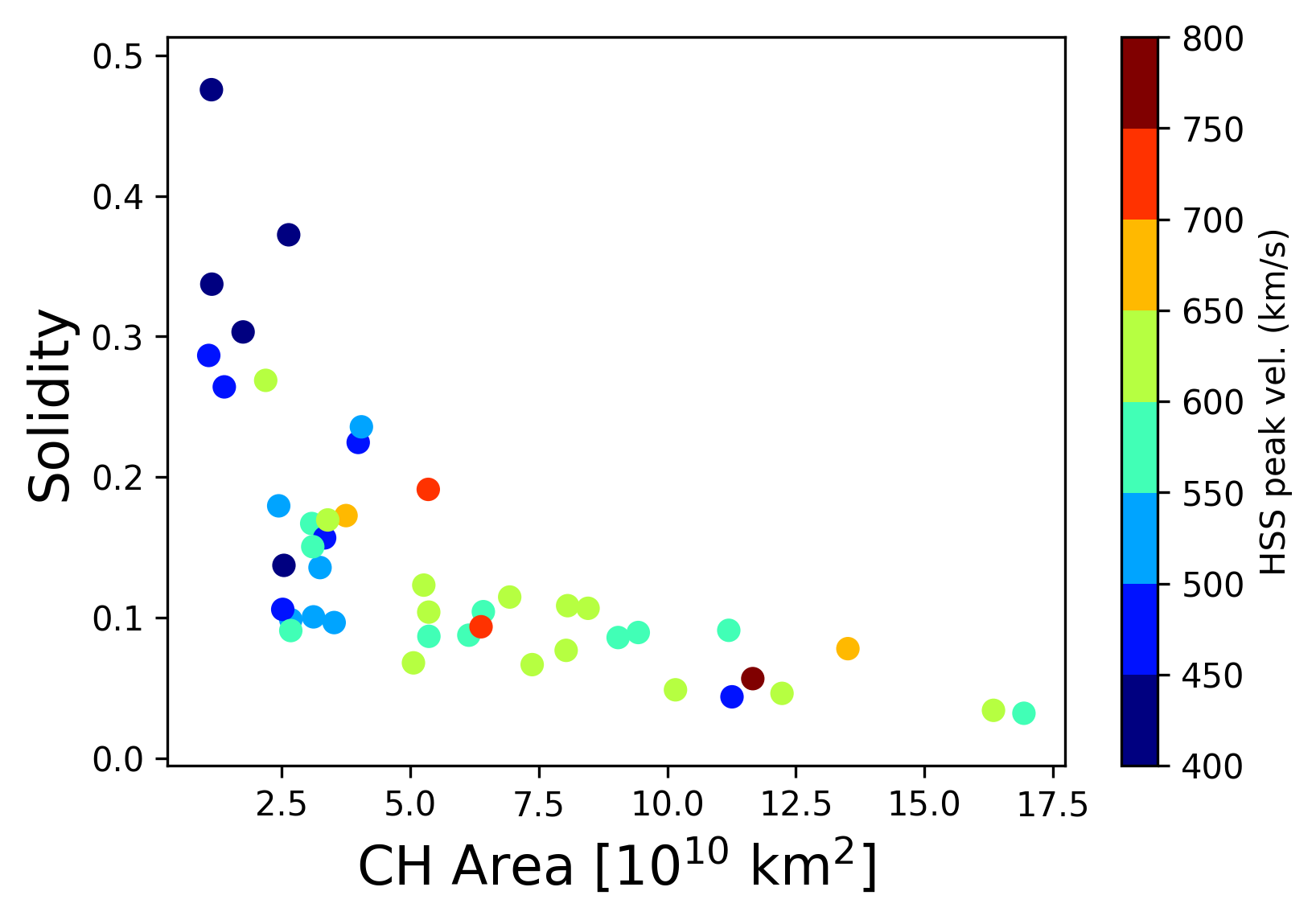}
\caption{}
\end{subfigure}
%hspace*{\fill}
\begin{subfigure}[]{0.99\linewidth}
\includegraphics[width=\linewidth]{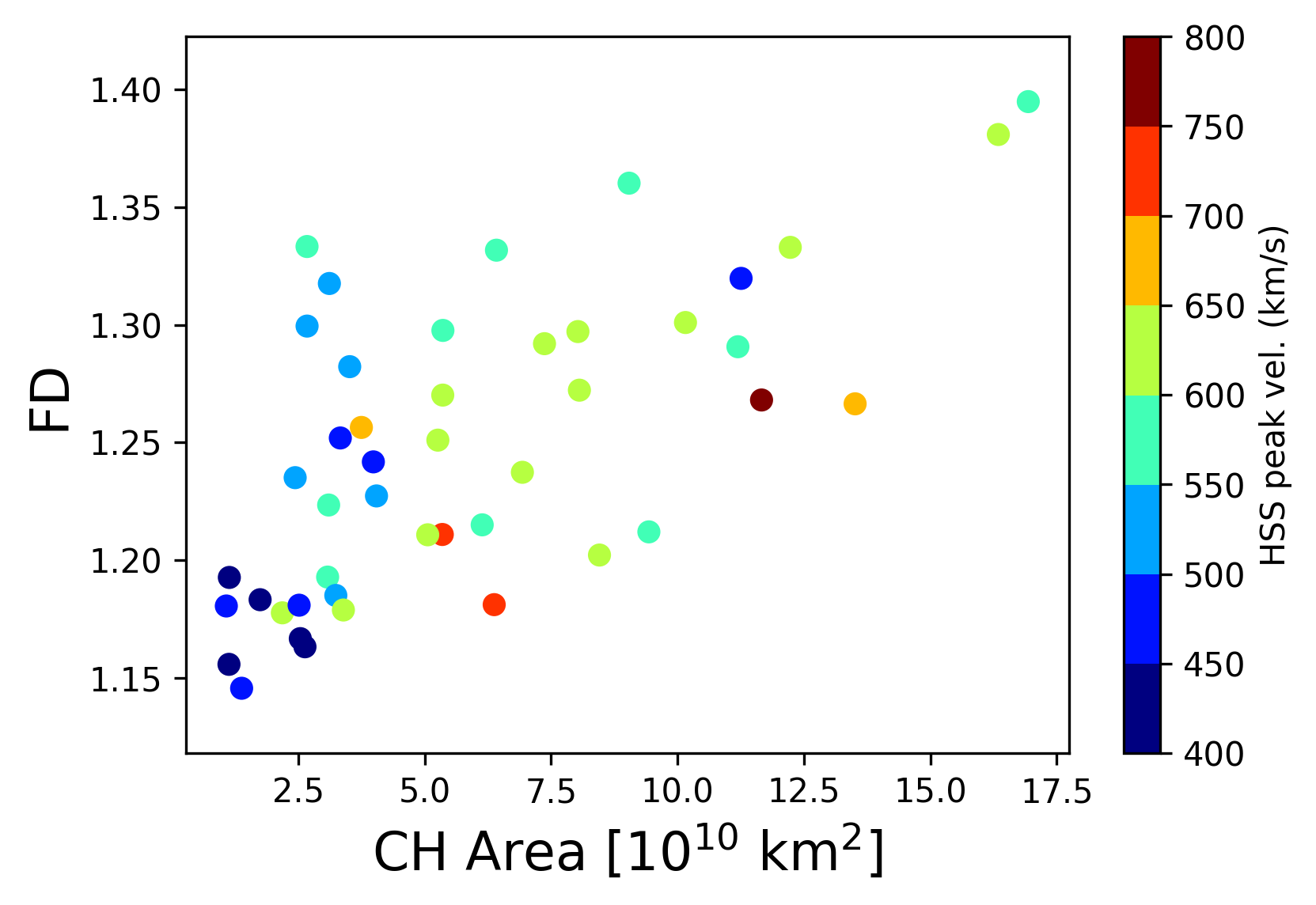}
\caption{}
\end{subfigure}%
\caption{Solidity (a) and FD (b) as a function of the CH area. The HSS peak velocity is color-coded.}
\label{Fig:FD/Ratio_area_speed}
\end{figure}

Figure~\ref{Fig:FD_vs_Ratio} presents the relation between the Solidity and the FD. We note that the two parameters are inverted quantities. Namely, less complex CHs correspond to higher values of the Solidity and lower values of FD. On the other hand, more complex CHs correspond to lower values of the Solidity, and higher values of FD. Figure~\ref{Fig:FD/Ratio_area_speed} presents the Solidity and the FD (panels (a) and (b), respectively), as a function of the CH area. The HSS peak velocity is color-coded. We find that small CHs are statistically less complex, while large CHs show a high degree of complexity. From the same plots, we also deduce that smaller HSS peak speeds, namely between 400 - 500 km/s, stem from smaller, less complex CHs with areas below 4 $\times$ 10$^{10}$ km$^{2}$ (dark blue-shaded circles). Faster HSSs stem from progressively more complex, but importantly larger, CHs.
 
%%%%%%%%%%%%%%%%%%%%%%%%%%%%%%%%%%%%%

%%%%%%%%%%%%%%%%%%%%%%%%%%%%%%%%%%

\begin{figure*}[h!]
\centering
\begin{subfigure}[]{0.31\linewidth}
\includegraphics[width=\linewidth]{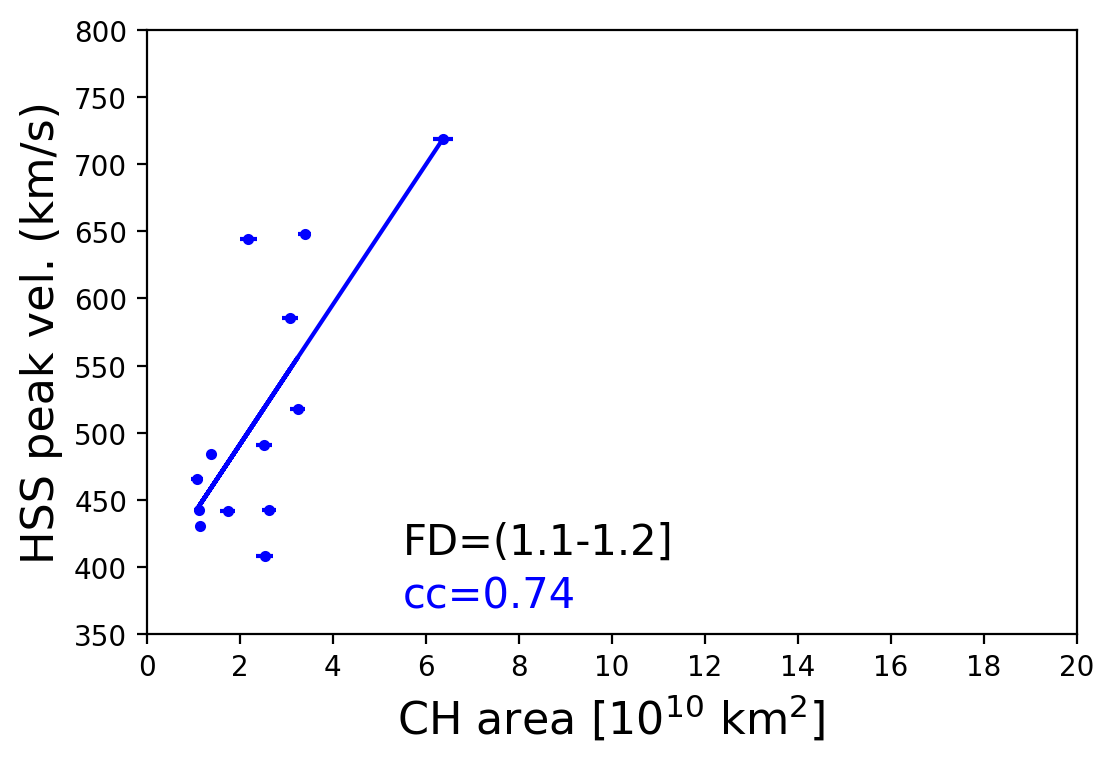}
\caption{}
\end{subfigure}
\hspace*{\fill}
\begin{subfigure}[]{0.31\linewidth}
\includegraphics[width=\linewidth]{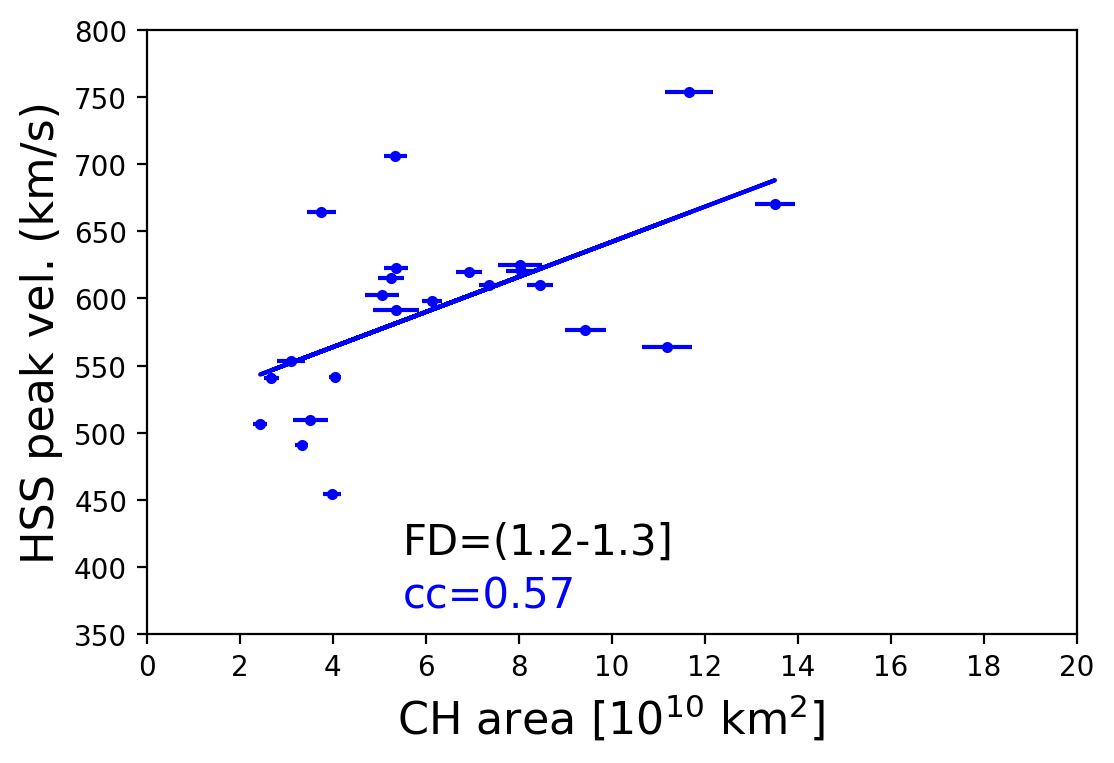}
\caption{}
\end{subfigure}%
\hspace*{\fill}
\begin{subfigure}[]{0.31\linewidth}
\includegraphics[width=\linewidth]{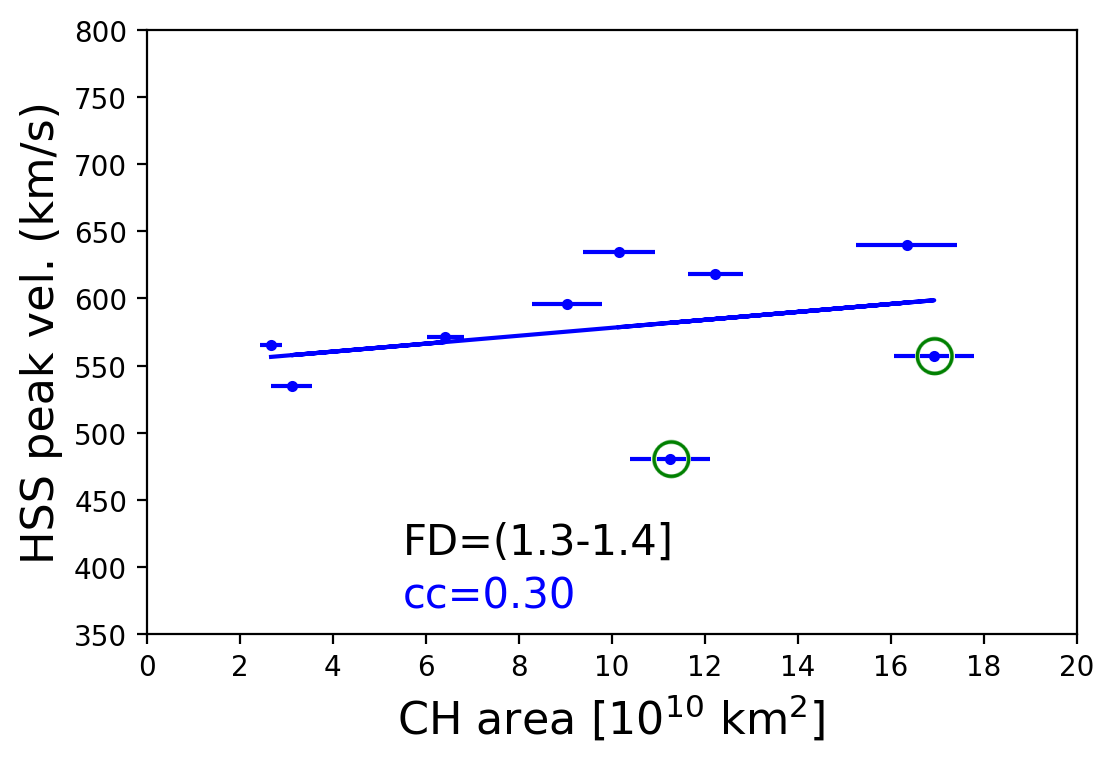}
\caption{}
\end{subfigure}%

\caption{HSS peak velocity as a function of the CH area for different FD ranges, going from the least complex (panel a) to the most complex (panel c) CHs. Green circles in panel (c) indicate the points responsible for the second peak observed in the Pearson's cc distribution after the application of the bootstrapping technique (Fig.~\ref{Fig:FD_appendix}c). For more details, see the main text.} 

\label{Fig:Speed_area_FDingroups}
\end{figure*}

\begin{figure*}[h!]
\centering
\begin{subfigure}[]{0.31\linewidth}
\includegraphics[width=\linewidth]{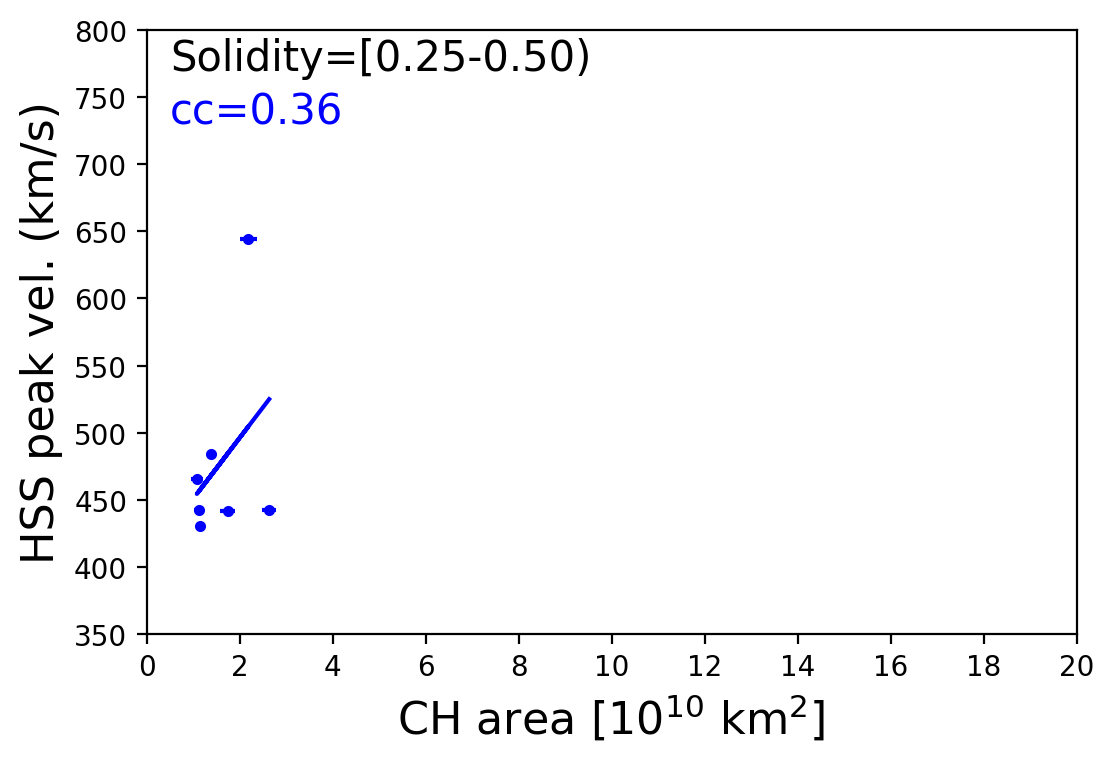}
\caption{}
\end{subfigure}
\hspace*{\fill}
\begin{subfigure}[]{0.31\linewidth}
\includegraphics[width=\linewidth]{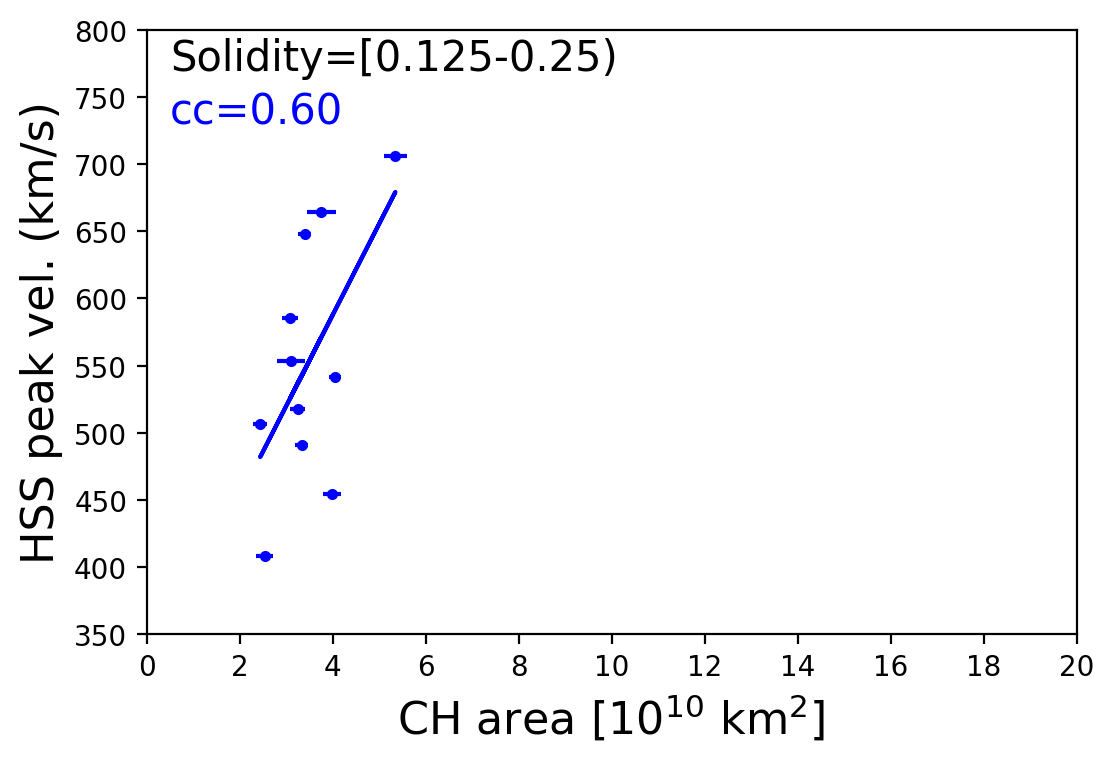}
\caption{}
\end{subfigure}%
\hspace*{\fill}
\begin{subfigure}[]{0.31\linewidth}
\includegraphics[width=\linewidth]{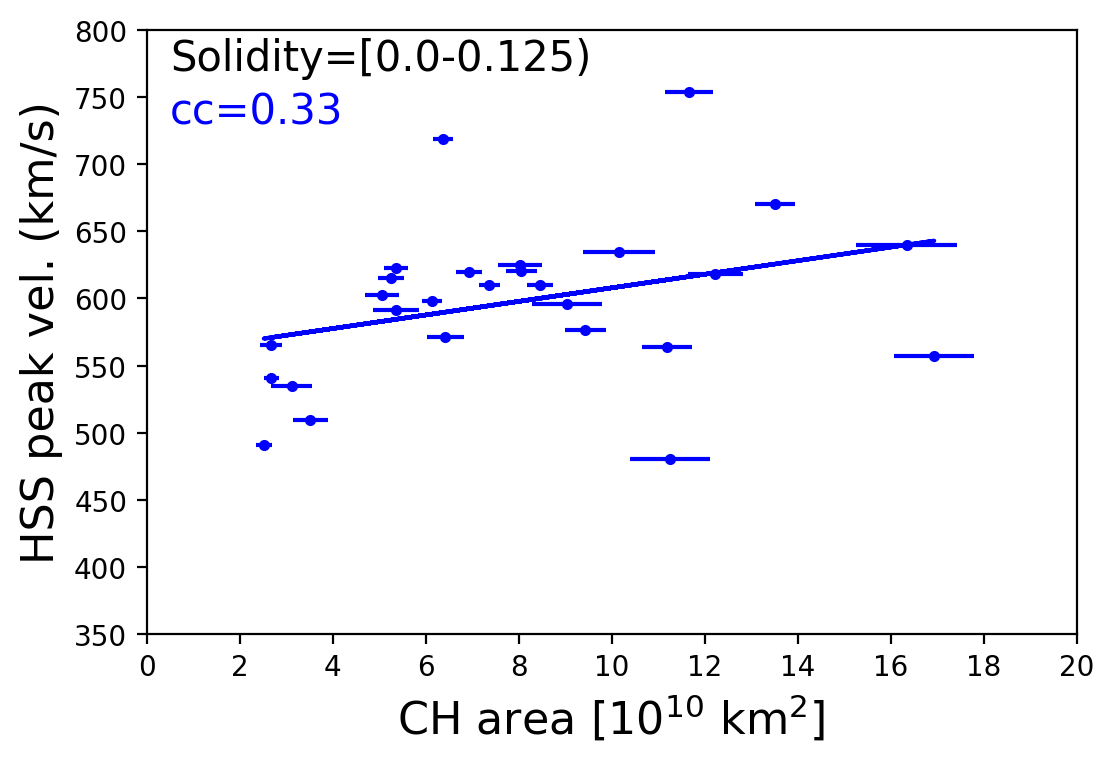}
\caption{}
\end{subfigure}%

\caption{HSS peak velocity as a function of the CH area for different Solidity ranges, going from the least complex (panel a) to the most complex (panel c) CHs. }
\label{Fig:Speed_area_RATIOgroups}
\end{figure*}

\begin{table*}[ht]
\centering
\caption{Comparison of Pearson's cc for groups of different geometrical complexities, going from the least to the most complex CHs. The subscript $boot$ refers to the Pearson's cc median value as well as to the first (Q1) and third (Q3) quartiles shown within brackets as
[Q1, Q3], after the application of the bootstrapping technique.}
\begin{center}
 \begin{tabular}{|l | c| c | c  |} 
 \hline
  Groups &  \makecell{Pearson's cc \\ (Least complex CHs)} & \makecell{Pearson's cc \\ (Medium complex CHs)}  & \makecell{Pearson's cc \\ (Most complex CHs)} \\
 \hline\hline

FD      & 0.74 & 0.57 & 0.30 \\
\hline
FD$_{boot}$     & 0.75 [0.61, 0.83] & 0.58 [0.47, 0.68] & 0.30 [0.11, 0.49] \\
\hline
Solidity & 0.36 & 0.60 & 0.33 \\
\hline
Solidity$_{boot}$ & 0.33 [-0.03 0.65] & 0.60 [0.40, 0.74] & 0.34 [0.21, 0.47] \\
\hline

\end{tabular}
\end{center}
\label{Table:bootstrapping_patchiness}
\end{table*}

To understand how the CH complexity influences the HSS velocity at Earth, we studied the relation between the HSS peak velocity and CH area by splitting the CHs into groups based on their FD and Solidity. For our sample, the FD ranges between 1.10 and 1.40. Therefore, three different groups were taken into account, namely, FD = (1.1, 1.2], FD = (1.2, 1.3], and FD = (1.3, 1.4], beginning from the least complex to the most complex structures, respectively. Figure \ref{Fig:Speed_area_FDingroups} shows how the HSS peak velocity depends on the CH area for these three categories. For the least complex CHs (see Fig. \ref{Fig:Speed_area_FDingroups}a), we obtain a good correlation (cc = 0.74) between the two parameters. We note that this subset contains CHs of small sizes (mostly below 4 $\times$ 10$^{10}$ km$^{2}$). As the geometric complexity increases (see Fig. \ref{Fig:Speed_area_FDingroups}b), the CH size becomes bigger and the correlation between the peak velocity of the HSS at Earth and the CH area decreases (cc = 0.57). For the most complex CHs (see Fig.~\ref{Fig:Speed_area_FDingroups}c), 
a weak, potentially insignificant correlation is obtained (see Table \ref{Table:bootstrapping_patchiness} for a summary of our results). We repeated the same procedure by splitting our points based on the Solidity. The latter parameter ranges from between 0 - 0.5. We divided this interval into three groups as well based on the ranges [0, 0.125), [0.125, 0.25), and [0.25, 0.50), which go from the most complex to the least complex structures, respectively. Our results in Fig.~\ref{Fig:Speed_area_RATIOgroups} show that the cc in the HSS peak velocity -- CH area plots for the different Solidity subsets follow the same pattern found based on the FD grouping, but only for the medium and most complex CHs. Namely, the cc for CHs of medium complexity (see Fig.~\ref{Fig:Speed_area_RATIOgroups}b) is 0.60 and drops for the very complex CHs to cc = 0.33 (see Fig.~\ref{Fig:Speed_area_RATIOgroups}c). For the first subgroup (see Fig.~\ref{Fig:Speed_area_RATIOgroups}a) cc = 0.36, which is lower than the one estimated for the medium-complexity group. The noticeably less significant association seen in Fig.~\ref{Fig:Speed_area_RATIOgroups}a, compared to the one seen in Fig.~\ref{Fig:Speed_area_FDingroups}a, possibly happens due to the definition of the Solidity which, as explained earlier, is a rougher complexity measure compared to the FD. 

To verify the robustness of our results since the three subgroups, created either based on the FD or the Solidity, contain a small number of points, we applied the pair bootstrapping technique similarly to the previous section. The second row of Table \ref{Table:bootstrapping_patchiness} summarizes the median, the Q1 and Q3 values of the Pearson's cc distributions after redrawing 100.000 data sets with replacement for each subgroup of different FD complexities. We notice that the median values decrease when going from the least complex to the most complex CHs, similarly to the behavior identified for our original sample (Fig.~\ref{Fig:Speed_area_FDingroups}).
However, bootstrapping for the most complex CH subset revealed a double-peak distribution of Pearson's cc (see Fig.~\ref{Fig:FD_appendix}c in Appendix \ref{appendix:Bootstrapping} for more details), unlike the distributions of other subsets. The second peak toward very high cc in Fig.~\ref{Fig:FD_appendix}c is caused when the two data points below the linear regression line of Fig.~\ref{Fig:Speed_area_FDingroups}c (circled in green) are not taken into account during the bootstrapping resampling. However, we should not assume them to be outliers or remove them from the statistics because they correspond to big and patchy CHs according to the criteria of the subsample. Applying the same bootstrapping technique to the CH subgroups based on the Solidity, we conclude that the median Pearson's cc drops while going from the medium complex toward the most complex CHs (see Table \ref{Table:bootstrapping_patchiness}, fourth row). The results are ambiguous regarding the least complex CH group as the scattering of the points is large and the Pearson's cc distribution produced after the bootstrapping has a very big interquartile range\footnote{This is the range between the first and the third quarantiles.} (see Fig.~\ref{Fig:Solidity_Appendix}a).

%Although, a detailed study of the evolutionary profile of these two CHs is somewhat out of the scope of this work, 

\subsection{Recurrent coronal holes}

Within our sample of 45 nonpolar CHs, there are two recurrent structures. The lifetime of the first one extends for 19 Carrington rotations and of the second one, for nine Carrington rotations.
The peak HSS speed, CH area, and geometric complexity for the 14 out of the 19 Carrington rotations regarding the first recurrent CH are presented in Fig.~\ref{Fig:RecurrentCHs}a. The specific CH is labeled as CH2, CH7, CH8, CH9, CH10, CH12, CH13, CH16, CH19, CH22, CH25, CH27, CH31, and CH34 for different Carrington rotations (see  Fig.~\ref{Fig:EUV_images},\ref{Fig:EUV_images2}, \ref{Fig:EUV_images3} for more details). For the first out of the five excluded cases, the HSS arrived to Earth simultaneously with a CME. Therefore, it was not possible to distinguish between the two structures in the in situ observations. During two other rotations, the CH was connected with the southern polar CH, so it was not possible to extract it with CATCH. During two other rotations, the CH was very patchy and the extraction of its characteristics was not reliable. For the case of the second recurrent CH, which is labeled as CH18, CH21, CH24, CH26, CH29, CH30, CH32, CH33, and CH35 for different Carrington rotations, we show the characteristics during all nine Carrington rotations in Fig.~\ref{Fig:RecurrentCHs}b.

Figure~\ref{Fig:RecurrentCHs} shows the FD (blue), CH area (red), and HSS peak velocity observed at Earth (green) as a function of time. Both profiles show that the complexity roughly follows the area, as already suggested by Fig.~\ref{Fig:FD/Ratio_area_speed}b. However, in the case of the first recurrent CH (Fig.~\ref{Fig:RecurrentCHs}a), both the complexity and CH area are simultaneously and constantly rising. The HSS velocity also rises concurrently with the complexity and the CH area, but it reaches a plateau of $\approx$ 600 km/s at approximately half of the life time of the CH. It stays at this level until the end of the CH life cycle. The behavior shown in Fig.~\ref{Fig:RecurrentCHs}a is fully consistent with the behavior of nonrecurrent CHs in our data set. Namely, the small and compact CHs are associated with the slow HSSs, and as the area and complexity of the CH grows, the velocity of the associated HSS grows. Once the velocity of the HSS reaches the plateau value, it does not further respond to the increase in the CH area and complexity (see also Fig. \ref{Fig:CHarea_vs_speed}).

The second CH in Fig.~\ref{Fig:RecurrentCHs}b shows a growing phase, a maximum phase, and a decaying phase in the CH area. As suggested by \citet{heinemann18}, this type of behavior indicates that we have observed the CH for most of its lifetime. The same 3-phase evolution was also observed for the FD. The HSS peak speed plateaus, and it decreases again as the CH complexity decreases. The drop in speed at the end of November 2018 is probably due to the increase in patchiness and the general irregular shape of that particular CH (CH 29). If that specific CH was more compact, while maintaining the same area, an increase to 600 km/s (levels of neighboring points) or more, could have been observed. In that case, the 3-phase behavior in flow speed would be more obvious as well. We note that only some of the recurrent CHs show such a well-defined 3-phase evolution. The first CH does not seem to be one of them, although it was analyzed for its entire lifetime\footnote{The first recurrent CH starts on Novermber 24, 2017 with CH 2 and ends on February 25, 2019 with CH 34. In the Carrington rotations following CH 34, the CHs observed were either very fragmented (so their extraction with CATCH was not possible), or no clear signatures of associated HSSs were observed at Earth, based on the criteria we set in Section 2. As a result, CH structures from the next few rotations after CH 34 could not be included in the study.}.

%%%%%%%%%%%%%%%%%%%%%%%%%%%%%%%%%

\section{Summary and conclusions}

%Our results show that the morphological parameters of CHs such as aspect ratio, orientation angle and complexity are very important. Therefore, they need to be taken into consideration for empirical models that aim to forecast the fast solar wind at Earth based on the CH properties on the Sun.

In this work, we studied how the morphological characteristics of CHs influence the HSS peak velocity and the HSS duration observed at Earth. We believe that a better understanding of the association between these parameters can help us improve modeling and forecasting of the HSSs in the vicinity of Earth. 

Our data sample consists of 45 CHs, out of which we have two recurrent CHs appearing with 28 labels and 17 nonrecurrent CHs. We inspected CH characteristics such as the CH area, CH longitudinal and latitudinal extent, CH CoM latitude, and their relationship with the maximum velocity and duration of the HSS at Earth. We divided our sample into different groups based on geometric criteria such as their aspect ratio, orientation angle, and geometric complexity. Our main conclusions are summarized as follows:

%The latter was quantified by means of (a) the ratio of the maximum inscribed rectangle to the convex hull area and (b) the fractal dimension of CHs. 

\begin{figure}[ht]
\centering
\begin{subfigure}[]{0.99\linewidth}
\includegraphics[width=\linewidth]{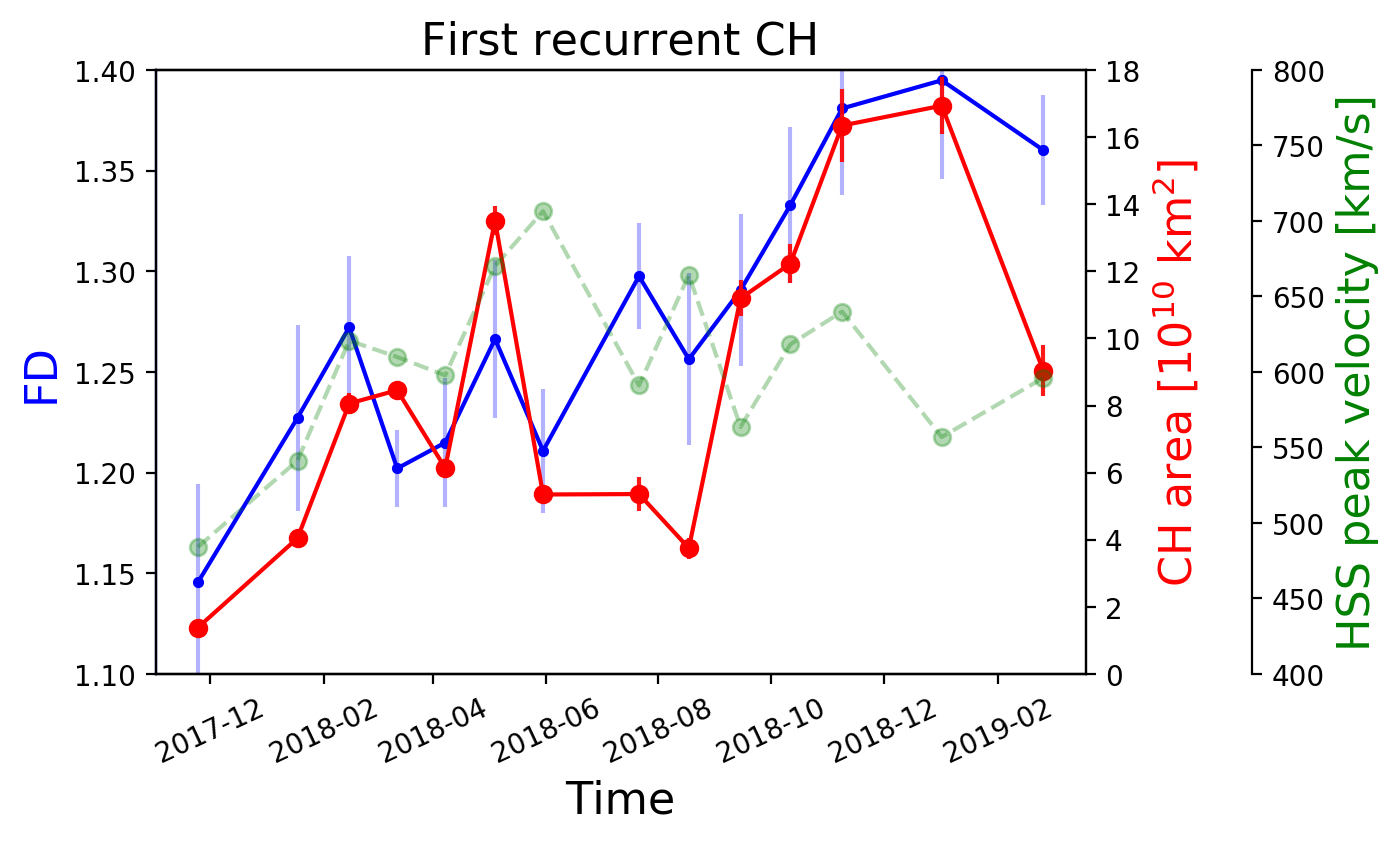}
\caption{}
\end{subfigure}%

\begin{subfigure}[]{0.99\linewidth}
\includegraphics[width=\linewidth]{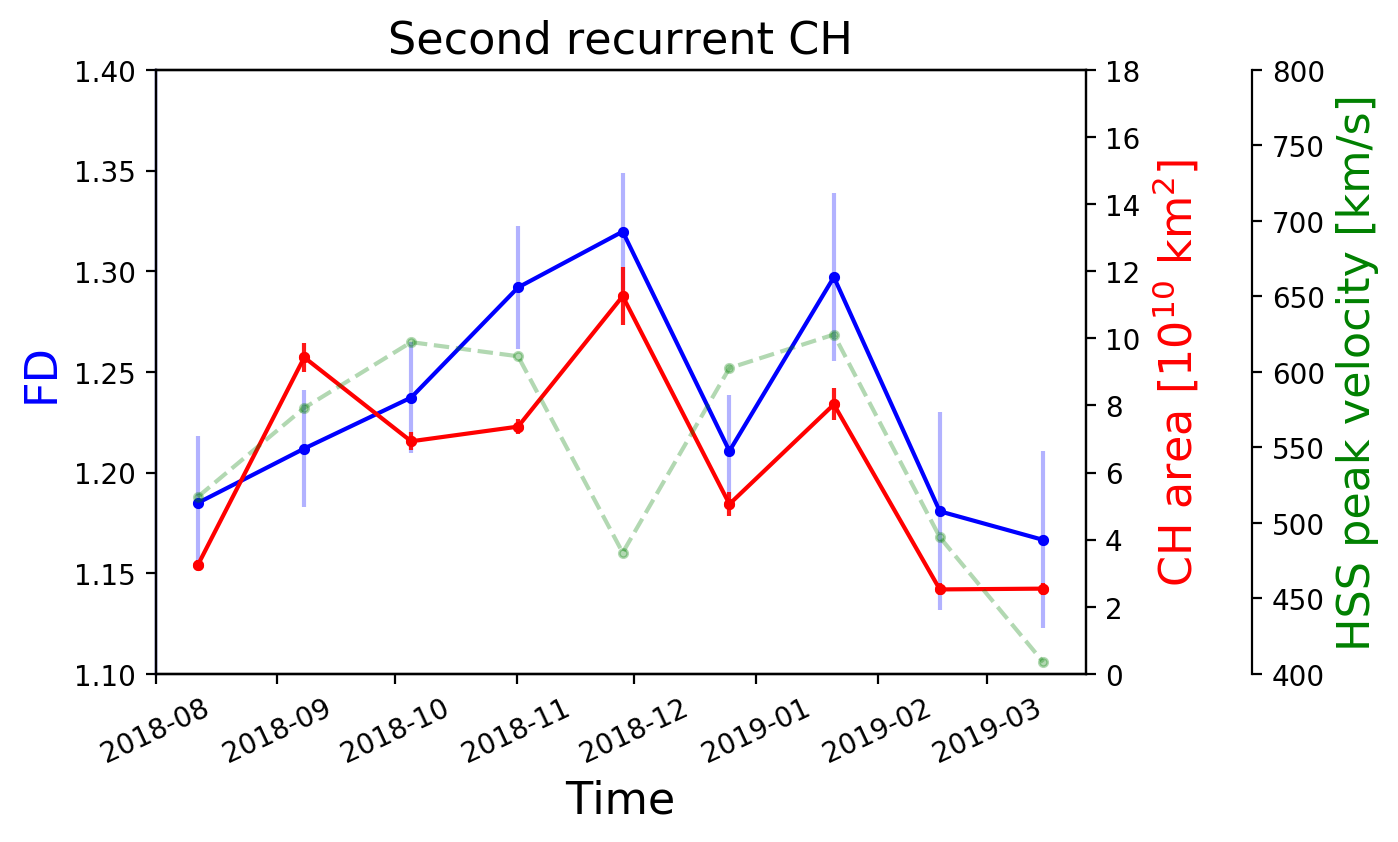}
\caption{}
\end{subfigure}%

\caption{Evolutionary profile for the first and second recurrent CH (panel (a) and (b), respectively). The first recurrent structure is labeled as follows: CH2, CH7, CH8, CH9, CH10, CH12, CH13, CH16, CH19, CH22, CH25, CH27, CH31, and CH34. The second recurrent CH is labeled as follows: CH18, CH21, CH24, CH26, CH29, CH30, CH32, CH33, and CH35.}
\label{Fig:RecurrentCHs}
\end{figure}

\begin{itemize}

 \item[\textbullet] When treating the full data set, the best correlation is obtained between the CH longitudinal extent and the HSS peak velocity in situ with a cc = 0.57. The better correlation between these two parameters, compared to what was estimated between the HSS peak velocity and the CH area or HSS peak velocity and CH latitudinal extent, is a consequence of the propagation effects that a HSS undergoes on the way from the Sun to Earth (see section \ref{subsection:CHlong_lat} for more details). CHs with larger longitudinal extents produce faster HSSs at Earth, up until the extent of 32 deg for our sample (cc = 0.75). Beyond this threshold, regardless of the increase in the longitudinal width of a CH, the HSS peak speed no longer increases. 
 
 \item[\textbullet] The CH area and latitudinal extent do not seem to be very strongly associated with the HSS duration observed in situ. As expected, the HSS duration has the best correlation with the CH longitudinal extent (cc = 0.68). 
 
 \item[\textbullet] When dividing our sample into two different groups based on their aspect ratio (longitudinal over latitudinal extent), we find that CHs with an aspect ratio $< 1$ show better correlations between the HSS peak velocity and the CH characteristics (area, longitudinal, and latitudinal extent), not only compared to the CHs with an aspect ratio higher than one, but also compared to the entire sample. In the case of an aspect ratio $< 1$, the associated HSSs more probably directly impact Earth, not just brush it off with their flanks. Furthermore, the majority of the CHs in this group have comparable latitudinal and longitudinal extents, which induce the longer impact of the HSS and assures arrival of the fastest part of the HSS.
 
 \item[\textbullet] We also divided the main data set into groups based on the CHs' orientation angle, with respect to the solar equator. The majority of CHs with negative orientation angles are situated in the northern solar hemisphere, and have their leading edge toward the equator. This subset shows very good correlations between the CH geometric characteristics and the HSS maximum velocity at Earth, which are better than the ones found for the entire sample. The same was obtained for the CHs with positive orientation angles, after the removal of outliers. The majority of CHs in this subset are situated in the southern solar hemisphere with their leading edge toward the equator. 
 
 \item[\textbullet] We also categorized our sample based on the degree of geometric complexity, a parameter, that, to our knowledge, is not usually considered in similar studies. We introduced two parameters to quantify the complexity: the ratio of the maximum inscribed rectangle over the convex hull CH area, which we call ``the Solidity,'' and the fractal dimension (FD) of the structures. Both methods take into account the irregularity of the CH boundary, but also the hollowness within the interior part of the CHs. %This similarity was not found for the third way of categorization, when the CH compactness is based on eq.~\ref{compactness_Reiss}. 
 We have found that the correlation between the HSS peak velocity at Earth and the CH area systematically decreases with the increase in the FD (increase in complexity). A good correlation (cc = 0.74) is identified for the least complex CHs, a medium correlation (cc = 0.57) for the medium-complex CHs, and a poor correlation (cc = 0.30) for the most complex structures. When categorizing our sample based on the Solidity, our results show a decrease in cc for the medium complex to the most complex CHs, whereas the results are more ambiguous for the least complex CHs.
 
 %Bootstrappingallows redrawing many datasets of the same length as the original dataset with replacement (i.e., a pair of datapoints from the original dataset can be selected multiple times). In the frame of this study, we redrew 100.000 datasets generating a distribution of 100.000 Pearson’s cc for each sample/subsample of interest. Then, we calculated the median, first (Q1) and third (Q3) quartiles values of these distributions. 

 \item[\textbullet] To confirm the reliability of the obtained results for the entire data sample and the different subsamples, we applied the bootstrapping method. This method is usually applied to small data sets in order to validate the statistical significance of the results obtained with other methods. Bootstrapping led to the same results for the entire sample compared to what we found based on the original data set. Namely, when treating the data set as a whole, the best correlation is obtained between the HSS peak velocity and the CH longitudinal extent, rather than the CH area or latitudinal extent. Bootstrapping also confirmed our results in the cases of the aspect ratio and orientation angle. The associations of the CH area, as well as longitudinal and latitudinal extents with the HSS peak speed in situ are stronger when we split our data set based on these two parameters, rather than when we dealt with the entire sample at once (see previous points for more details). Therefore, we conclude that both the aspect ratio and orientation angle are significant geometric factors that influence the way a HSS encounters Earth. They should be taken into account for empirical relationships and models that aim to predict the fast solar wind based on the CH properties observed on the Sun. Moreover, results with bootstrapping show that the median value of the Pearson's cc decreases while going from the least complex to the most complex CHs, based on the criterion of FD. This result agrees with our findings obtained from the original data set (Fig.~\ref{Fig:Speed_area_FDingroups}). The very low correlation found between the HSS peak speed and CH area for the most complex CHs is somewhat ambiguous, though. The bootstrapping method shows the existence of a second peak in the distribution, indicative of the possibility that some particular combinations of data points can give a very high cc (see Fig.~\ref{Fig:FD_appendix}c). Such high cc values mainly occur when the two data points, circled in green below the linear regression line of Fig.~\ref{Fig:Speed_area_FDingroups}c, are not taken into account during the resampling of the bootstrapping. Nevertheless, we should not assume that they are outliers or remove them from the statistics because they correspond to big and patchy CHs according to the criteria of that particular subsample. We note that this was the smallest subsample among the three considered. A bigger and more diverse sample of CHs would clarify if the Pearson's cc continues to decrease or not, in the case of the CHs of very high FD. Regarding the CHs' categorization based on the Solidity, the results with bootstrapping show that the median Pearson's cc drops when going from the medium complex to the most complex CHs. Our findings for the least complex group are ambiguous because of the large scattering of the points (see Fig.~\ref{Fig:Speed_area_FDingroups}a) and the large interquartile range of the cc, after the bootstrapping application. 

\end{itemize}

The current study highlights the need to better understand the CH characteristics. It helped us identify the most important parameters and correlations that need to be addressed and, ideally, understood to forecast the solar wind at Earth better. We have found that out of all the different CH characteristics, their aspect ratio, orientation angle, and complexity seem to be parameters that need to be taken into account for solar wind modeling and forecasting purposes. A follow-up study will aim to understand how the revealed correlations can help improve solar wind modeling, in particular the HSSs modeling with the European Heliospheric Forecasting Information Asset \citep[EUHFORIA; see][]{pomoell18}.

\begin{acknowledgements}
    The authors would like to acknowledge M. Temmer from the University of Graz, Austria, for the helpful discussions on the general concept of this paper. Many thanks also go to R. F. Pinto from IRAP, University of Toulouse, France and University Paris-Saclay, France for his ideas on this work. E.S. was supported by PhD grants awarded by the Royal Observatory of Belgium. S.J.H. acknowledges funding from the German Research Foundation (DFG), grant 448336908. This project has received funding from the European Union’s Horizon 2020 research and innovation program under grant agreements No 870405 (EUHFORIA 2.0). These results were also obtained in the framework of the projects C14/19/089 (C1 project Internal Funds KU Leuven), G.0D07.19N (FWO-Vlaanderen), SIDC Data Exploitation (ESA Prodex-12), and Belspo projects BR/165/A2/CCSOM and B2/191/P1/SWiM. For computations we used the infrastructure of the VSC – Flemish Supercomputer Center, funded by the Hercules foundation and the Flemish Government – department EWI.
\end{acknowledgements}

\bibliography{bibliography}

\begin{thebibliography}{50}
\expandafter\ifx\csname natexlab\endcsname\relax\def\natexlab#1{#1}\fi

\bibitem[{{Abramenko} {et~al.}(2009){Abramenko}, {Yurchyshyn}, \&
  {Watanabe}}]{Abramenko09}
{Abramenko}, V., {Yurchyshyn}, V., \& {Watanabe}, H. 2009, Sol. Phys, 260, 43

\bibitem[{{Billings} \& {Roberts}(1964)}]{BillingsRoberts64}
{Billings}, D.~E. \& {Roberts}, W.~O. 1964, Astrophys. Norveg, 9, 147

\bibitem[{{Bromage} {et~al.}(2000){Bromage}, {Alexander}, {Breen}, {Clegg},
  {Del Zanna}, {DeForest}, {Dobrzycka}, {Gopalswamy}, {Thompson}, \&
  {Browning}}]{Bromage2000}
{Bromage}, B.~J.~J., {Alexander}, D., {Breen}, A., {et~al.} 2000, \solphys,
  193, 181

\bibitem[{{Cane} \& {Richardson}(2003)}]{RClist2003}
{Cane}, H.~V. \& {Richardson}, I.~G. 2003, Journal of Geophysical Research
  (Space Physics), 108, 1156

\bibitem[{{Cranmer}(2002)}]{Cranmer2002}
{Cranmer}, S.~R. 2002, \ssr, 101, 229

\bibitem[{{Cranmer}(2009)}]{Cranmer2009}
{Cranmer}, S.~R. 2009, Living Reviews in Solar Physics

\bibitem[{{Cranmer} {et~al.}(2017){Cranmer}, {Gibson}, \& {Riley}}]{cranmer17}
{Cranmer}, S.~R., {Gibson}, S.~E., \& {Riley}, P. 2017, \ssr, 212, 1345

\bibitem[{{Davis}(1965)}]{Davis65}
{Davis}, L. 1965, Stellar and Solar Magnetic Fields, North Holland Publishing
  Co. Amsterdam, 204

\bibitem[{{Del Zanna} \& {Bromage}(1999)}]{DelZannaBromage99}
{Del Zanna}, G. \& {Bromage}, B.~J.~I. 1999, \jgr, 104, 9753

\bibitem[{{Falconer}(1990)}]{Falconer1990}
{Falconer}, K. 1990, {Fractal Geometry: Mathematical Foundations and
  Applications} (John Wiley \& Sons, New York)

\bibitem[{Fernández \& Jelinek(2001)}]{Fernandez2001}
Fernández, E. \& Jelinek, H.~F. 2001, Methods, 24, 309

\bibitem[{{Fludra} {et~al.}(1999){Fludra}, {Del Zanna}, {Alexander}, \&
  {Bromage}}]{Flundra99}
{Fludra}, A., {Del Zanna}, G., {Alexander}, D., \& {Bromage}, B.~J.~I. 1999,
  \jgr, 104, 9709

\bibitem[{{Garton} {et~al.}(2018){Garton}, {Murray}, \&
  {Gallagher}}]{Garton2018}
{Garton}, T.~M., {Murray}, S.~A., \& {Gallagher}, P.~T. 2018, \apjl, 869, L12

\bibitem[{{Georgoulis}(2005)}]{Georgoulis2005}
{Georgoulis}, M.~K. 2005, \solphys, 228, 5

\bibitem[{{Georgoulis}(2012)}]{Georgoulis2012}
{Georgoulis}, M.~K. 2012, \solphys, 276, 161

\bibitem[{{Habbal} {et~al.}(1997){Habbal}, {Woo}, {Fineschi}, {O'Neal}, {Kohl},
  {Noci}, \& {Korendyke}}]{Habbal97}
{Habbal}, S.~R., {Woo}, R., {Fineschi}, S., {et~al.} 1997, \apjl, 489, L103

\bibitem[{{Heinemann} {et~al.}(2020){Heinemann}, {Jer{\v{c}}i{\'c}}, {Temmer},
  {Hofmeister}, {Dumbovi{\'c}}, {Vennerstrom}, {Verbanac}, \&
  {Veronig}}]{Heinemann2020}
{Heinemann}, S.~G., {Jer{\v{c}}i{\'c}}, V., {Temmer}, M., {et~al.} 2020, \aap,
  638, A68

\bibitem[{{Heinemann} {et~al.}(2021){Heinemann}, {Saqri}, {Veronig},
  {Hofmeister}, \& {Temmer}}]{Heinemann2021a}
{Heinemann}, S.~G., {Saqri}, J., {Veronig}, A.~M., {Hofmeister}, S.~J., \&
  {Temmer}, M. 2021, \solphys, 296, 18

\bibitem[{{Heinemann} {et~al.}(2019){Heinemann}, {Temmer}, {Heinemann},
  {Dissauer}, {Samara}, {Jer{\v{c}}i{\'c}}, {Hofmeister}, \&
  {Veronig}}]{Heinemann2019}
{Heinemann}, S.~G., {Temmer}, M., {Heinemann}, N., {et~al.} 2019, \solphys,
  294, 144

\bibitem[{{Heinemann} {et~al.}(2018{\natexlab{a}}){Heinemann}, {Temmer},
  {Hofmeister}, {Veronig}, \& {Vennerstr{\o}m}}]{heinemann18}
{Heinemann}, S.~G., {Temmer}, M., {Hofmeister}, S.~J., {Veronig}, A.~M., \&
  {Vennerstr{\o}m}, S. 2018{\natexlab{a}}, \apj, 861, 151

\bibitem[{{Heinemann} {et~al.}(2018{\natexlab{b}}){Heinemann}, {Temmer},
  {Hofmeister}, {Veronig}, \& {Vennerstr{\o}m}}]{Heinemann2018}
{Heinemann}, S.~G., {Temmer}, M., {Hofmeister}, S.~J., {Veronig}, A.~M., \&
  {Vennerstr{\o}m}, S. 2018{\natexlab{b}}, \apj, 861, 151

\bibitem[{Hofmeister {et~al.}(2021)Hofmeister, Asvestari, Guo,
  Heidrich-Meisner, Heinemann, Magdalenic, Poedts, Samara, Temmer, Vennerstrom,
  {et~al.}}]{Hofmeister2021}
Hofmeister, S.~J., Asvestari, E., Guo, J., {et~al.} 2021, Astronomy \&
  Astrophysics

\bibitem[{{Hofmeister} {et~al.}(2018){Hofmeister}, {Veronig}, {Temmer},
  {Vennerstrom}, {Heber}, \& {Vr{\v{s}}nak}}]{Hofmeister2018}
{Hofmeister}, S.~J., {Veronig}, A., {Temmer}, M., {et~al.} 2018, Journal of
  Geophysical Research (Space Physics), 123, 1738

\bibitem[{{Hofmeister} {et~al.}(2020){Hofmeister}, {Veronig}, {Poedts},
  {Samara}, \& {Magdalenic}}]{Hofmeister2020}
{Hofmeister}, S.~J., {Veronig}, A.~M., {Poedts}, S., {Samara}, E., \&
  {Magdalenic}, J. 2020, \apjl, 897, L17

\bibitem[{{Jian} {et~al.}(2006){Jian}, {Russell}, {Luhmann}, \&
  {Skoug}}]{Jian2006}
{Jian}, L., {Russell}, C.~T., {Luhmann}, J.~G., \& {Skoug}, R.~M. 2006,
  \solphys, 239, 337

\bibitem[{{Karachik} \& {Pevtsov}(2011)}]{karachik11}
{Karachik}, N.~V. \& {Pevtsov}, A.~A. 2011, \apj, 735, 47

\bibitem[{Konatar {et~al.}(2020)Konatar, Popovic, \& Popovic}]{Konatar2020}
Konatar, I., Popovic, T., \& Popovic, N. 2020, in 2020 24th International
  Conference on Information Technology (IT), 1--4

\bibitem[{{Krieger} {et~al.}(1973){Krieger}, {Timothy}, \&
  {Roelof}}]{Krieger1972}
{Krieger}, A.~S., {Timothy}, A.~F., \& {Roelof}, E.~C. 1973, Solar Phys., 29,
  505

\bibitem[{Lemen {et~al.}(2012)Lemen, Title, Akin, Boerner, Chou, Drake, Duncan,
  Edwards, Friedlaender, Heyman, Hurlburt, Katz, Kushner, Levay, Lindgren,
  Mathur, McFeaters, Mitchell, Rehse, Schrijver, Springer, Stern, Tarbell,
  Wuelser, Wolfson, Yanari, Bookbinder, Cheimets, Caldwell, Deluca, Gates,
  Golub, Park, Podgorski, Bush, Scherrer, Gummin, Smith, Auker, Jerram, Pool,
  Soufli, Windt, Beardsley, Clapp, Lang, \& Waltham}]{Lemen2012}
Lemen, J.~R., Title, A.~M., Akin, D.~J., {et~al.} 2012, Solar Physics, 275, 17,
  the Solar Dynamics Observatory Guest Editors: W. Dean Pesnell, Phillip C.
  Chamberlin, and Barbara J. Thompson

\bibitem[{{Mandelbrot}(1983)}]{Mandelbrot1983}
{Mandelbrot}, B.~B. 1983, {The fractal geometry of nature /Revised and enlarged
  edition/}

\bibitem[{{McComas} {et~al.}(1998){McComas}, {Bame}, {Barker}, {Feldman},
  {Phillips}, {Riley}, \& {Griffee}}]{McComas98}
{McComas}, D.~J., {Bame}, S.~J., {Barker}, P., {et~al.} 1998, \ssr, 86, 563

\bibitem[{{Milo\v{s}evi\'c} {et~al.}(2013){Milo\v{s}evi\'c}, {Elston},
  {Krstono\v{s}i\'c}, \& {Rajkovi\'c}}]{Milosevic2013}
{Milo\v{s}evi\'c}, N., {Elston}, G.~N., {Krstono\v{s}i\'c}, B., \&
  {Rajkovi\'c}, N. 2013, in 2013 19th International Conference on Control
  Systems and Computer Science, 306--312

\bibitem[{{Nolte} {et~al.}(1976){Nolte}, {Krieger}, {Timothy}, {Gold},
  {Roelof}, {Vaiana}, {Lazarus}, {Sullivan}, \& {McIntosh}}]{Nolte76}
{Nolte}, J.~T., {Krieger}, A.~S., {Timothy}, A.~F., {et~al.} 1976, Sol. Phys.,
  46:2

\bibitem[{{Ogilvie} {et~al.}(1995){Ogilvie}, {Chornay}, {Fritzenreiter},
  {Hunsaker}, {Keller}, {Lobell}, {Miller}, {Scudder}, \& {Sittler}}]{WIND}
{Ogilvie}, K.~W., {Chornay}, D.~J., {Fritzenreiter}, R.~J., {et~al.} 1995,
  Space Science Reviews, 71

\bibitem[{{Pesnell} {et~al.}(2012){Pesnell}, {Thompson}, \& {Chamberlin}}]{SDO}
{Pesnell}, W.~D., {Thompson}, B.~J., \& {Chamberlin}, P.~C. 2012, \solphys,
  275, 3

\bibitem[{{Pomoell} \& {Poedts}(2018)}]{pomoell18}
{Pomoell}, J. \& {Poedts}, S. 2018, J Space Weather Space Climate, 8, A35

\bibitem[{{Rajkovi\'c} {et~al.}(2017){Rajkovi\'c}, {Krstono\v{s}i\'c}, \&
  {Milo\v{s}evi\'c}}]{Rajkovic2017}
{Rajkovi\'c}, N., {Krstono\v{s}i\'c}, B., \& {Milo\v{s}evi\'c}, N. 2017, Comput
  Math Methods Med

\bibitem[{{Reiss} {et~al.}(2016){Reiss}, {Temmer}, {Veronig}, {Nikolic},
  {Vennerstrom}, {Sch{\"o}ngassner}, \& {Hofmeister}}]{reiss16}
{Reiss}, M.~A., {Temmer}, M., {Veronig}, A.~M., {et~al.} 2016, Space Weather,
  14, 495

\bibitem[{{Richardson} \& {Cane}(2010)}]{RClist2010}
{Richardson}, I.~G. \& {Cane}, H.~V. 2010, \solphys, 264, 189

\bibitem[{{Rotter} {et~al.}(2012{\natexlab{a}}){Rotter}, {Veronig}, {Temmer},
  \& {Vr{\v s}nak}}]{rotter12}
{Rotter}, T., {Veronig}, A.~M., {Temmer}, M., \& {Vr{\v s}nak}, B.
  2012{\natexlab{a}}, \solphys, 281, 793

\bibitem[{{Rotter} {et~al.}(2012{\natexlab{b}}){Rotter}, {Veronig}, {Temmer},
  \& {Vr{\v{s}}nak}}]{Rotter2012}
{Rotter}, T., {Veronig}, A.~M., {Temmer}, M., \& {Vr{\v{s}}nak}, B.
  2012{\natexlab{b}}, \solphys, 281, 793

\bibitem[{{Schwenn}(1990)}]{schwenn90}
{Schwenn}, R. 1990, Physics of the Inner Heliosphere: 1. Large-Scale Phenomena,
  99

\bibitem[{{Schwenn}(2006)}]{Schwenn2006}
{Schwenn}, R. 2006, \ssr, 124, 51

\bibitem[{{Seibold}(2020)}]{MaxInscribedRect_MATLAB}
{Seibold}, P. 2020, {Largest inscribed rectangle square or circle, MATLAB
  Central File Exchange.}

\bibitem[{{Smith} {et~al.}(1996){Smith}, {Lange}, \& {Marks}}]{Smith1996}
{Smith}, T.~G.~J., {Lange}, G.~D., \& {Marks}, W.~B. 1996, Journal of
  Neuroscience Methods, 69, 123

\bibitem[{{Snyder} \& {Neugebauer}(1966)}]{Snyder&Neugebauer66}
{Snyder}, C.~W. \& {Neugebauer}, M. 1966, R. J. Mackin, Pergamon Press

\bibitem[{Tibshirani \& Efron(1993)}]{bootstrap2}
Tibshirani, R.~J. \& Efron, B. 1993, Monographs on statistics and applied
  probability, 57, 1

\bibitem[{{Vr{\v s}nak} {et~al.}(2007){Vr{\v s}nak}, {Temmer}, \&
  {Veronig}}]{vrsnak07}
{Vr{\v s}nak}, B., {Temmer}, M., \& {Veronig}, A.~M. 2007, \solphys, 240, 315

\bibitem[{{Wang} \& {Sheeley}(1990)}]{WangSheeley90}
{Wang}, Y.~M. \& {Sheeley}, N.~R., J. 1990, \apj, 355, 726

\bibitem[{{Wilcox}(1968)}]{Wilcox1968}
{Wilcox}, J.~M. 1968, Space Sci. Rev., 8, 258

\end{thebibliography}
\bibliographystyle{aa}
%\bibliography{sola_bibliography}

%\clearpage
     
% %\clearpage
% %%%%%%%%%%%%%%%%%%%%%%%%%%%%%%%%%%%%%%%%%%%%%%%%%%%%%%%%%%%%%%%%%%%%%%%%%%%%%%%%%%%%%%%%%%%

\begin{appendix}
\section{Images of all CHs}
\label{appendix:Images_of_CHs}

\begin{figure*}[h!]
\centering
\begin{subfigure}[]{0.24\linewidth}
\includegraphics[width=\linewidth]{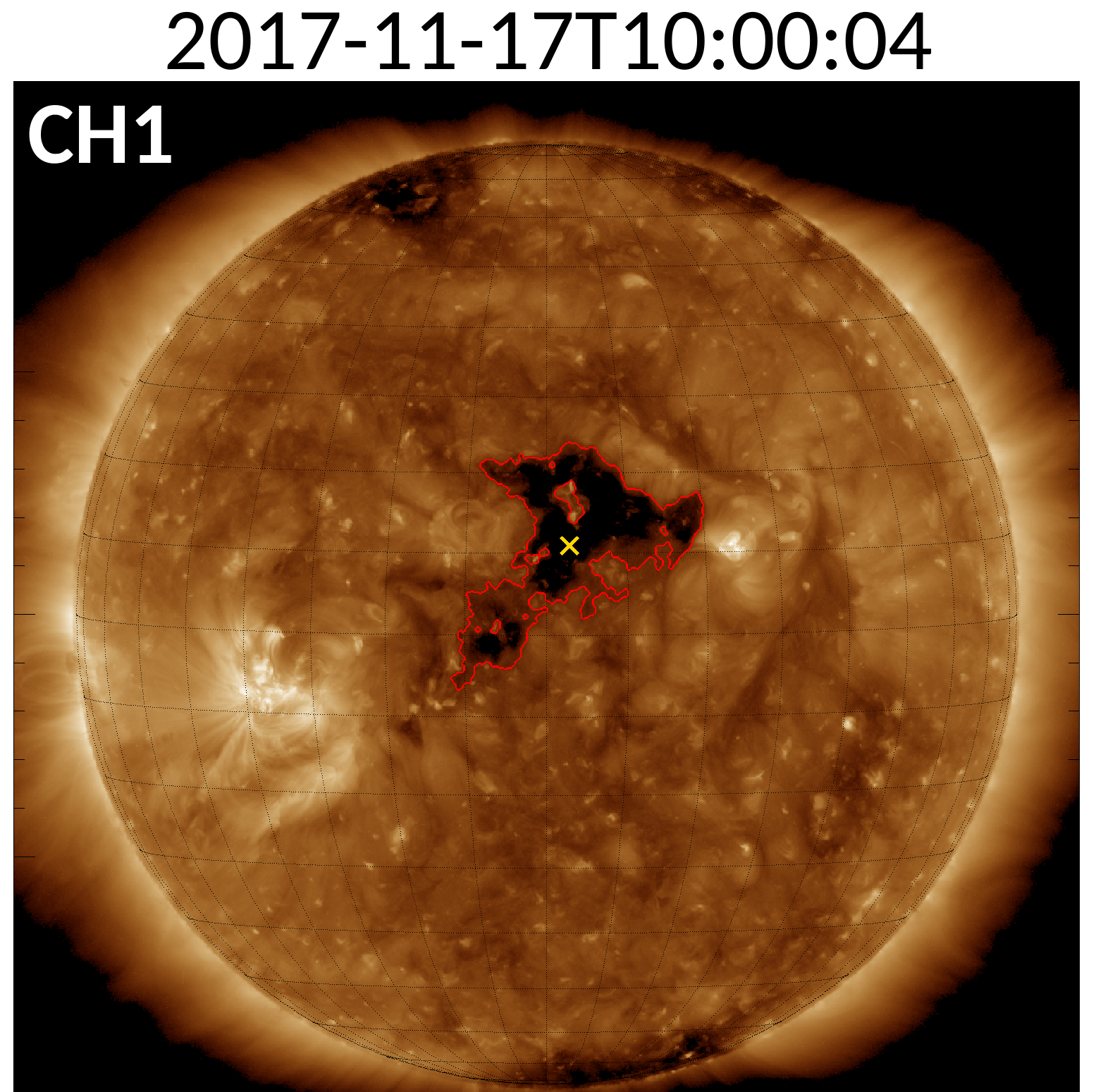}
\end{subfigure}
 \hspace*{0.1\fill}
\begin{subfigure}[]{0.24\linewidth}
\includegraphics[width=\linewidth]{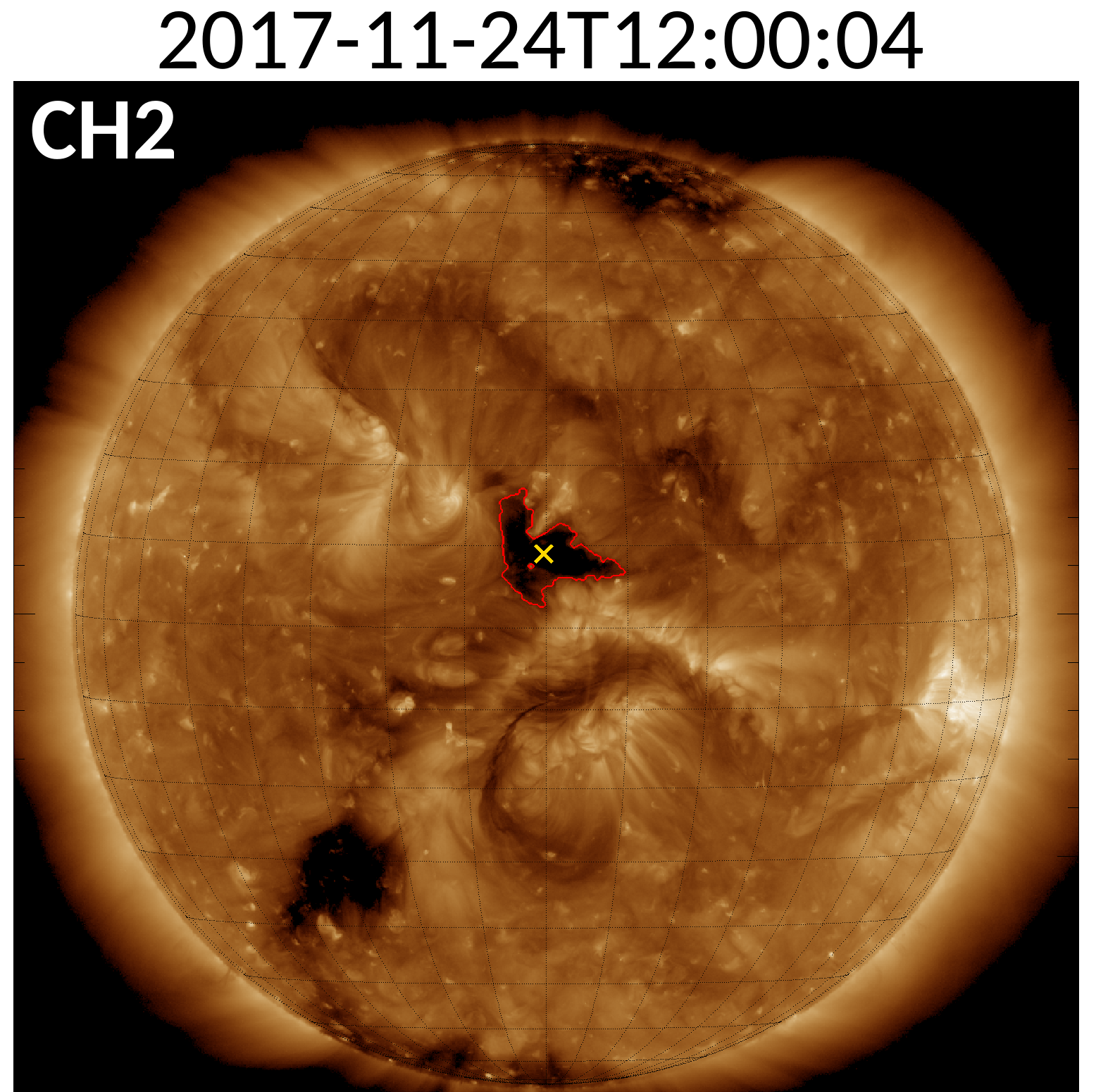}
\end{subfigure}%
 \hspace*{0.1\fill}
\begin{subfigure}[]{0.24\linewidth}
\includegraphics[width=\linewidth]{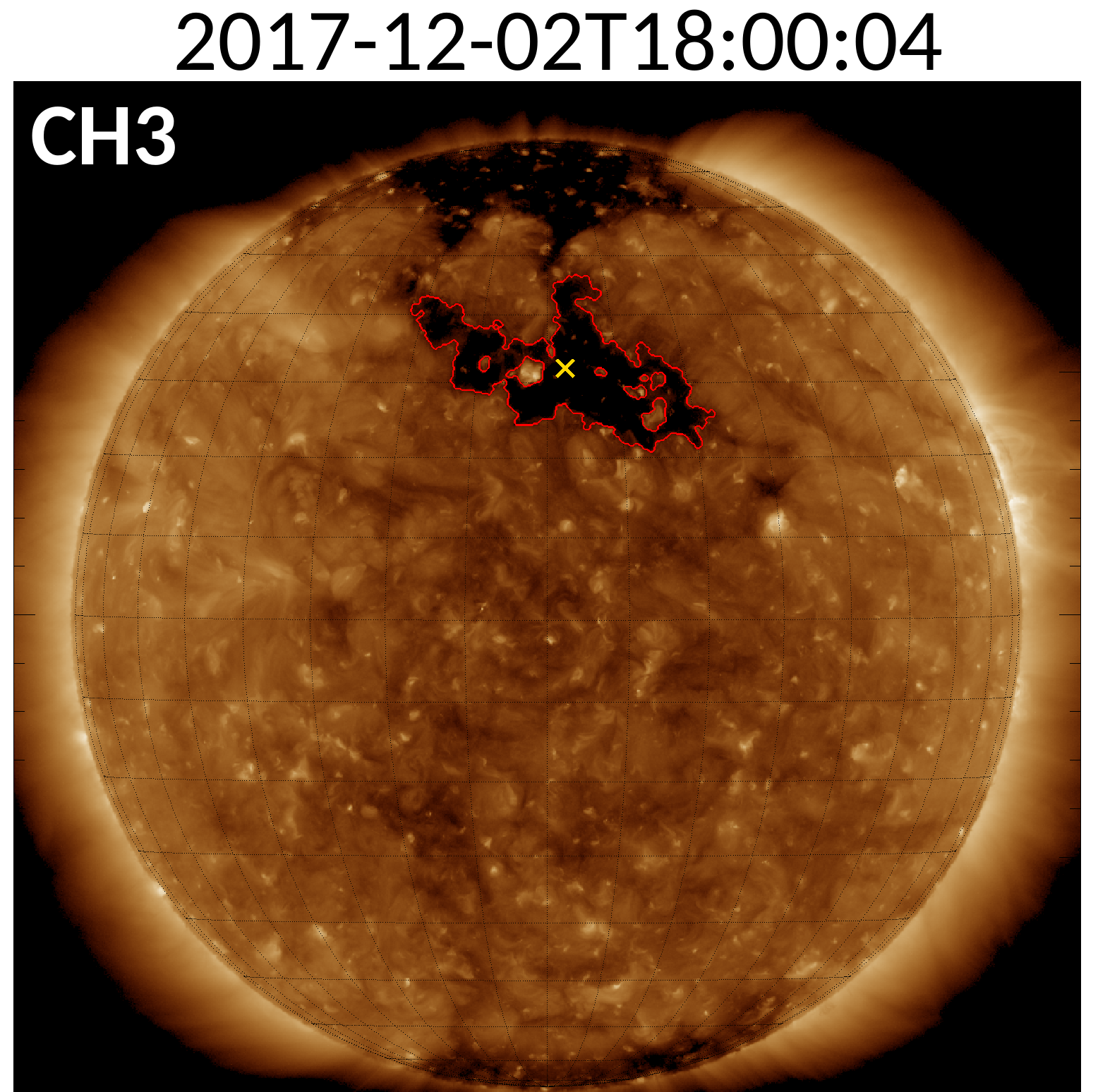}
\end{subfigure}%
 \hspace*{0.1\fill}
\begin{subfigure}[]{0.24\linewidth}
\includegraphics[width=\linewidth]{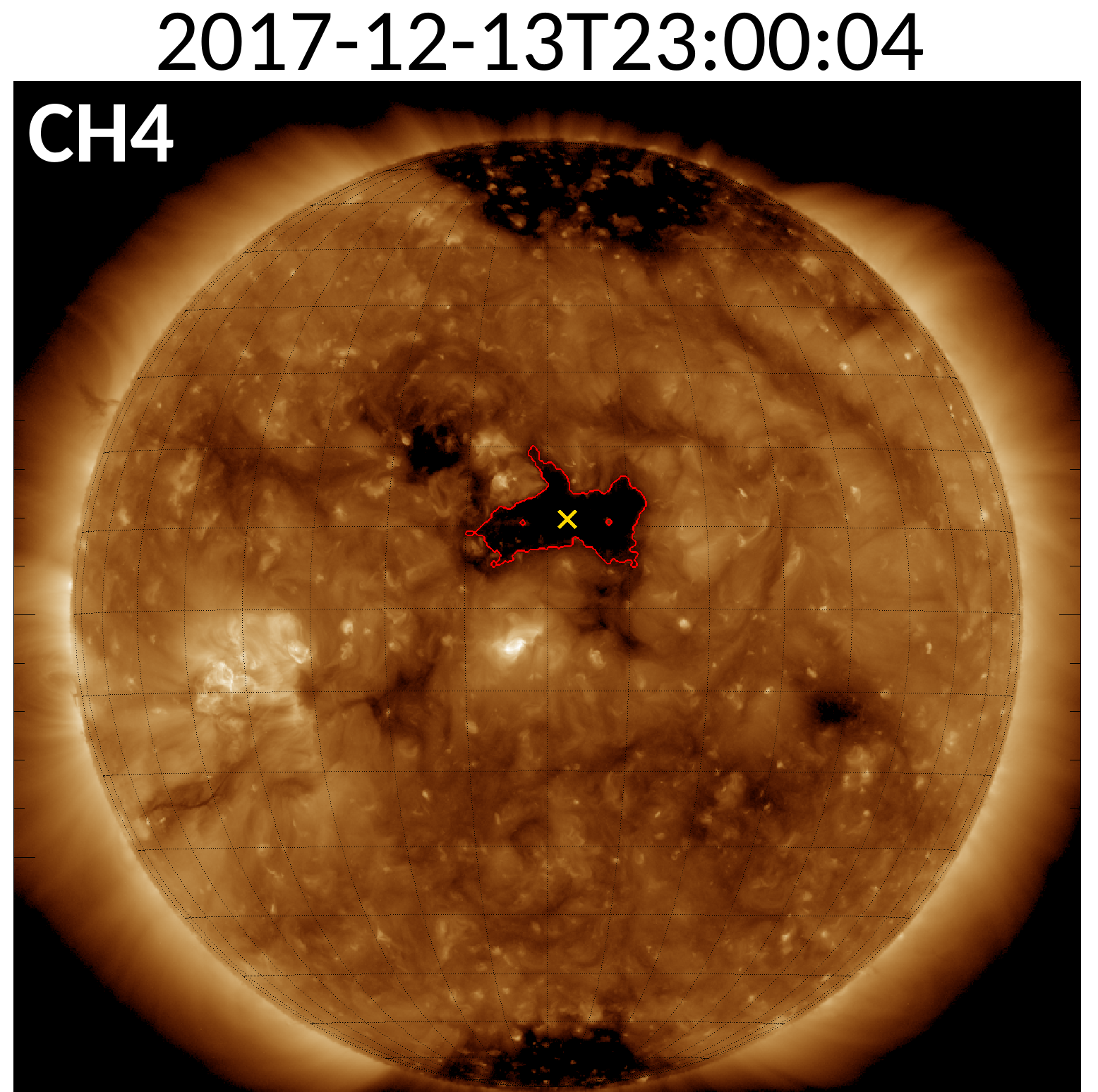}
\end{subfigure}%

\begin{subfigure}[]{0.24\linewidth}
\includegraphics[width=\linewidth]{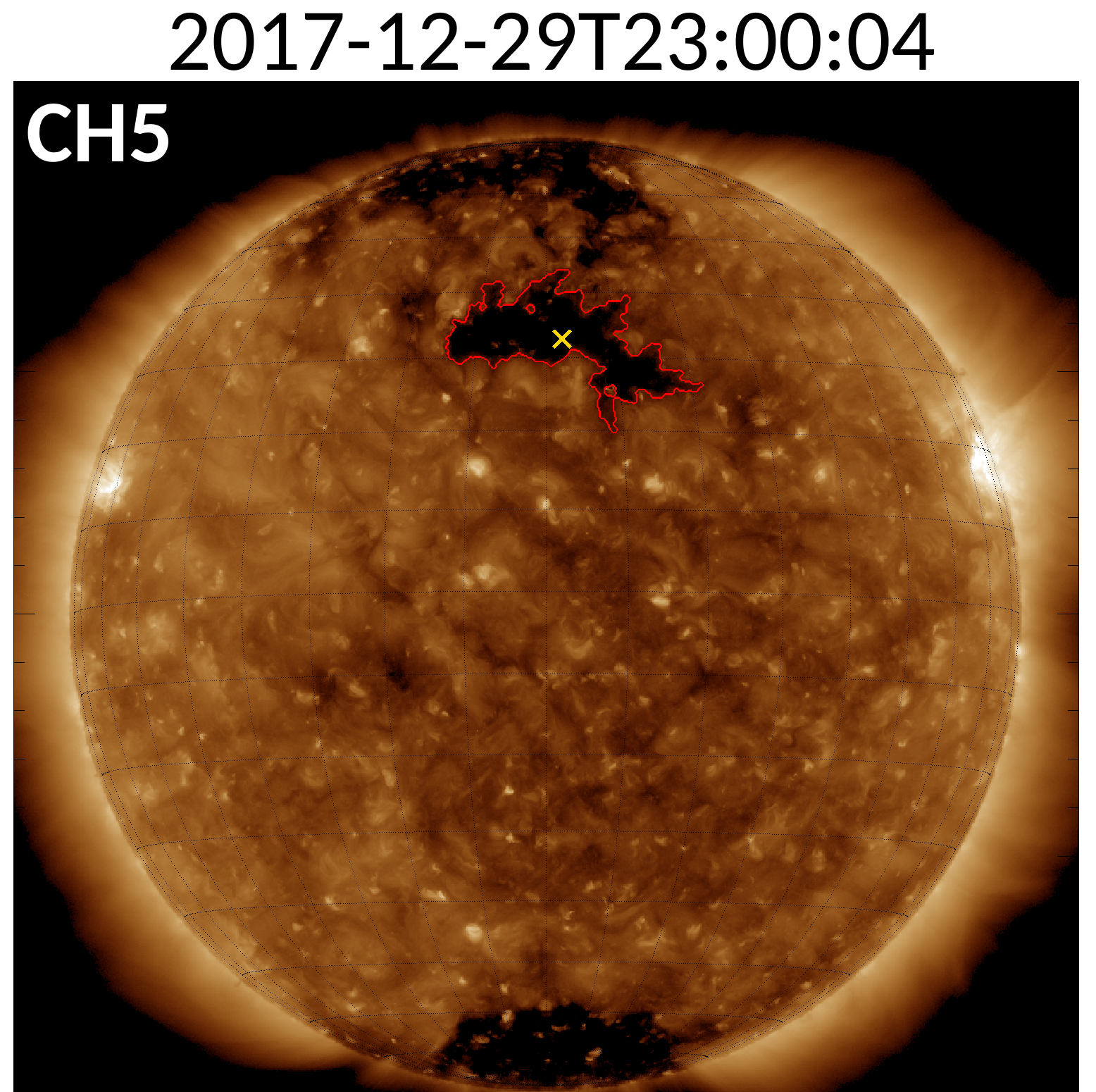}
\end{subfigure}%
 \hspace*{0.1\fill}
\begin{subfigure}[]{0.24\linewidth}
\includegraphics[width=\linewidth]{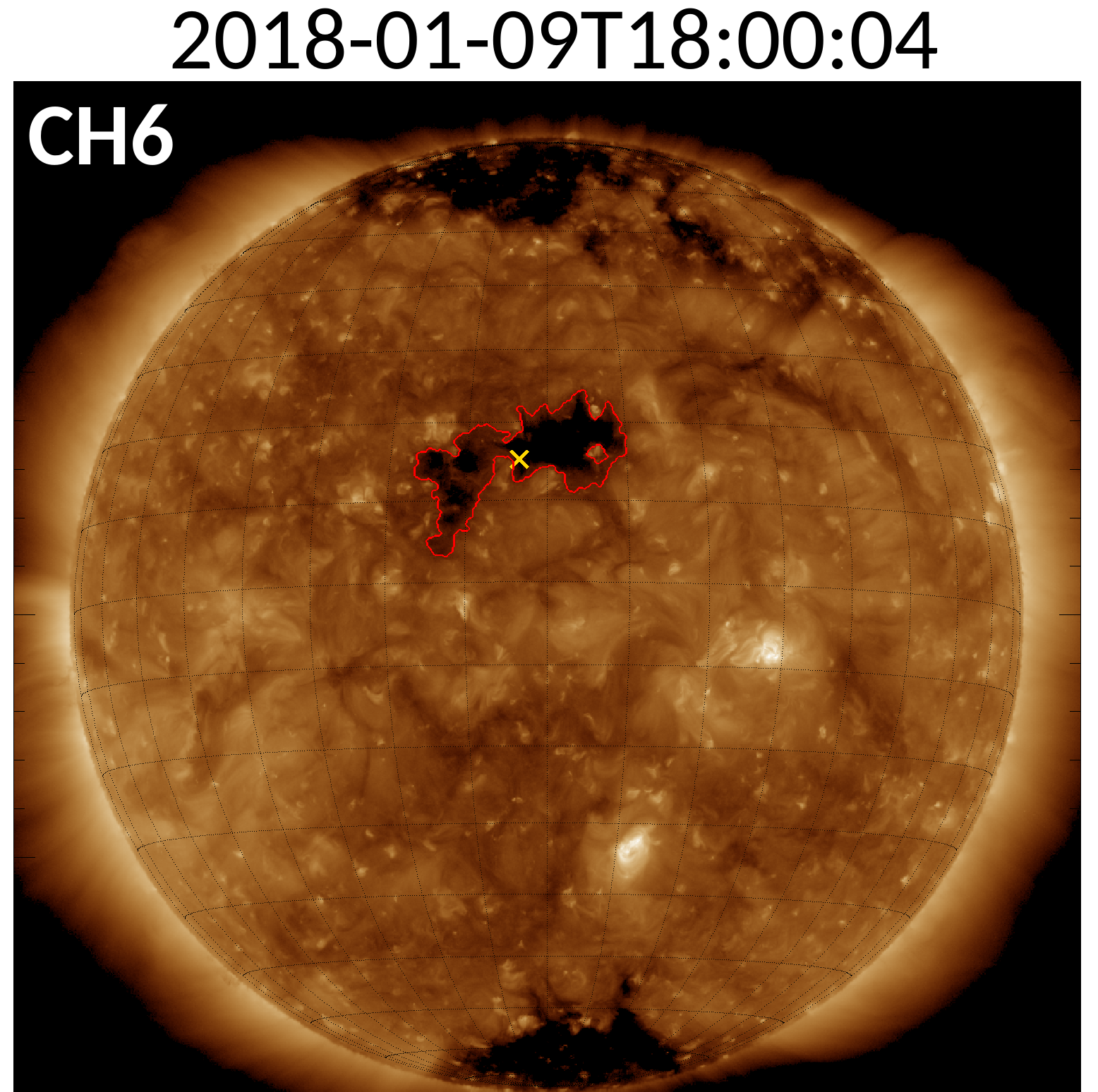}
\end{subfigure}%
 \hspace*{0.1\fill}
\begin{subfigure}[]{0.24\linewidth}
\includegraphics[width=\linewidth]{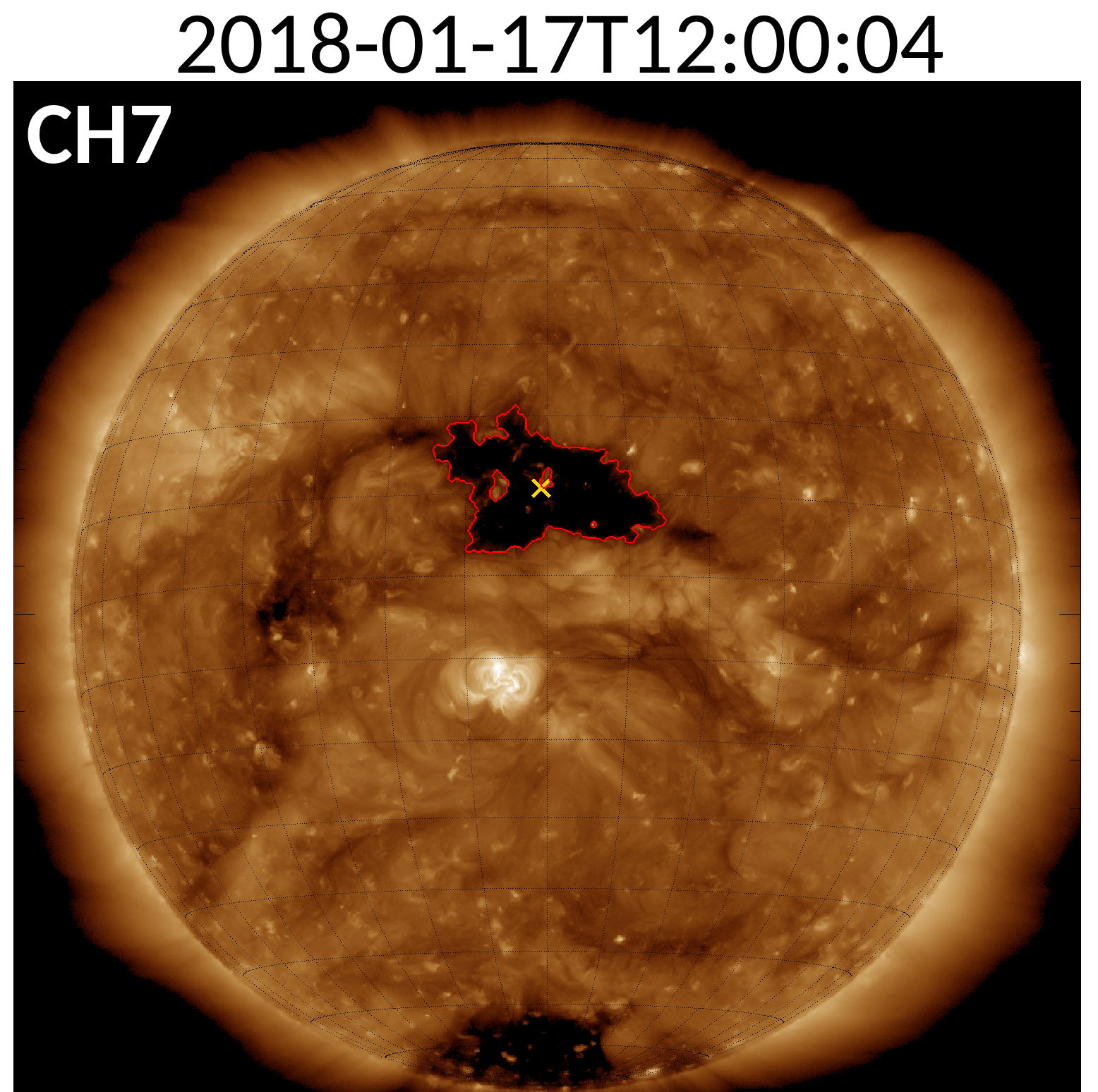}
\end{subfigure}%
 \hspace*{0.1\fill}
\begin{subfigure}[]{0.24\linewidth}
\includegraphics[width=\linewidth]{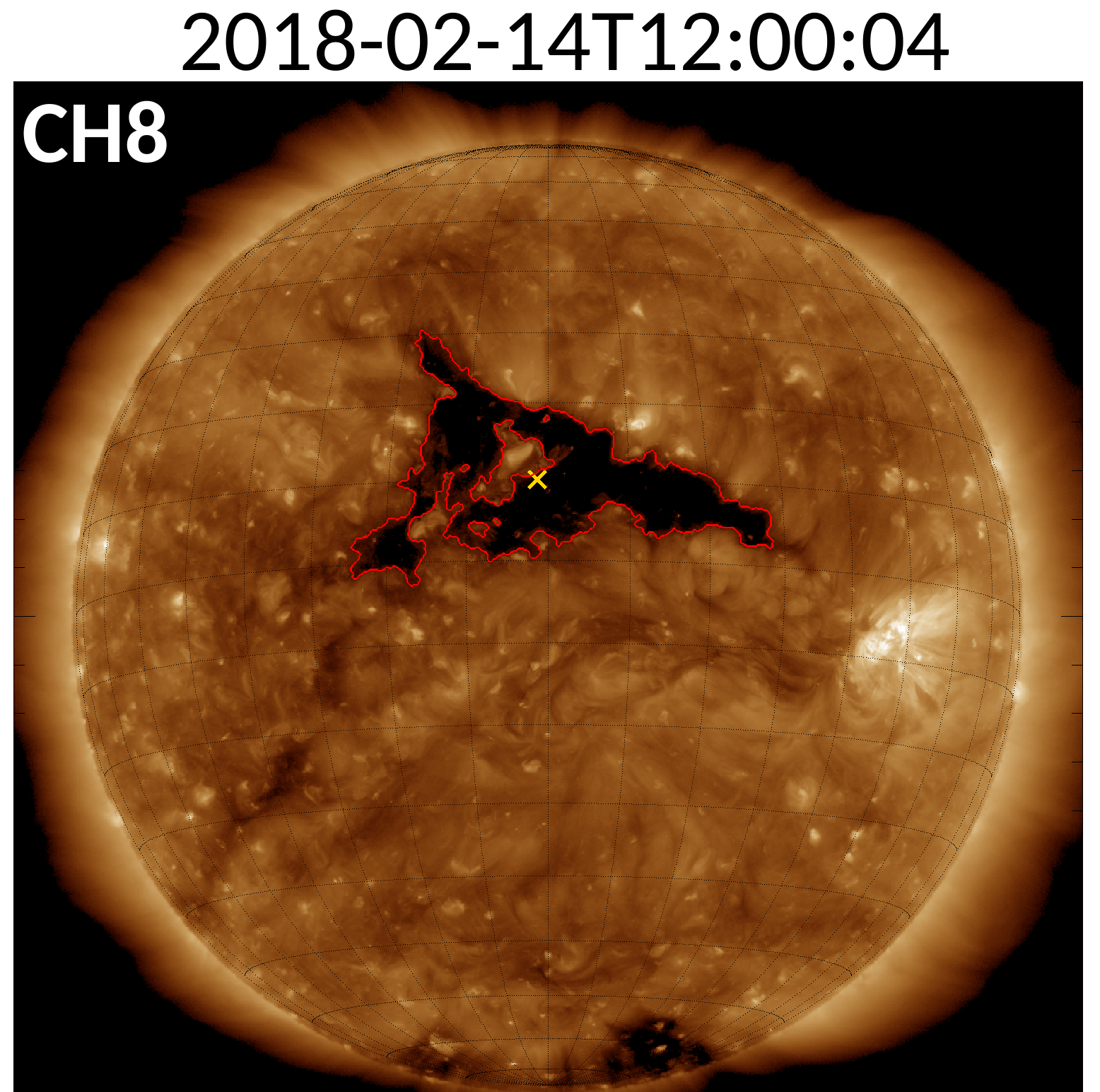}
\end{subfigure}%

\begin{subfigure}[]{0.24\linewidth}
\includegraphics[width=\linewidth]{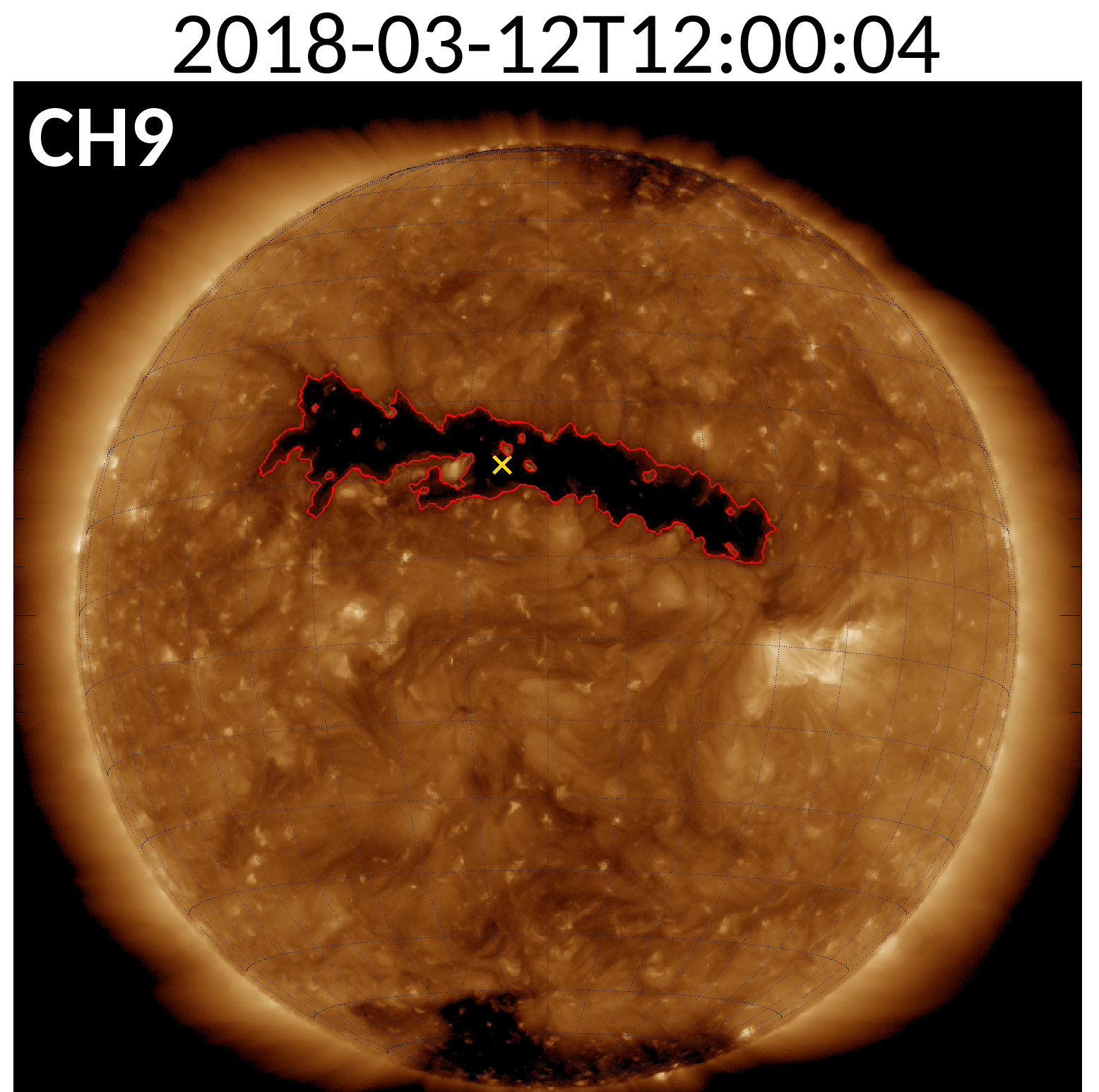}
\end{subfigure}%
\hspace*{0.1\fill}
\begin{subfigure}[]{0.24\linewidth}
\includegraphics[width=\linewidth]{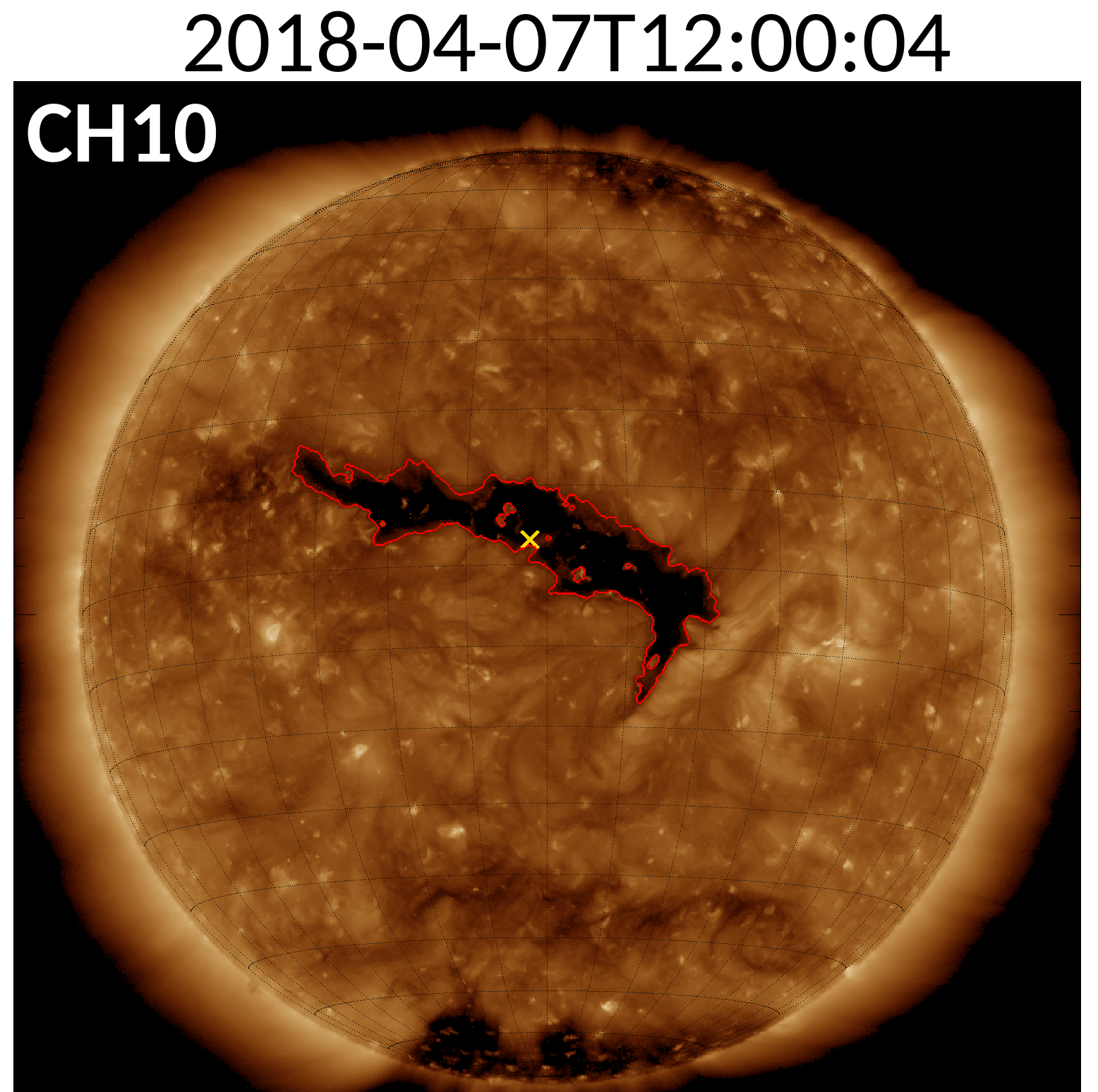}
\end{subfigure}%
\hspace*{0.1\fill}
\begin{subfigure}[]{0.24\linewidth}
\includegraphics[width=\linewidth]{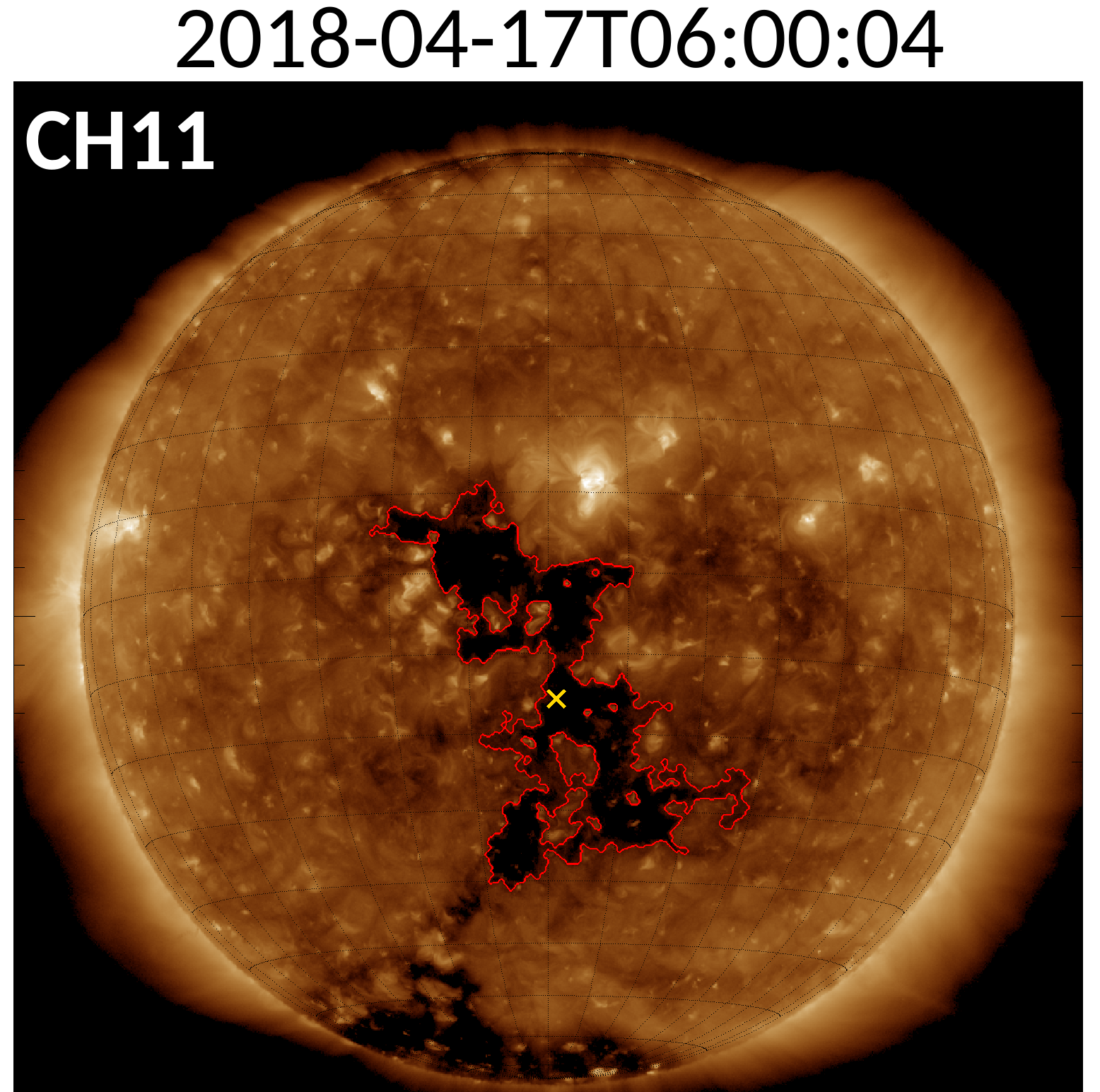}
\end{subfigure}%
 \hspace*{0.1\fill}
\begin{subfigure}[]{0.24\linewidth}
\includegraphics[width=\linewidth]{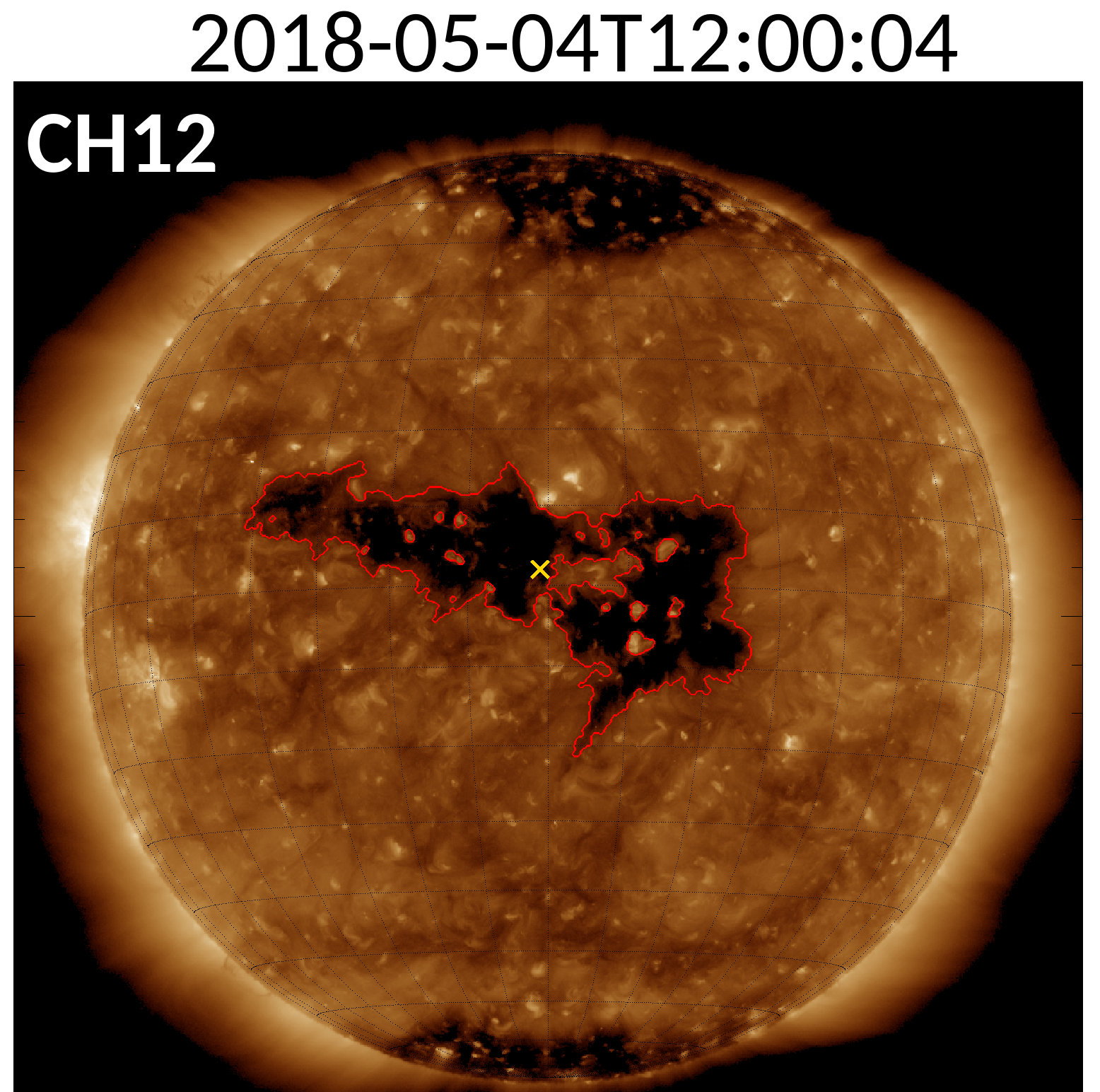}
\end{subfigure}%

\begin{subfigure}[]{0.24\linewidth}
\includegraphics[width=\linewidth]{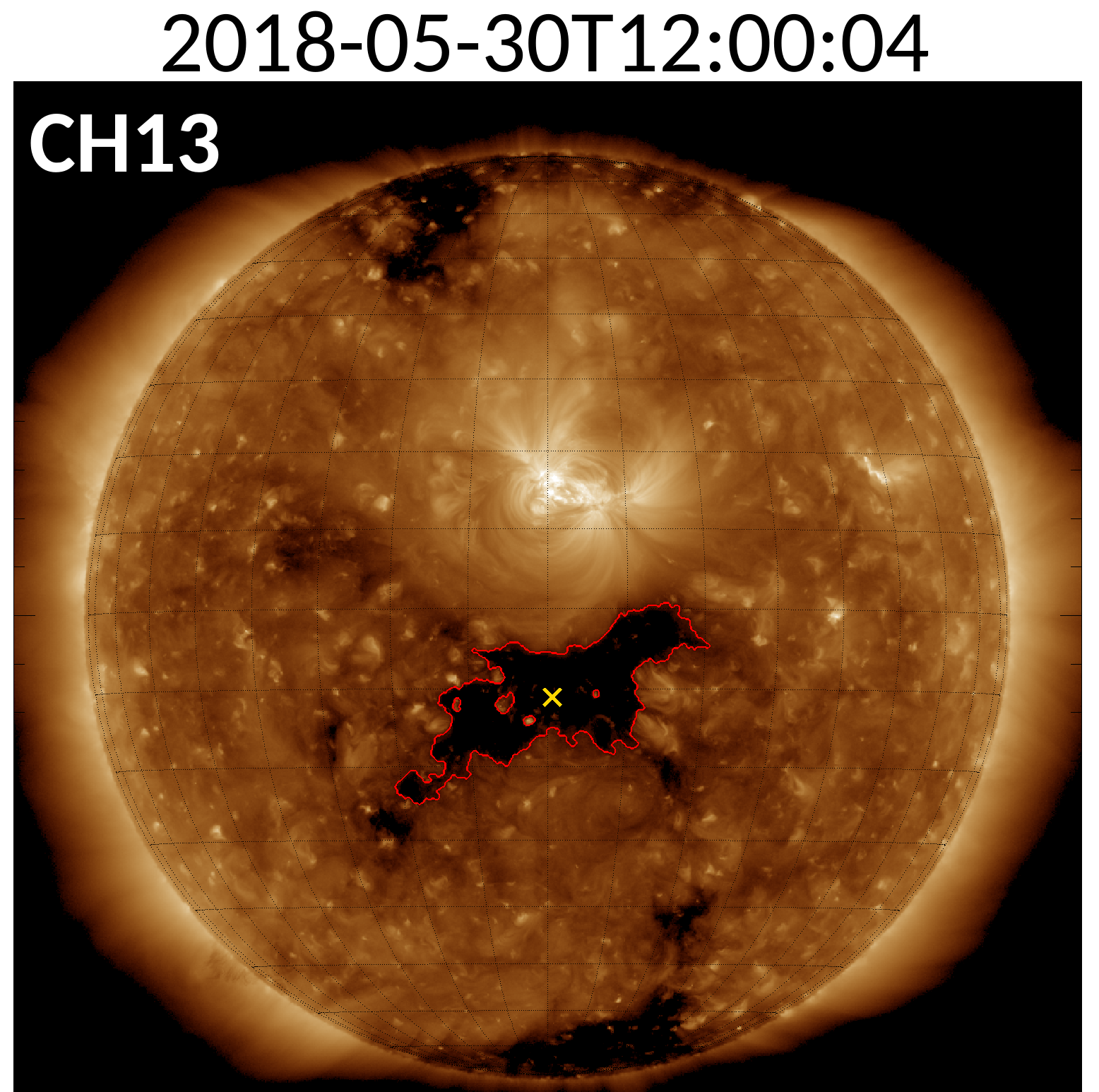}
\end{subfigure}%
 \hspace*{0.1\fill}
\begin{subfigure}[]{0.24\linewidth}
\includegraphics[width=\linewidth]{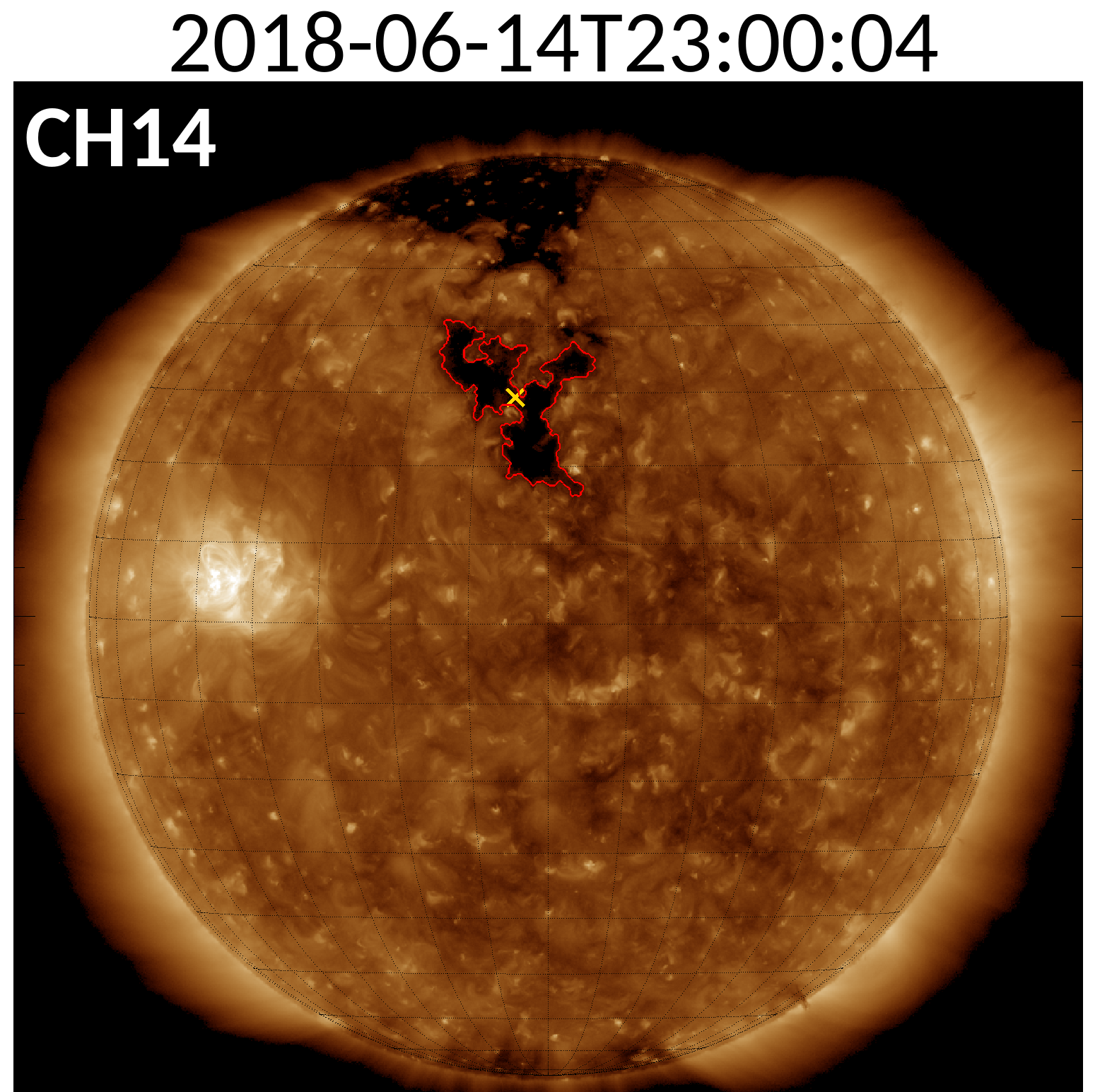}
\end{subfigure}%
\hspace*{0.1\fill}
\begin{subfigure}[]{0.24\linewidth}
\includegraphics[width=\linewidth]{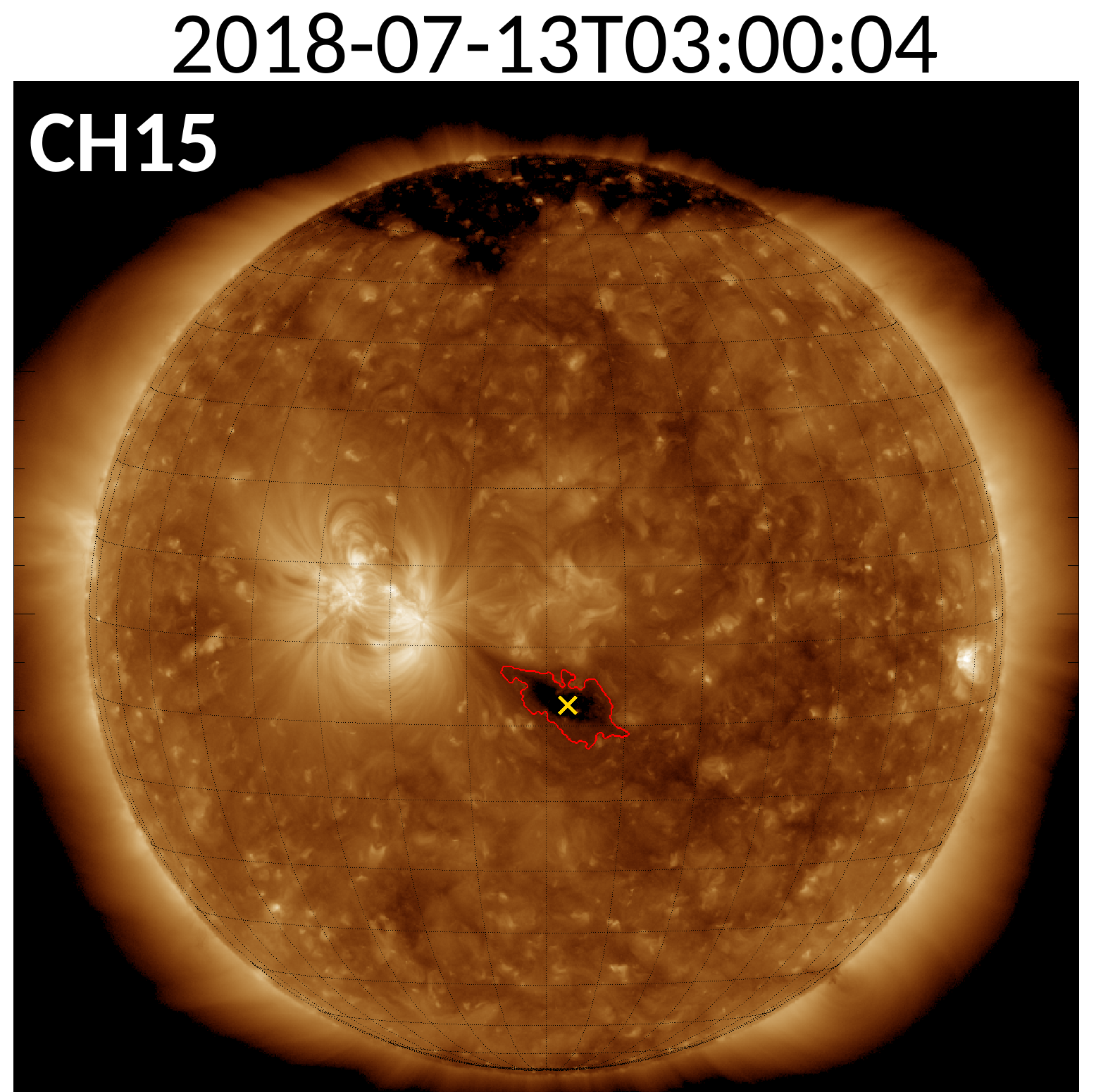}
\end{subfigure}%
 \hspace*{0.1\fill}
\begin{subfigure}[]{0.24\linewidth}
\includegraphics[width=\linewidth]{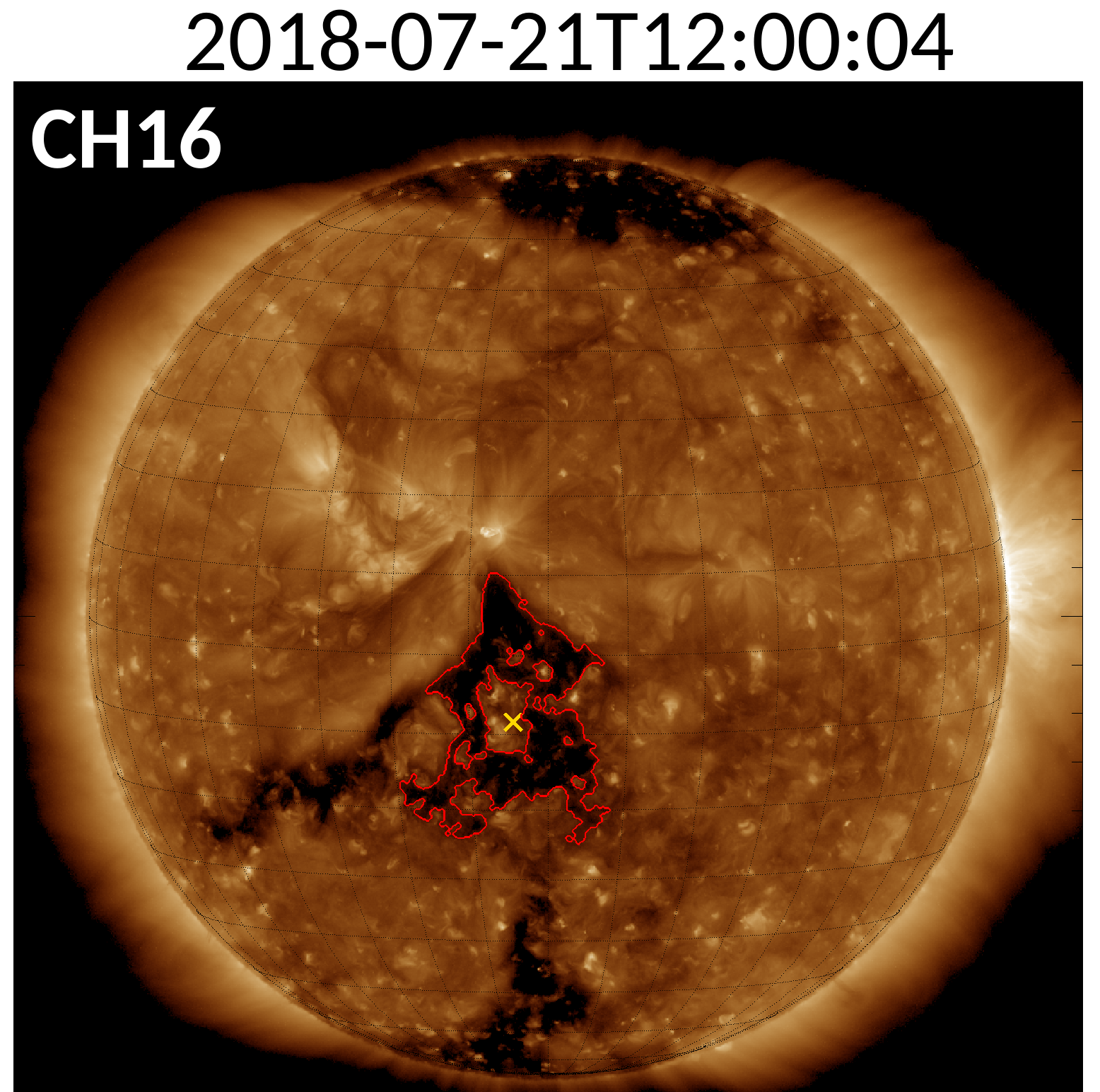}
\end{subfigure}%
 
\begin{subfigure}[]{0.24\linewidth}
\includegraphics[width=\linewidth]{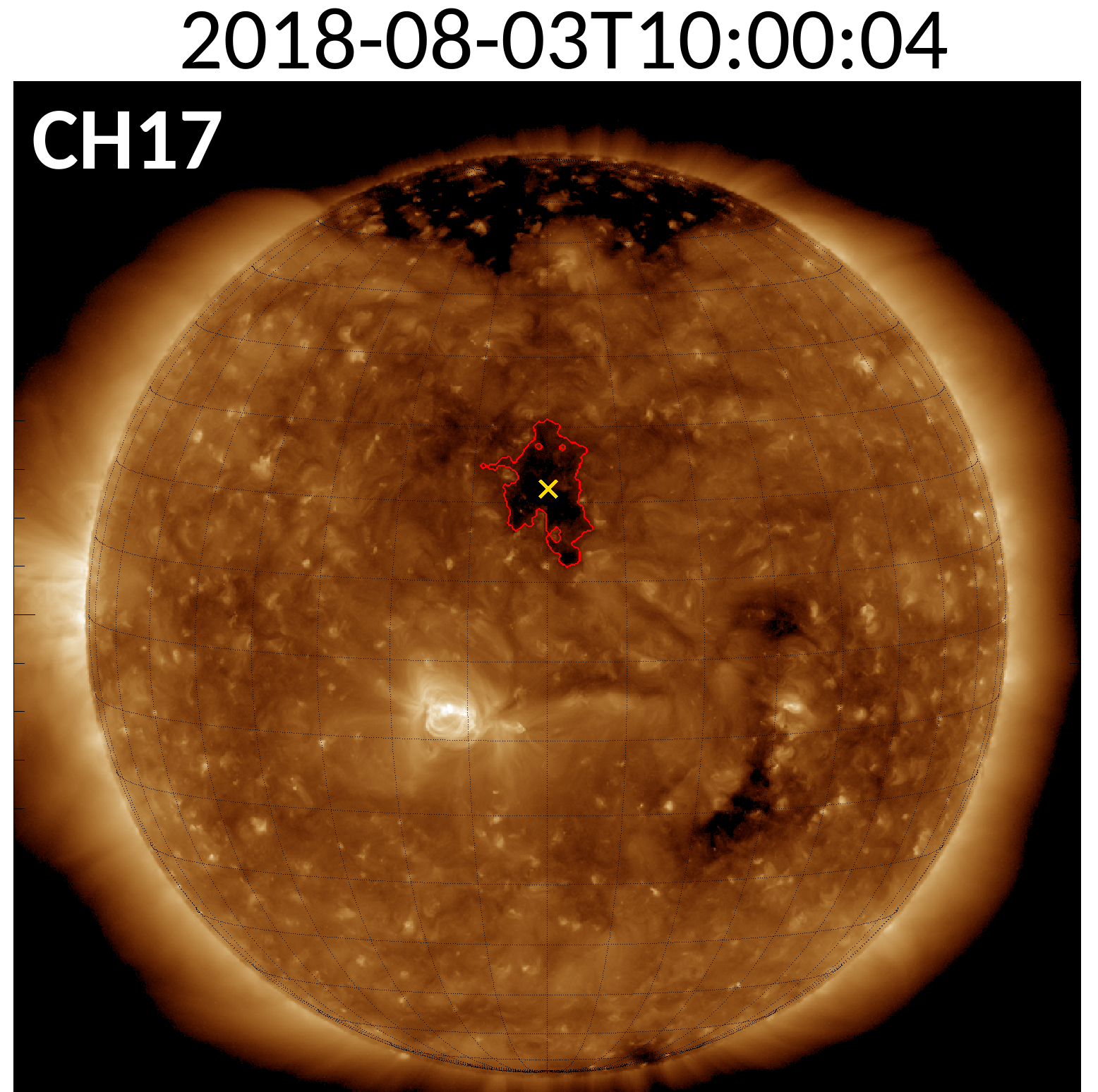}
\end{subfigure}%
 \hspace*{0.1\fill}
\begin{subfigure}[]{0.24\linewidth}
\includegraphics[width=\linewidth]{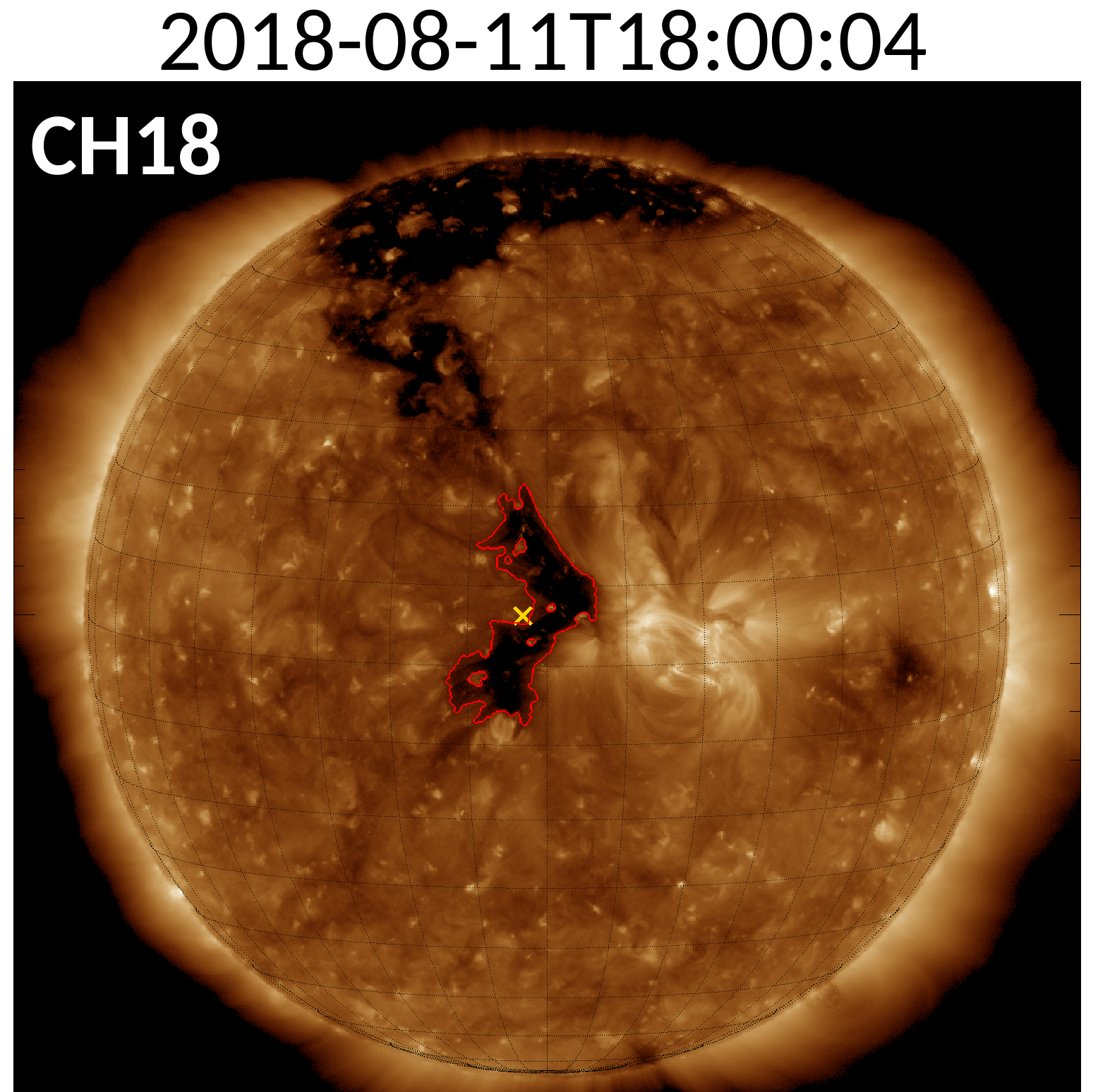}
\end{subfigure}%
 \hspace*{0.1\fill}
\begin{subfigure}[]{0.24\linewidth}
\includegraphics[width=\linewidth]{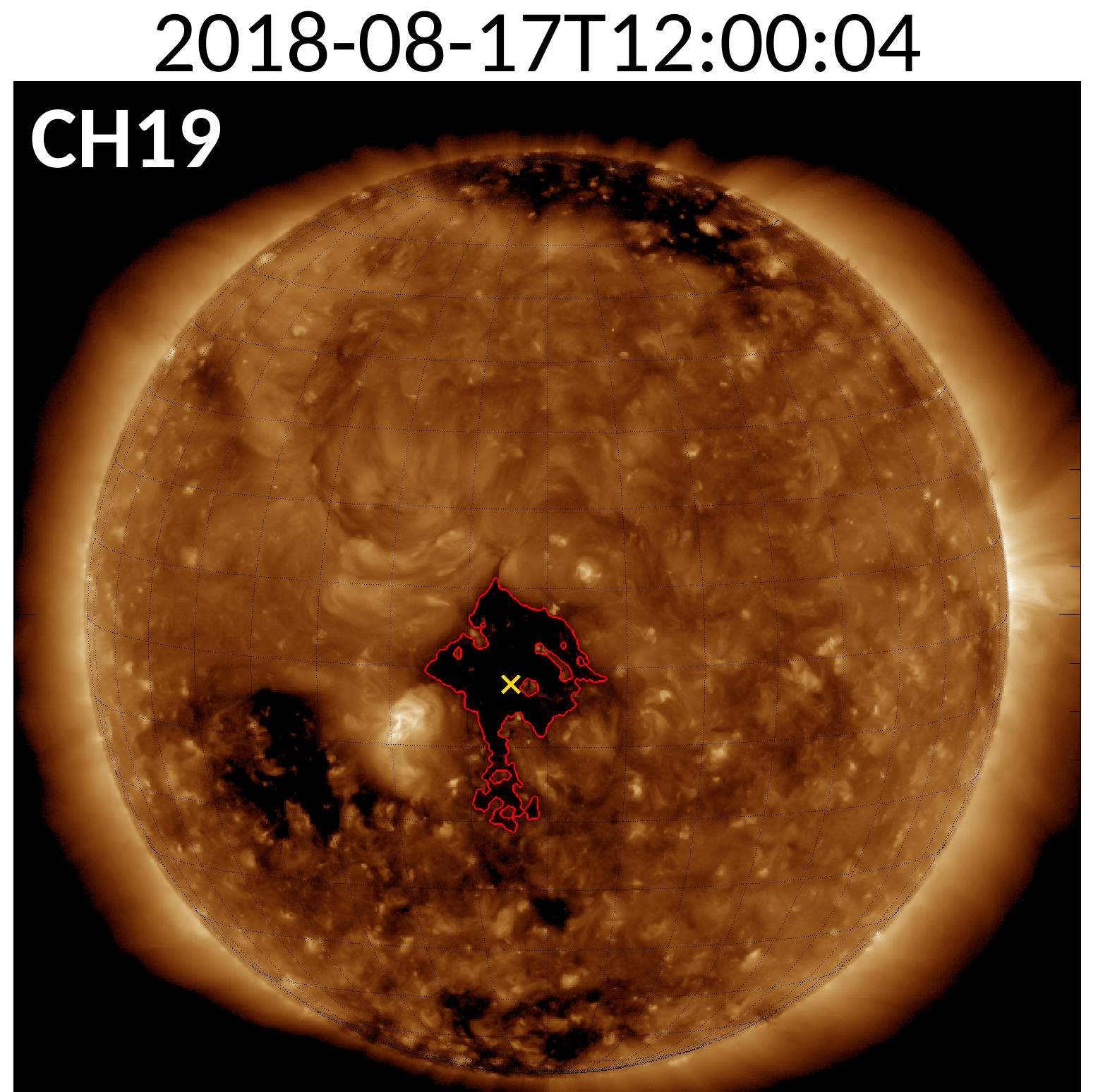}
\end{subfigure}%
\hspace*{0.1\fill}
\begin{subfigure}[]{0.24\linewidth}
\includegraphics[width=\linewidth]{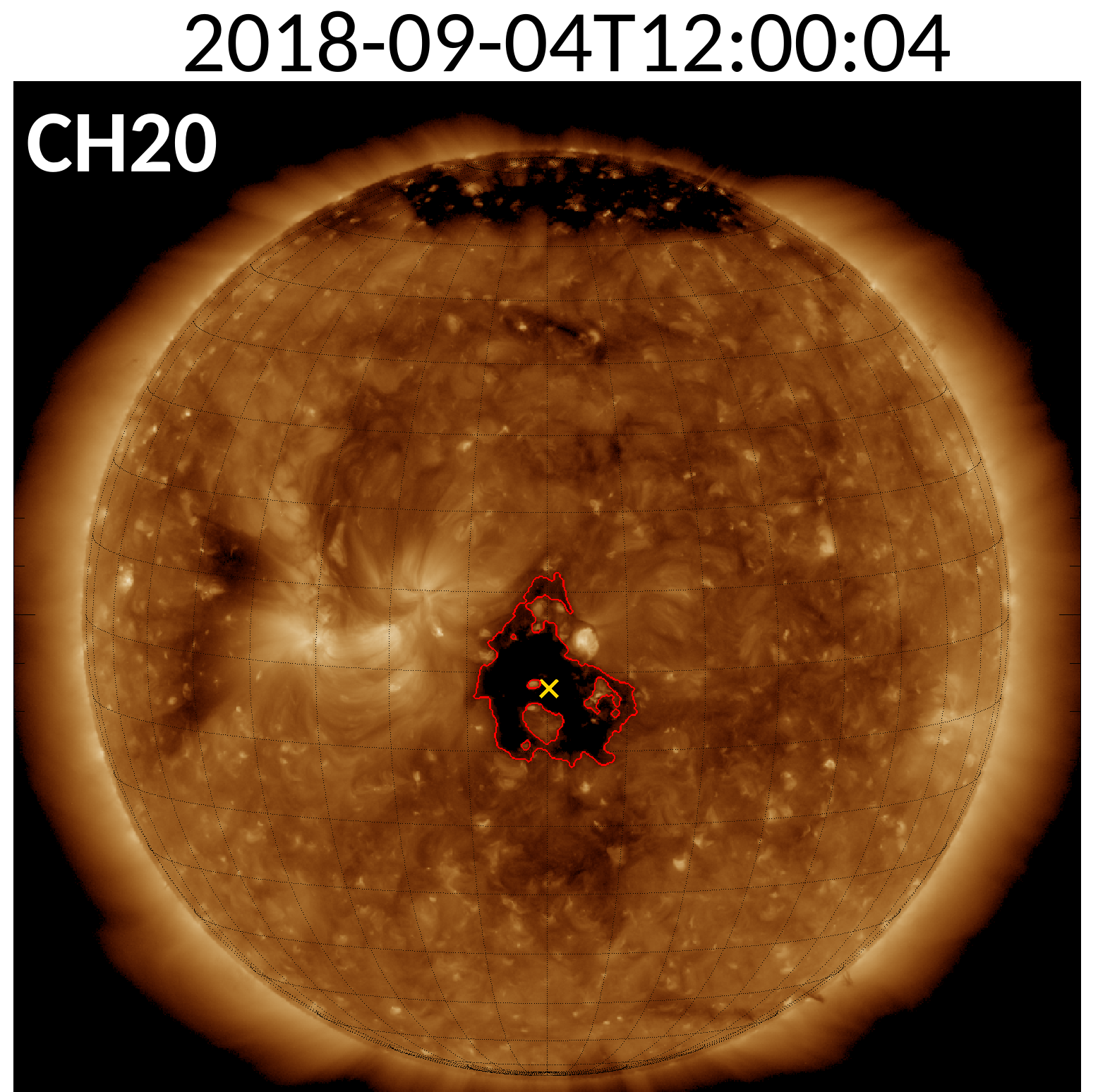}
\end{subfigure}%

\caption{Full-Sun EUV images at 193$\r{A}$ showing CHs 1-20 as extracted by CATCH. CH boundaries are outlined by red curves. The point represented by a cross in each CH corresponds to its CoM.}
\label{Fig:EUV_images}
\end{figure*}

\begin{figure*}[h!]
\centering

\begin{subfigure}[]{0.24\linewidth}
\includegraphics[width=\linewidth]{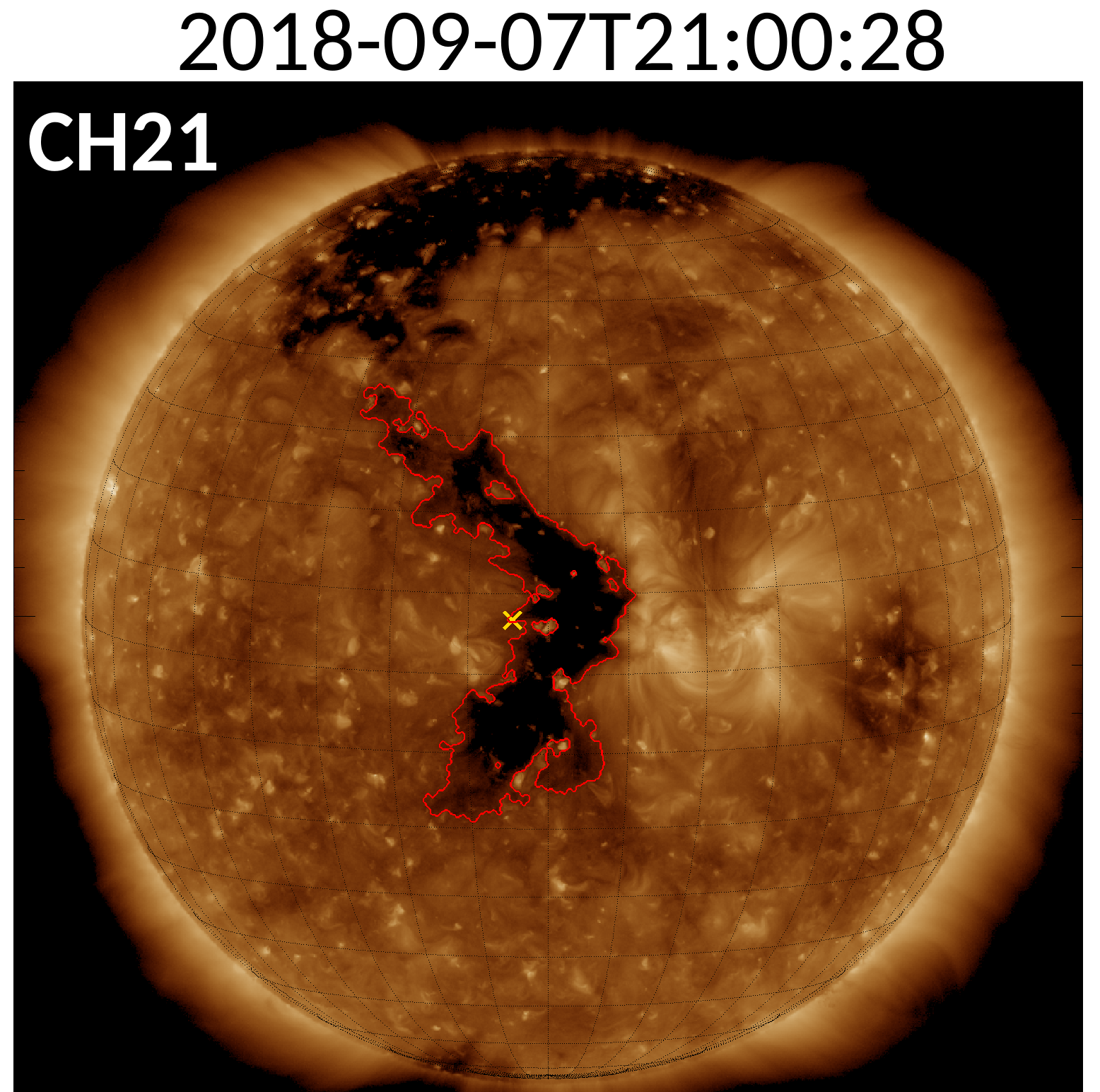}
\end{subfigure}%
 \hspace*{0.1\fill}
\begin{subfigure}[]{0.24\linewidth}
\includegraphics[width=\linewidth]{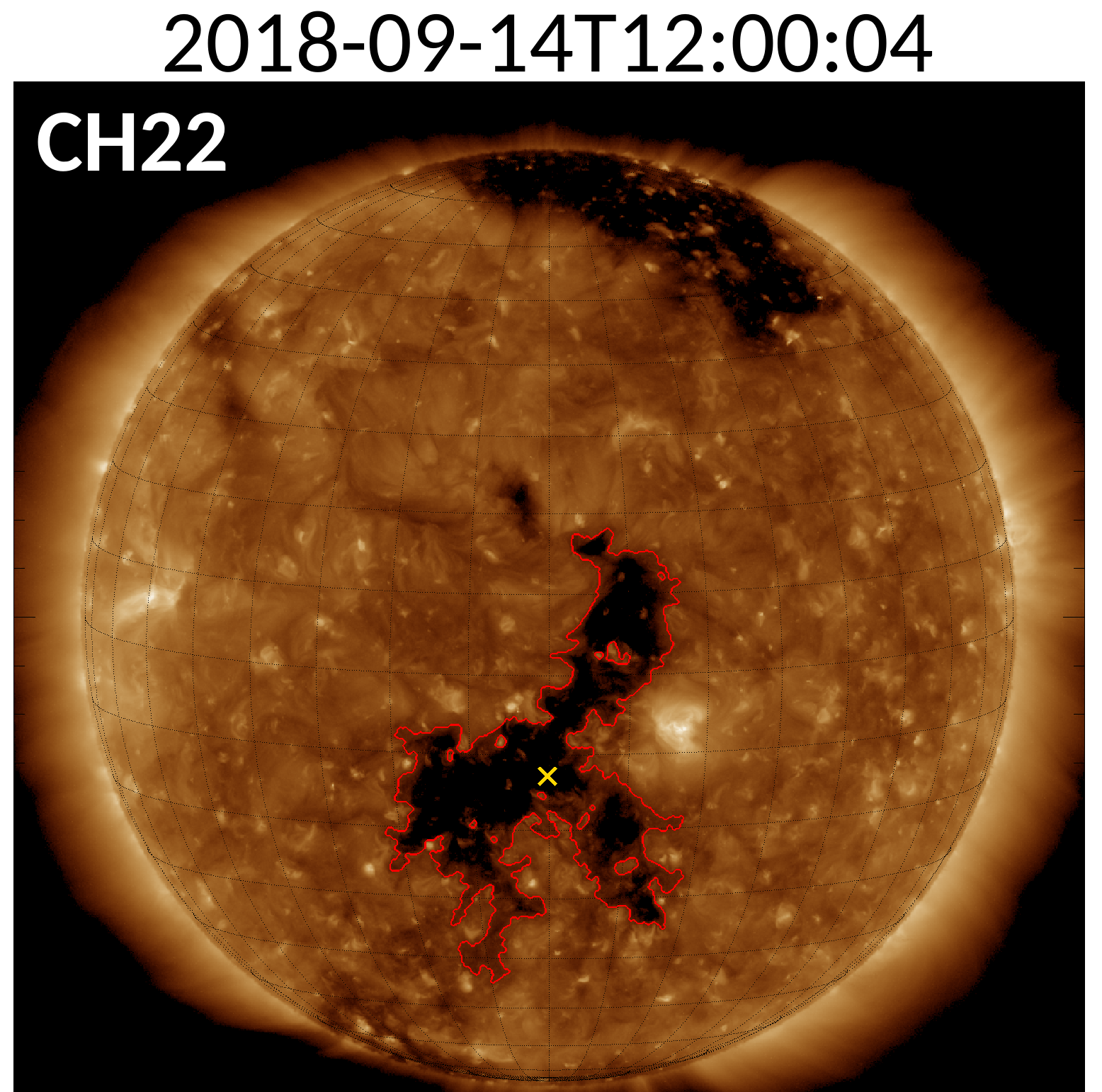}
\end{subfigure}%
 \hspace*{0.1\fill}
\begin{subfigure}[]{0.24\linewidth}
\includegraphics[width=\linewidth]{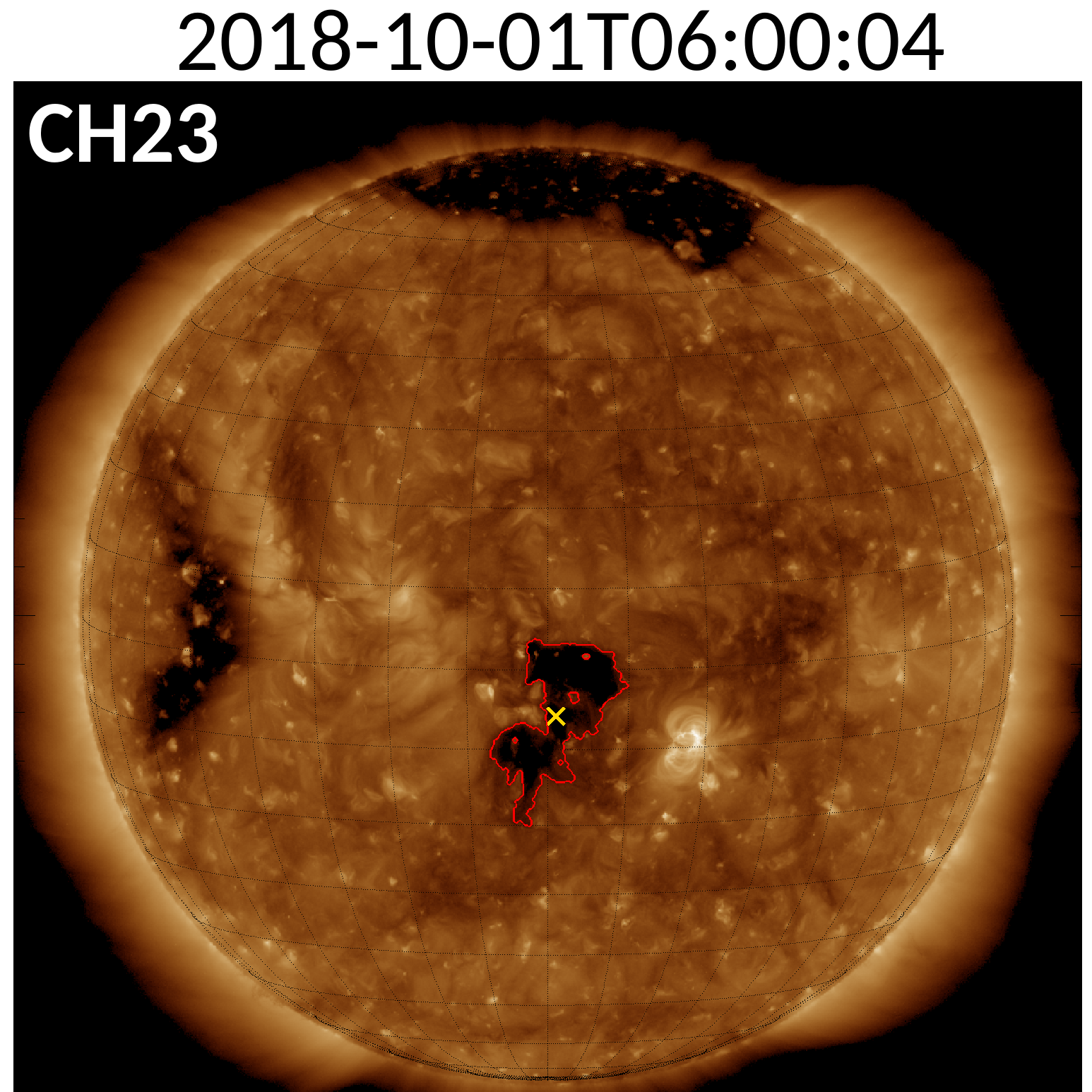}
\end{subfigure}%
 \hspace*{0.1\fill}
\begin{subfigure}[]{0.24\linewidth}
\includegraphics[width=\linewidth]{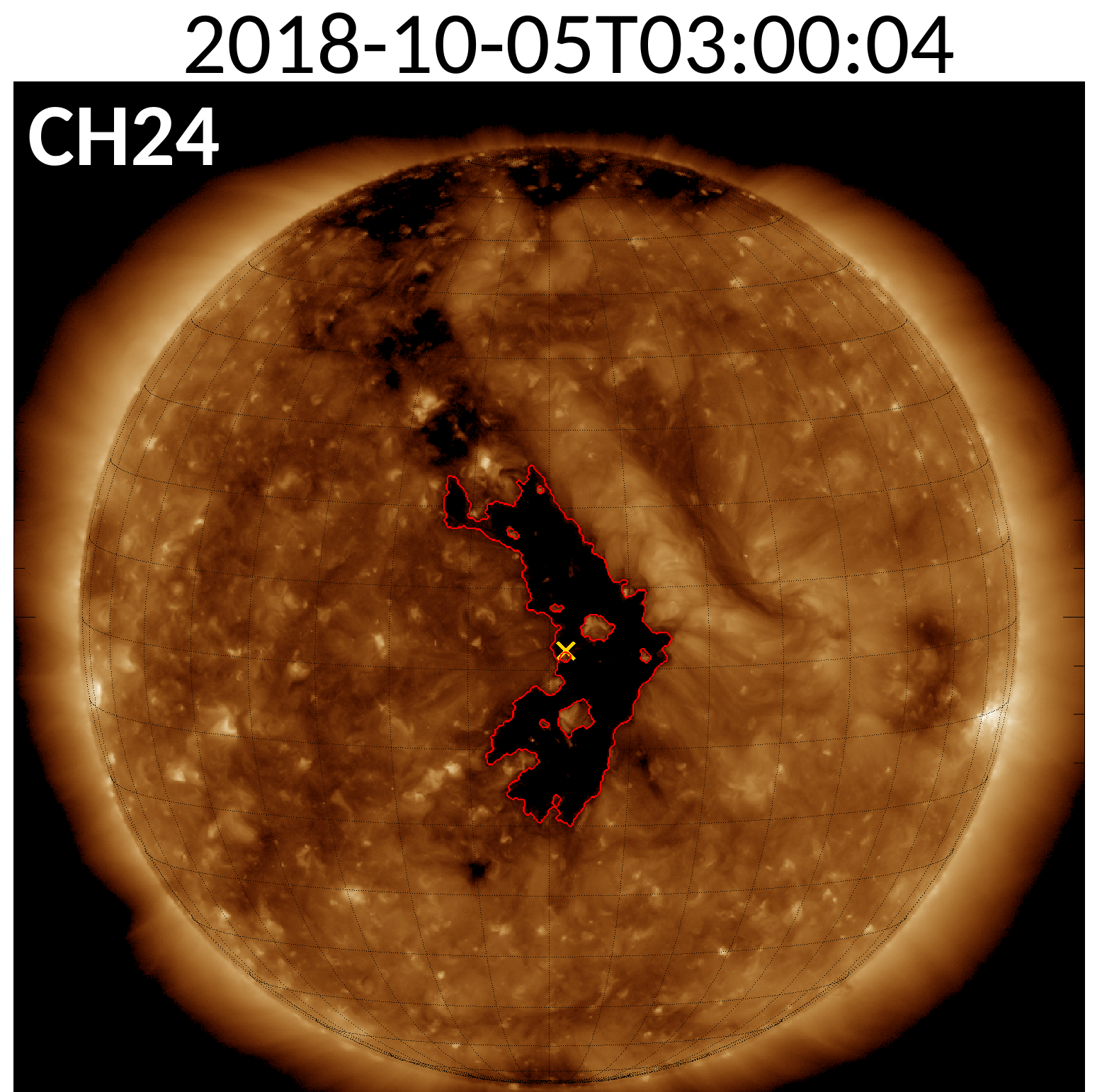}
\end{subfigure}%

\begin{subfigure}[]{0.24\linewidth}
\includegraphics[width=\linewidth]{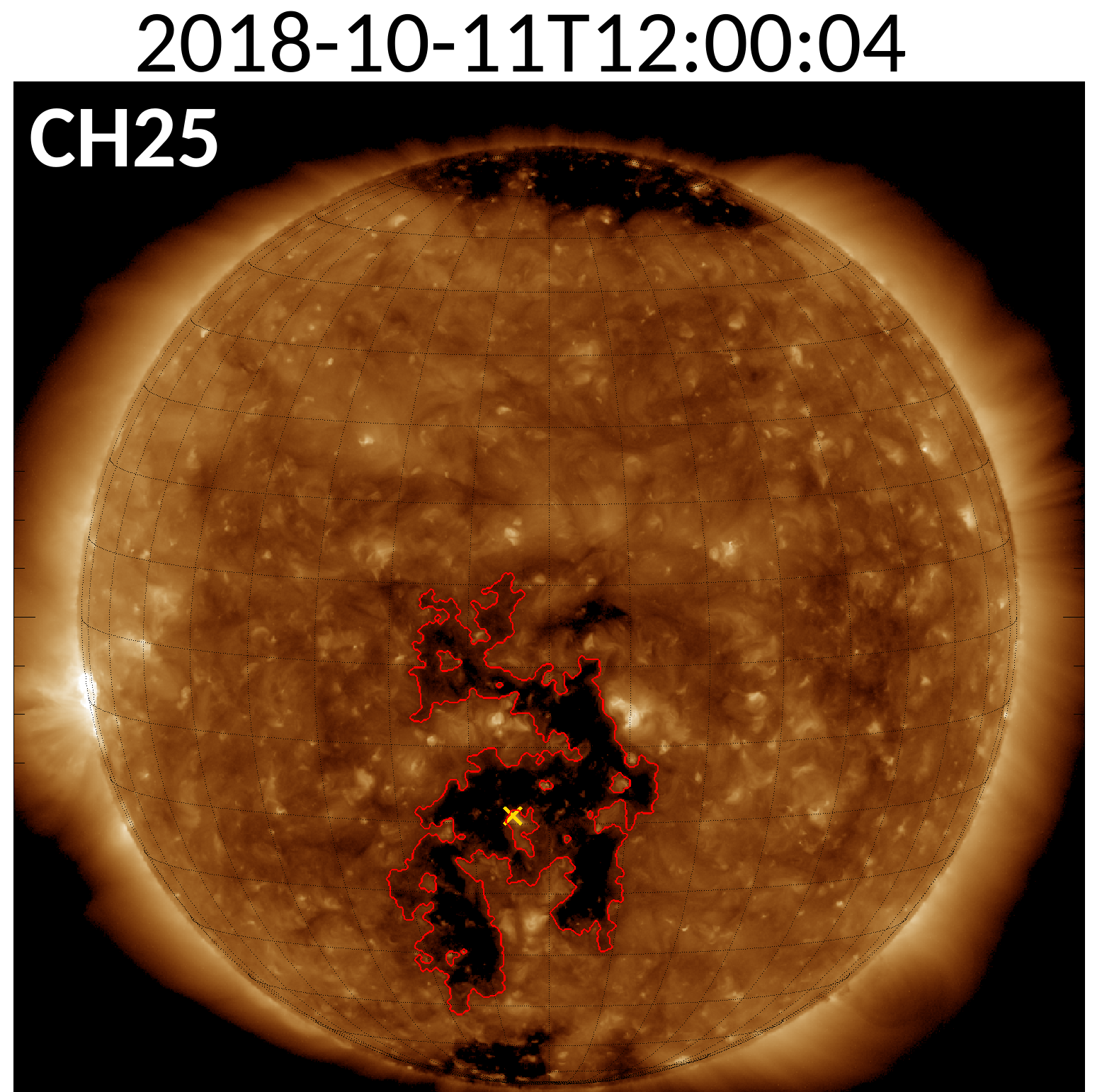}
\end{subfigure}%
\hspace*{0.1\fill}
\begin{subfigure}[]{0.24\linewidth}
\includegraphics[width=\linewidth]{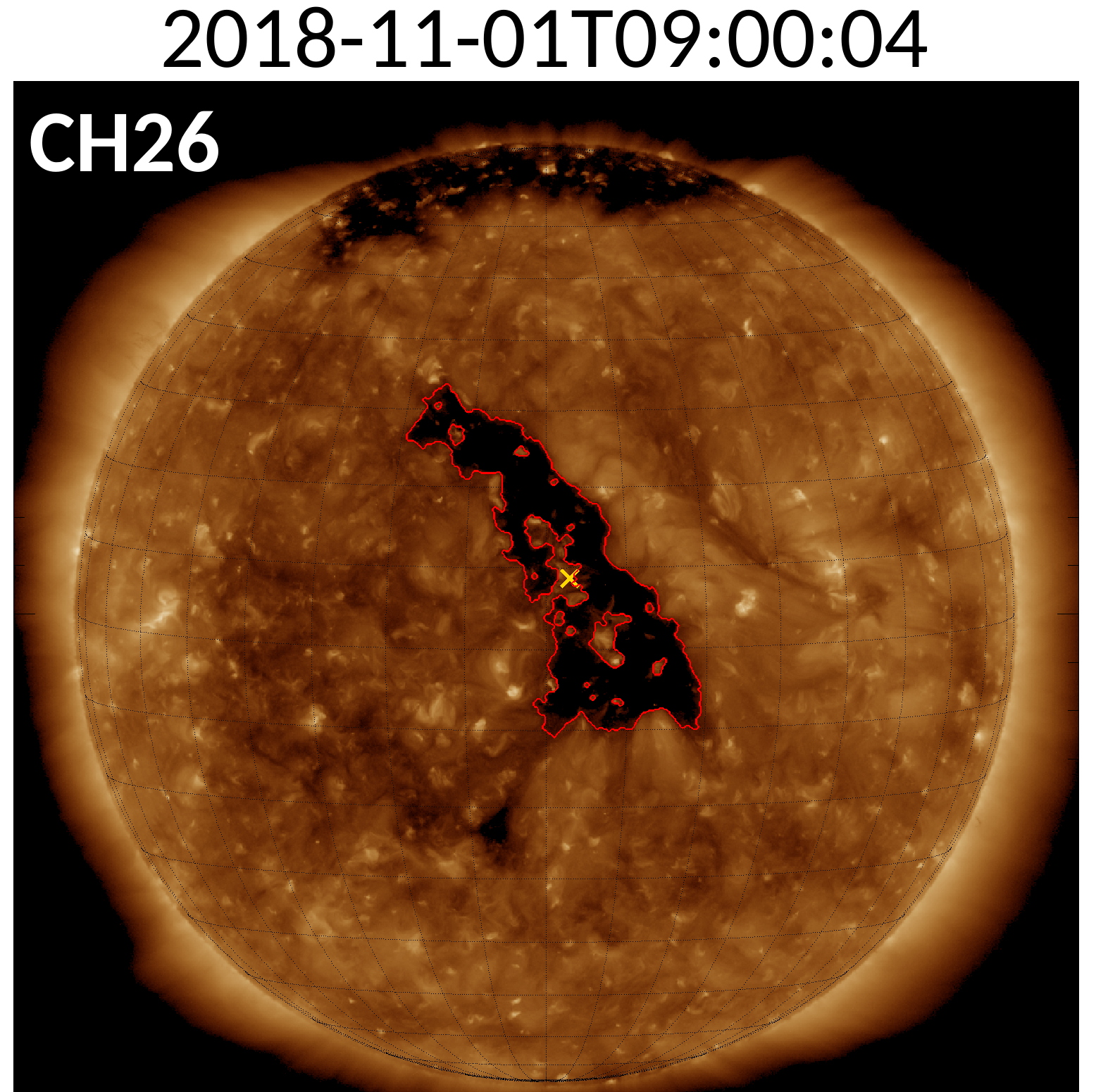}
\end{subfigure}%
 \hspace*{0.1\fill}
\begin{subfigure}[]{0.24\linewidth}
\includegraphics[width=\linewidth]{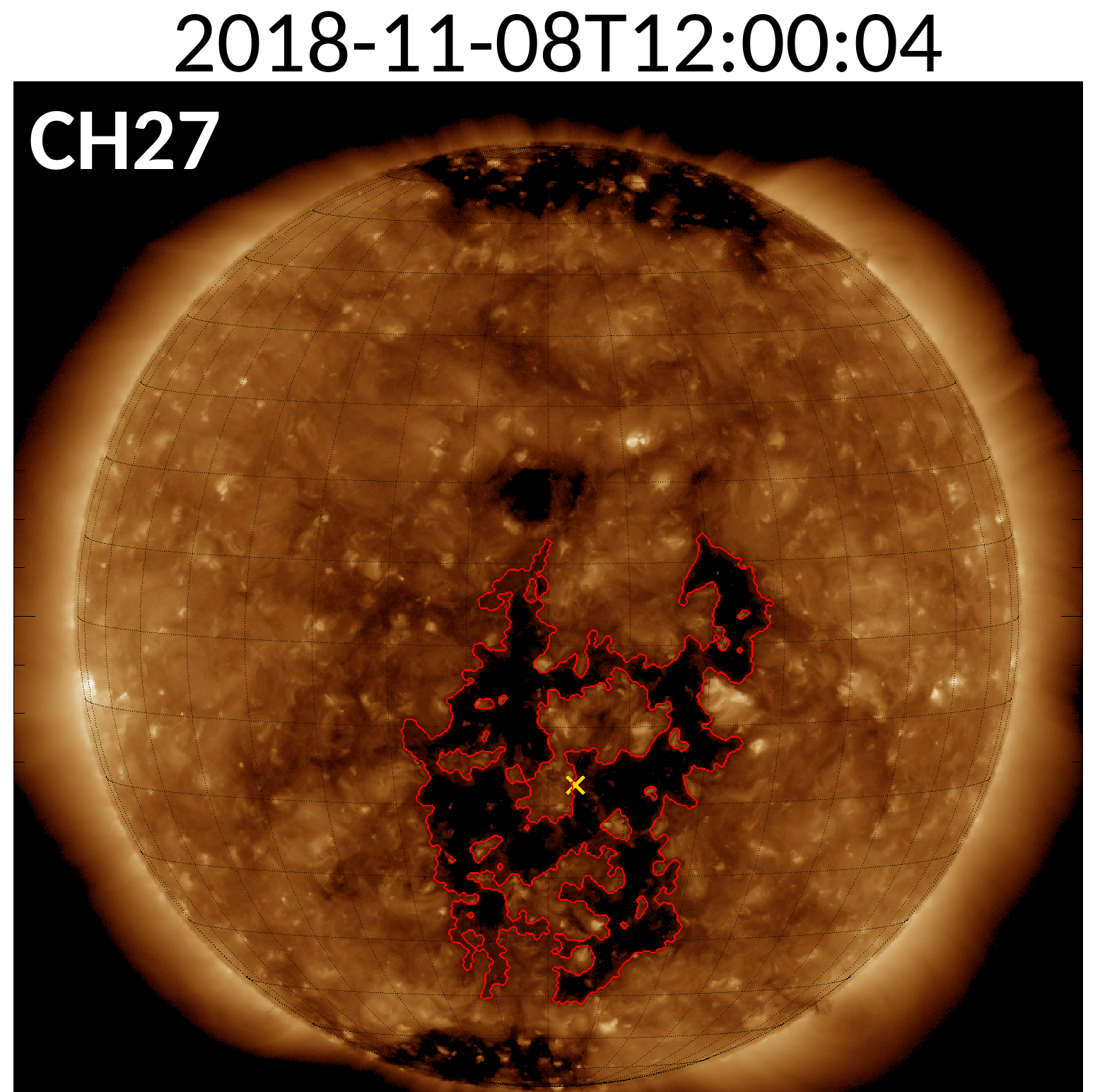}
\end{subfigure}%
 \hspace*{0.1\fill}
\begin{subfigure}[]{0.24\linewidth}
\includegraphics[width=\linewidth]{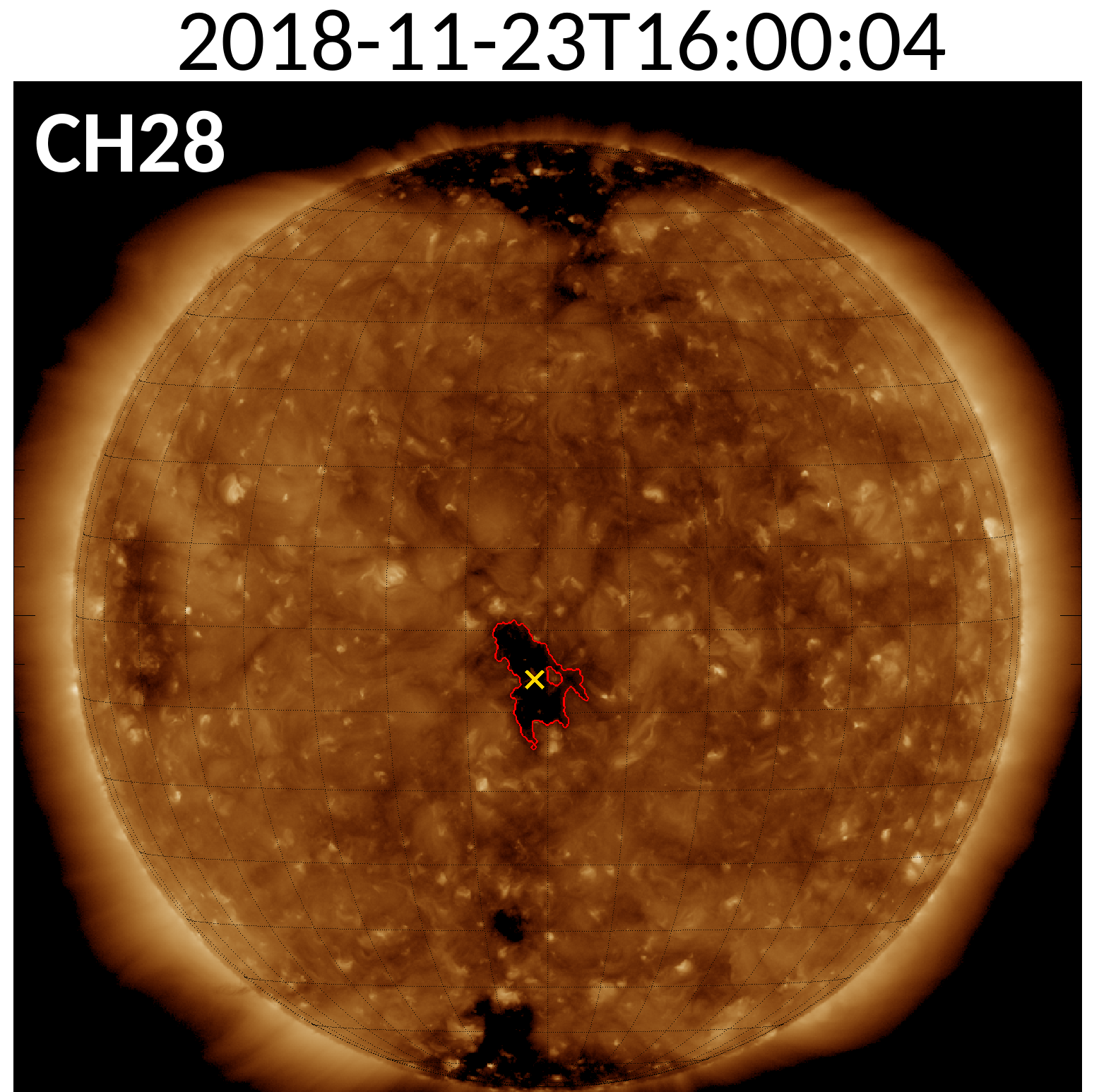}
\end{subfigure}%

\begin{subfigure}[]{0.24\linewidth}
\includegraphics[width=\linewidth]{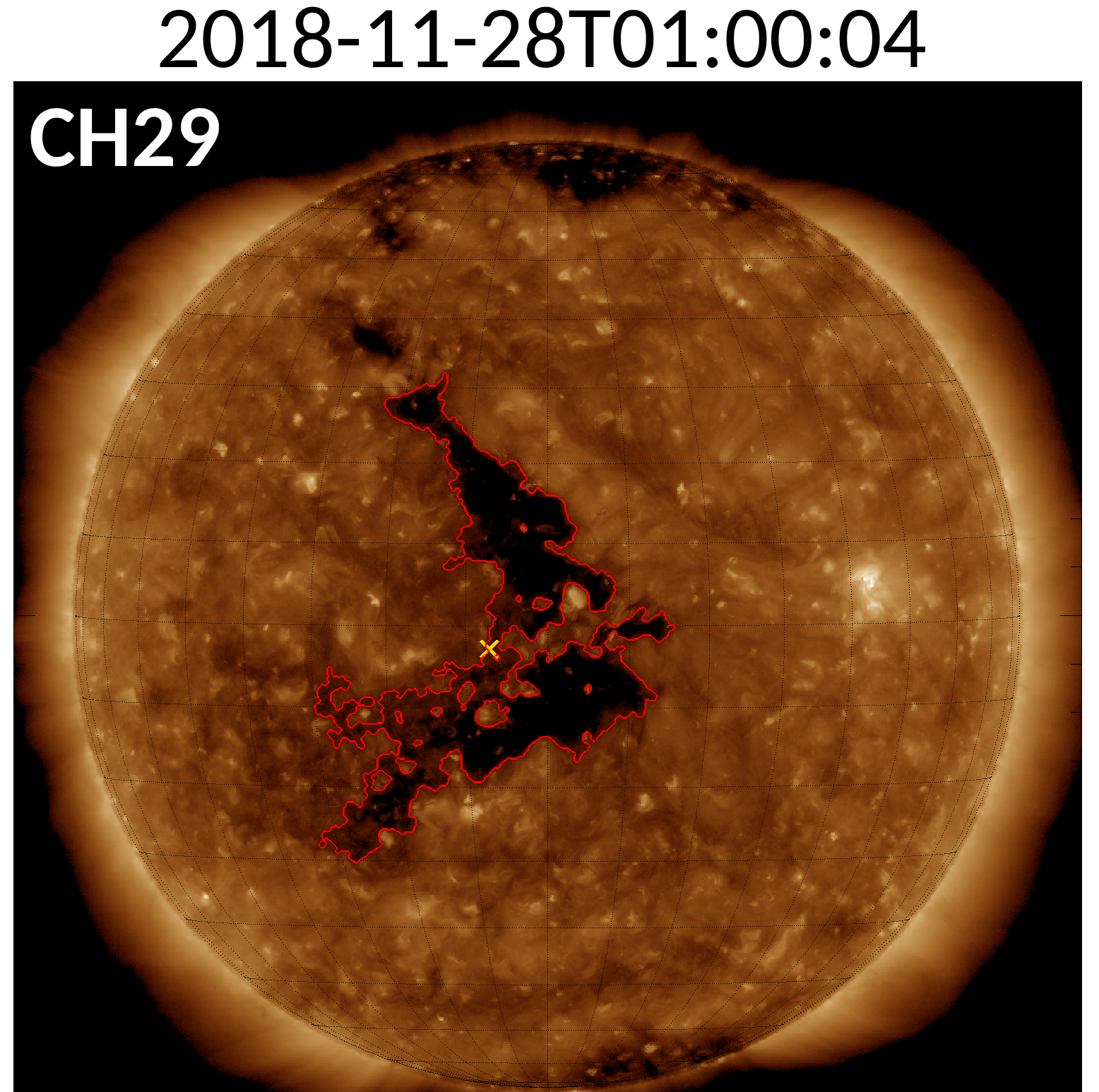}
\end{subfigure}%
\hspace*{0.1\fill}
\begin{subfigure}[]{0.24\linewidth}
\includegraphics[width=\linewidth]{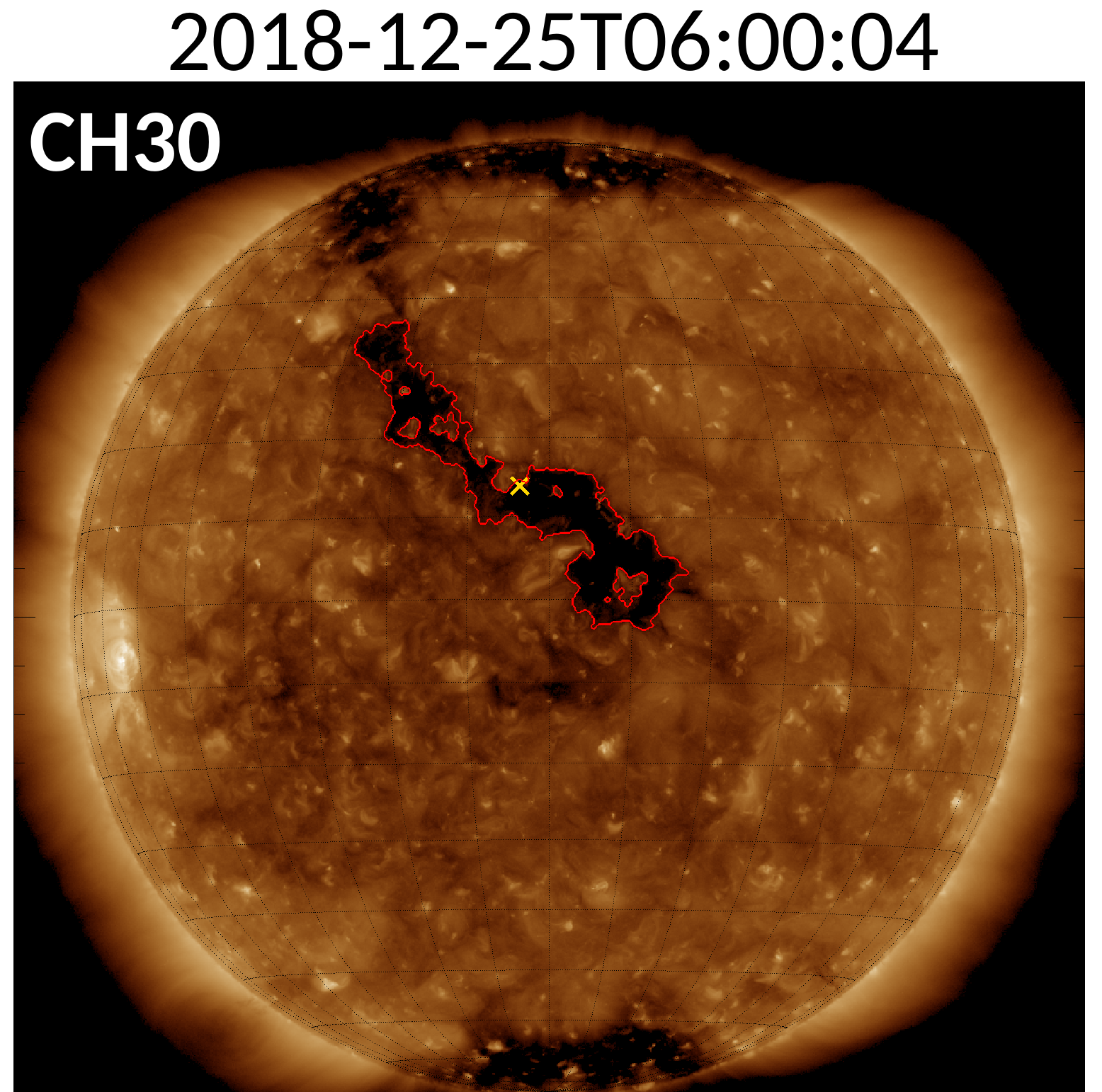}
\end{subfigure}%
\hspace*{0.1\fill}
\begin{subfigure}[]{0.24\linewidth}
\includegraphics[width=\linewidth]{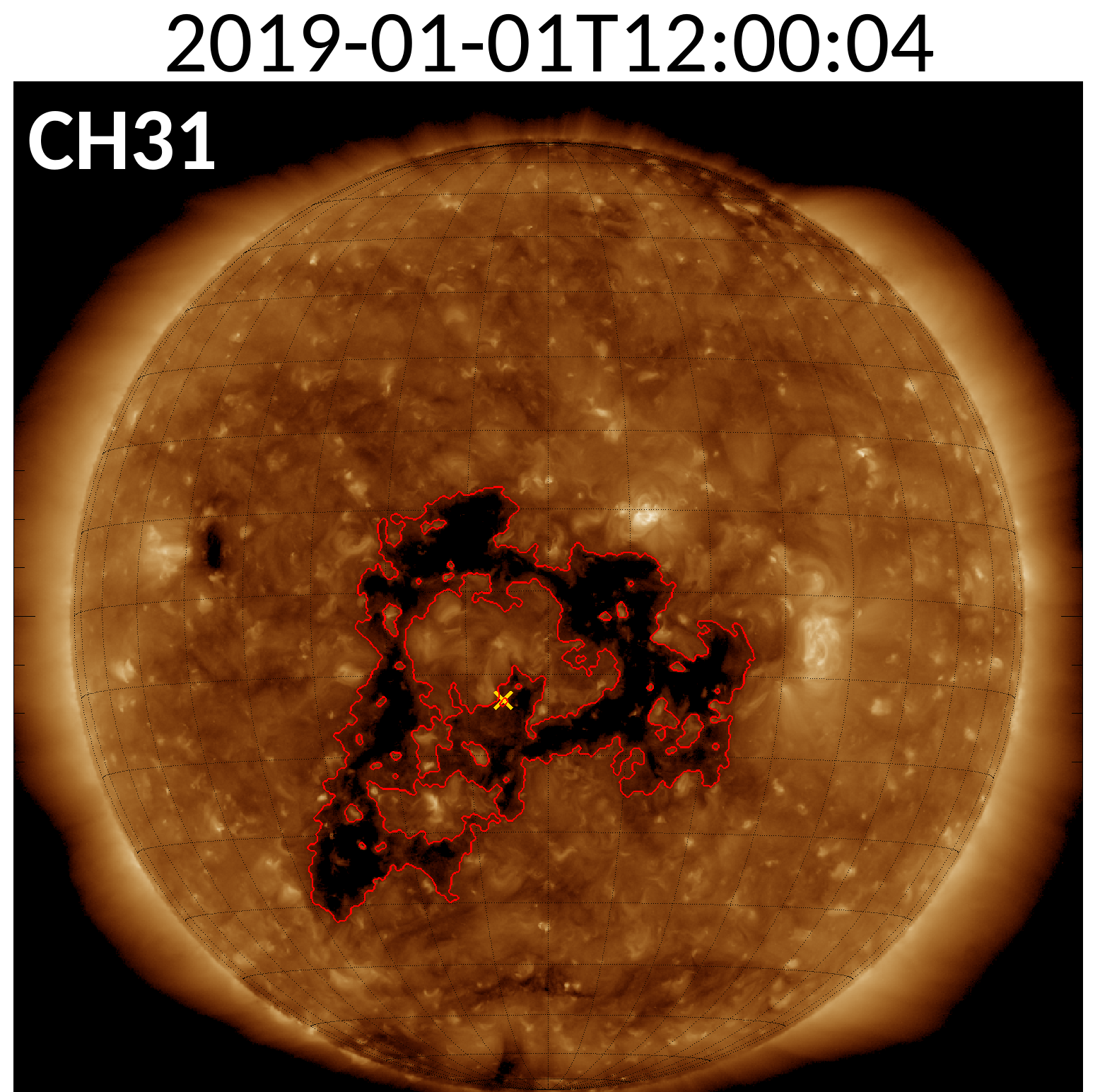}
\end{subfigure}%
 \hspace*{0.1\fill}
\begin{subfigure}[]{0.24\linewidth}
\includegraphics[width=\linewidth]{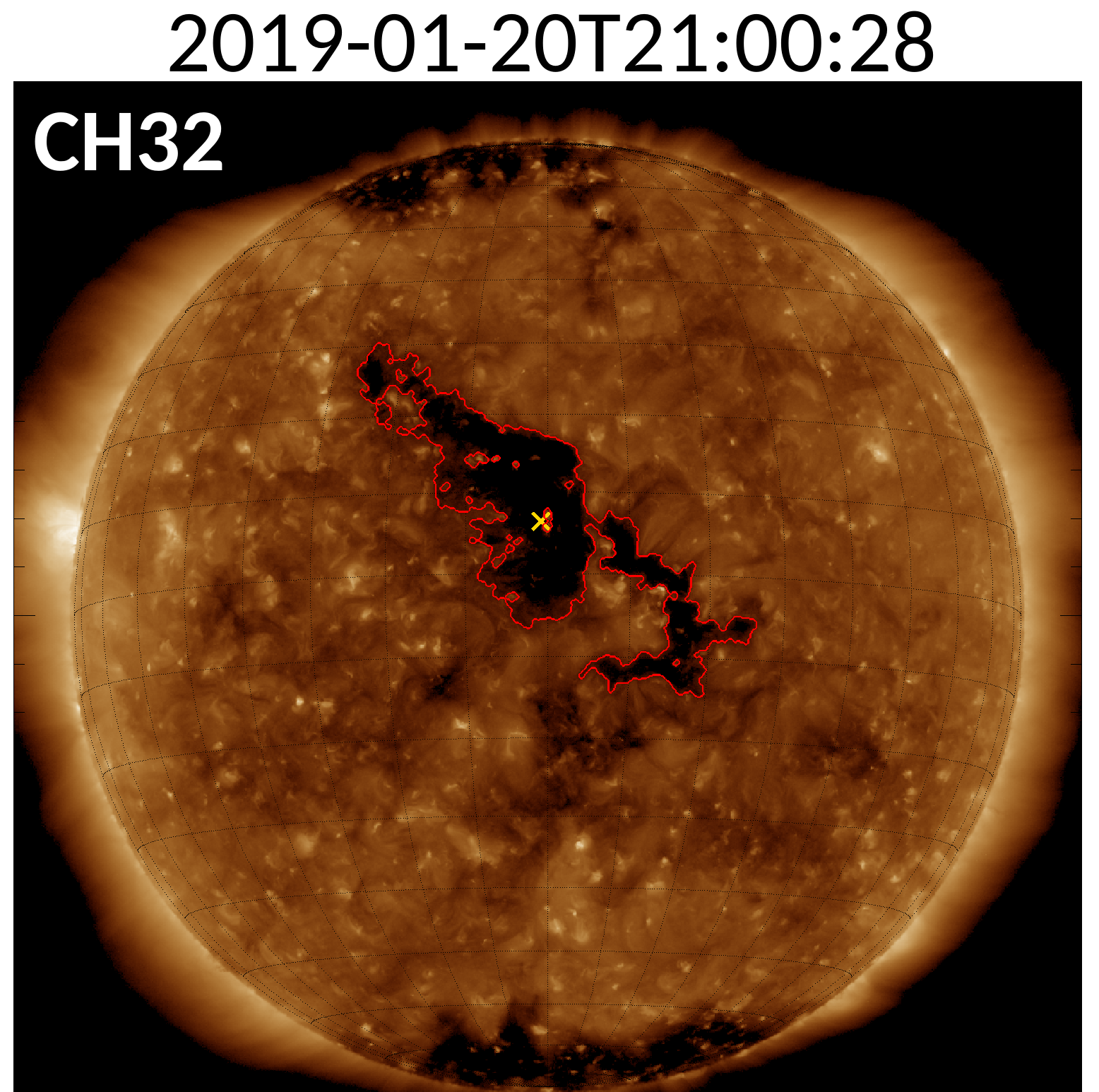}
\end{subfigure}%

\begin{subfigure}[]{0.24\linewidth}
\includegraphics[width=\linewidth]{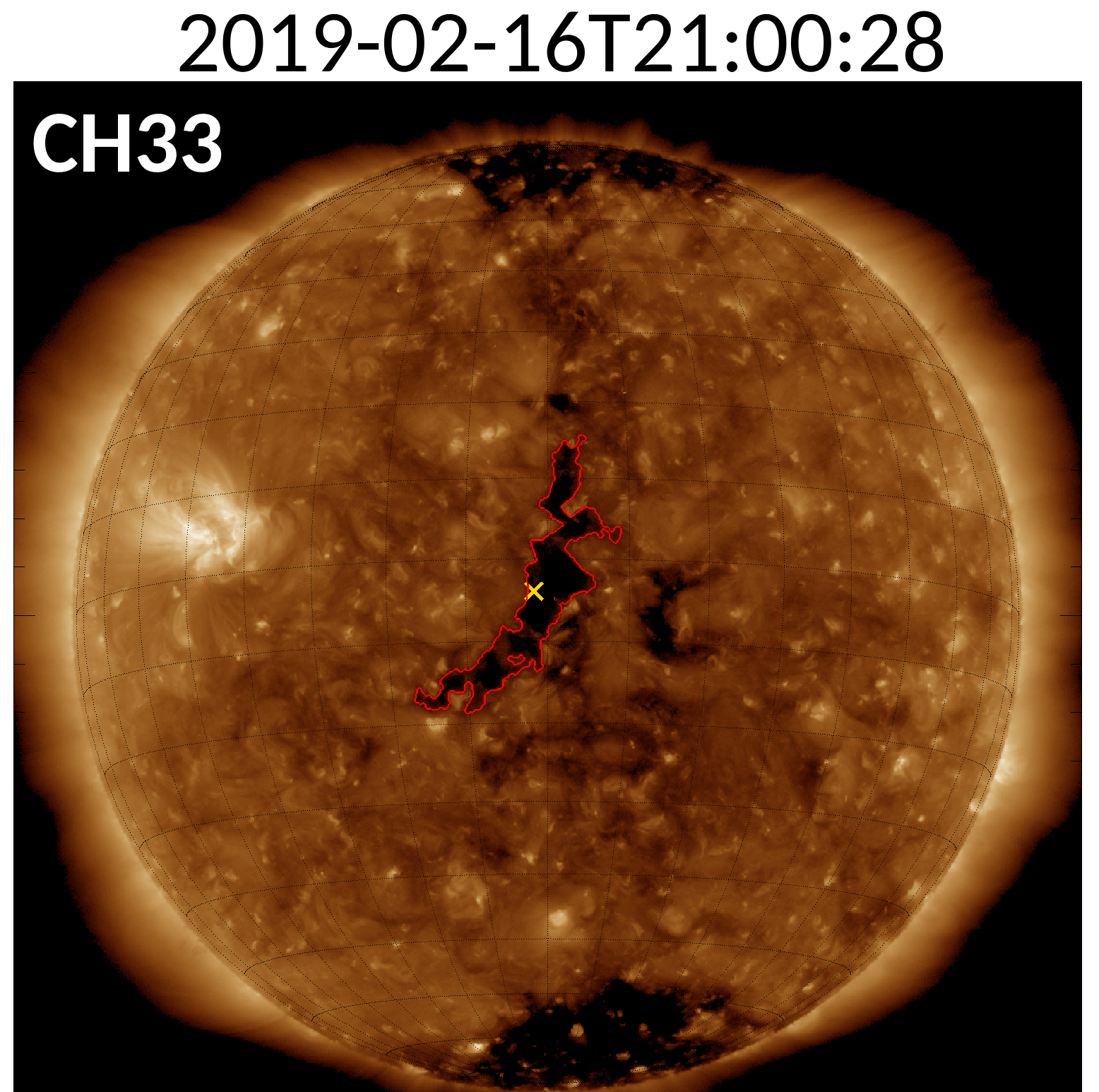}
\end{subfigure}%
\hspace*{0.1\fill}
\begin{subfigure}[]{0.24\linewidth}
\includegraphics[width=\linewidth]{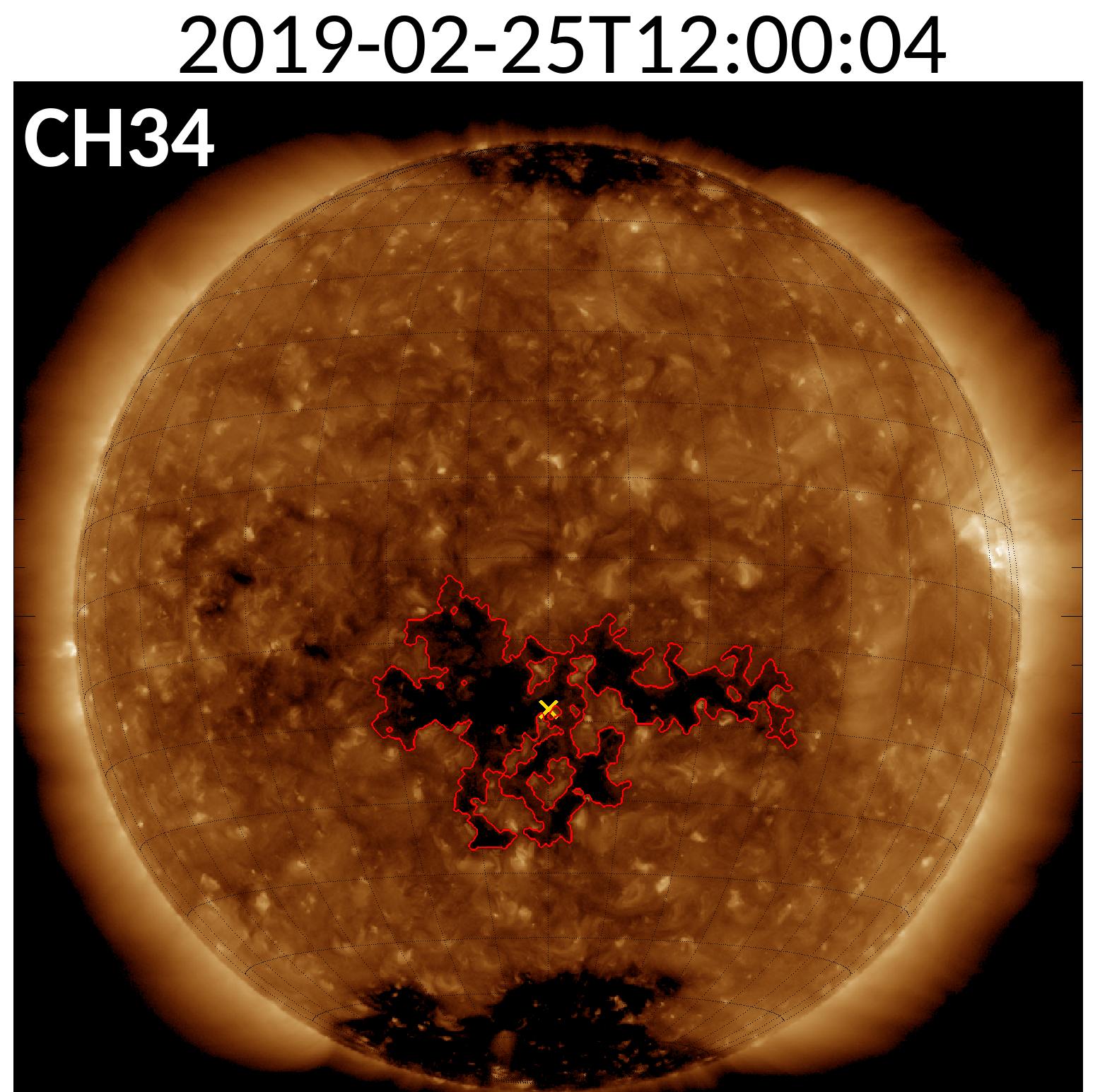}
\end{subfigure}%
\hspace*{0.1\fill}
\begin{subfigure}[]{0.24\linewidth}
\includegraphics[width=\linewidth]{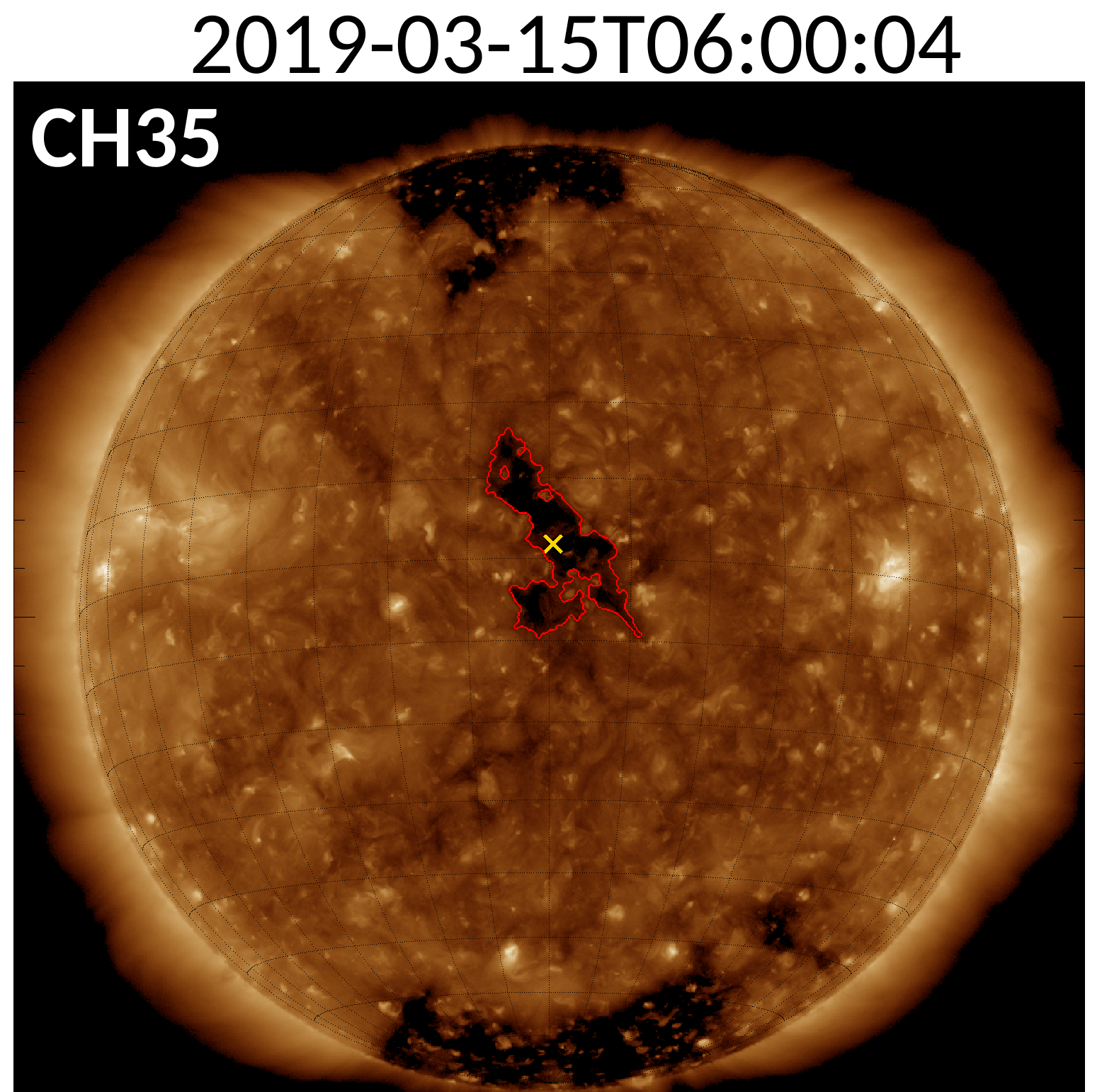}
\end{subfigure}%
\hspace*{0.1\fill}
\begin{subfigure}[]{0.24\linewidth}
\includegraphics[width=\linewidth]{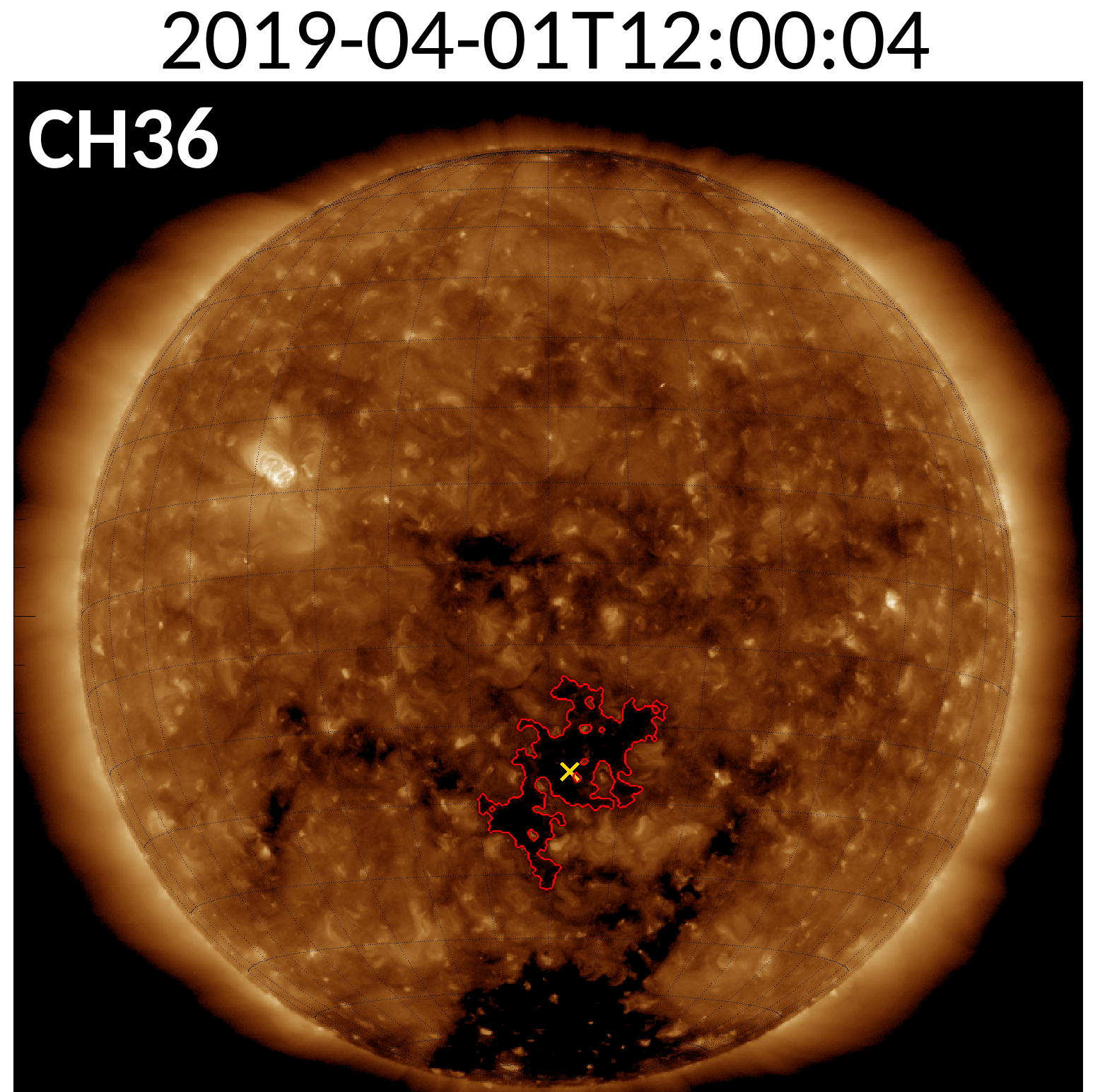}
\end{subfigure}
 
\begin{subfigure}[]{0.24\linewidth}
\includegraphics[width=\linewidth]{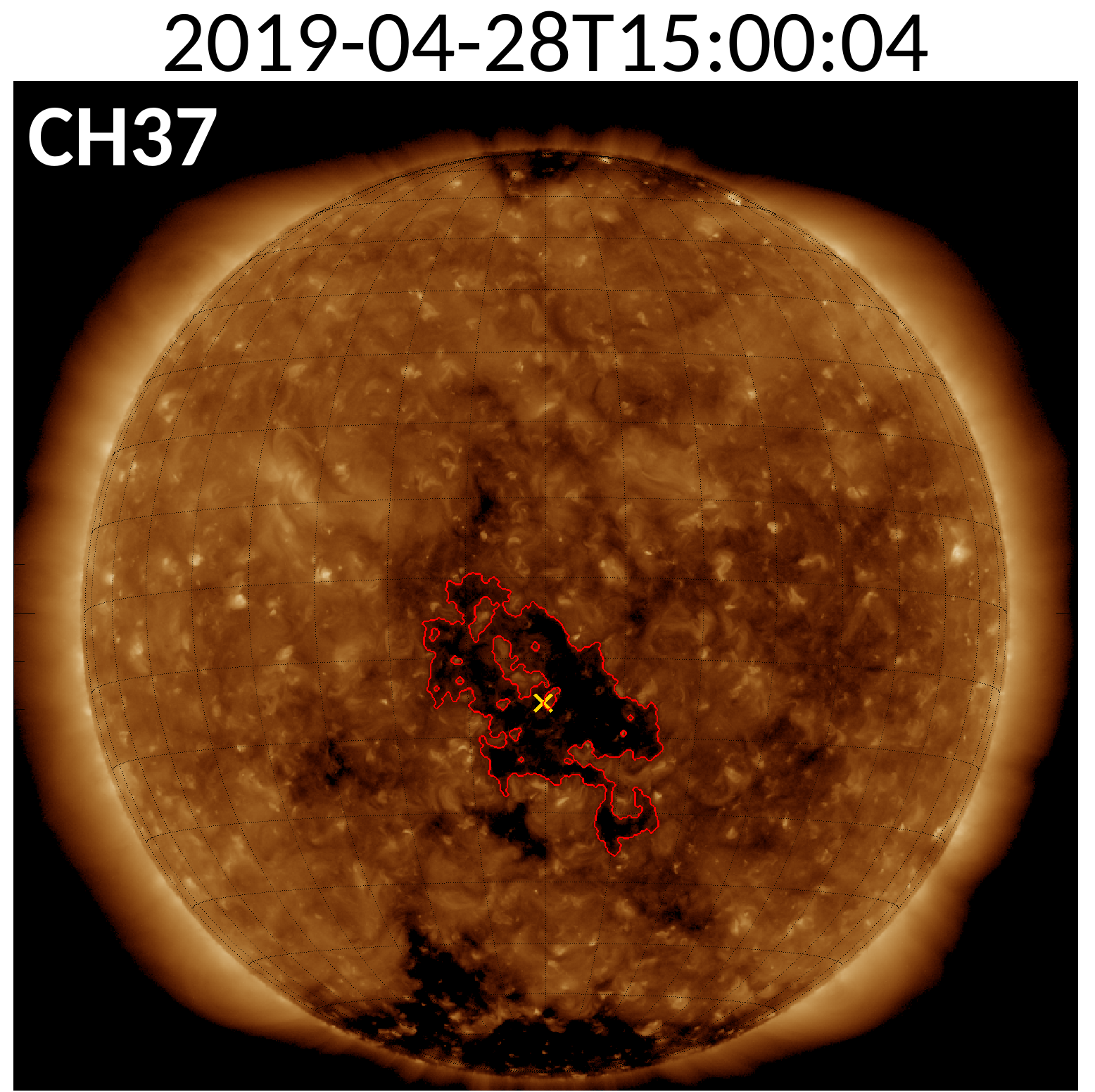}
\end{subfigure}%
 \hspace*{0.1\fill}
\begin{subfigure}[]{0.24\linewidth}
\includegraphics[width=\linewidth]{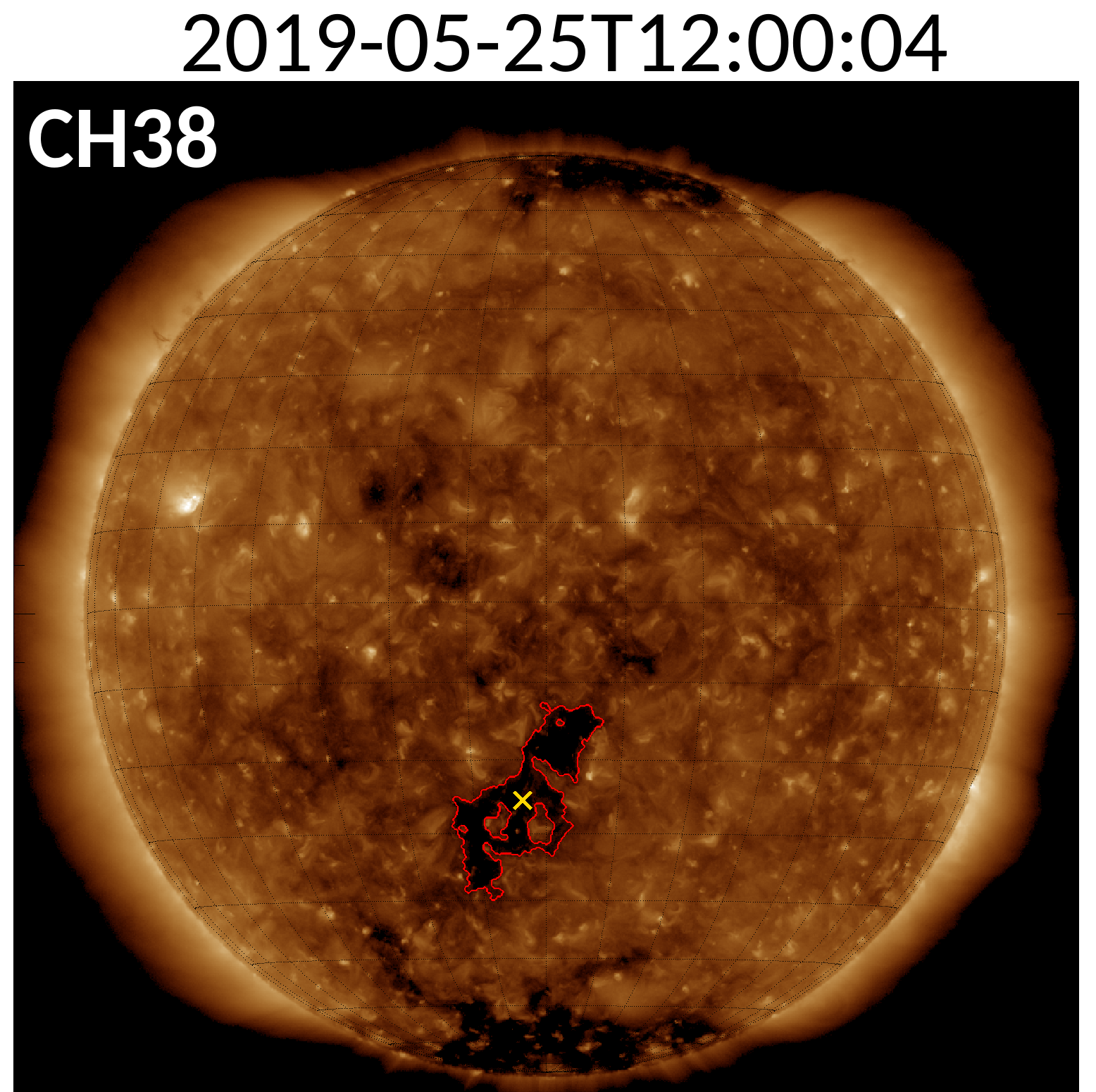}
\end{subfigure}%
 \hspace*{0.1\fill}
\begin{subfigure}[]{0.24\linewidth}
\includegraphics[width=\linewidth]{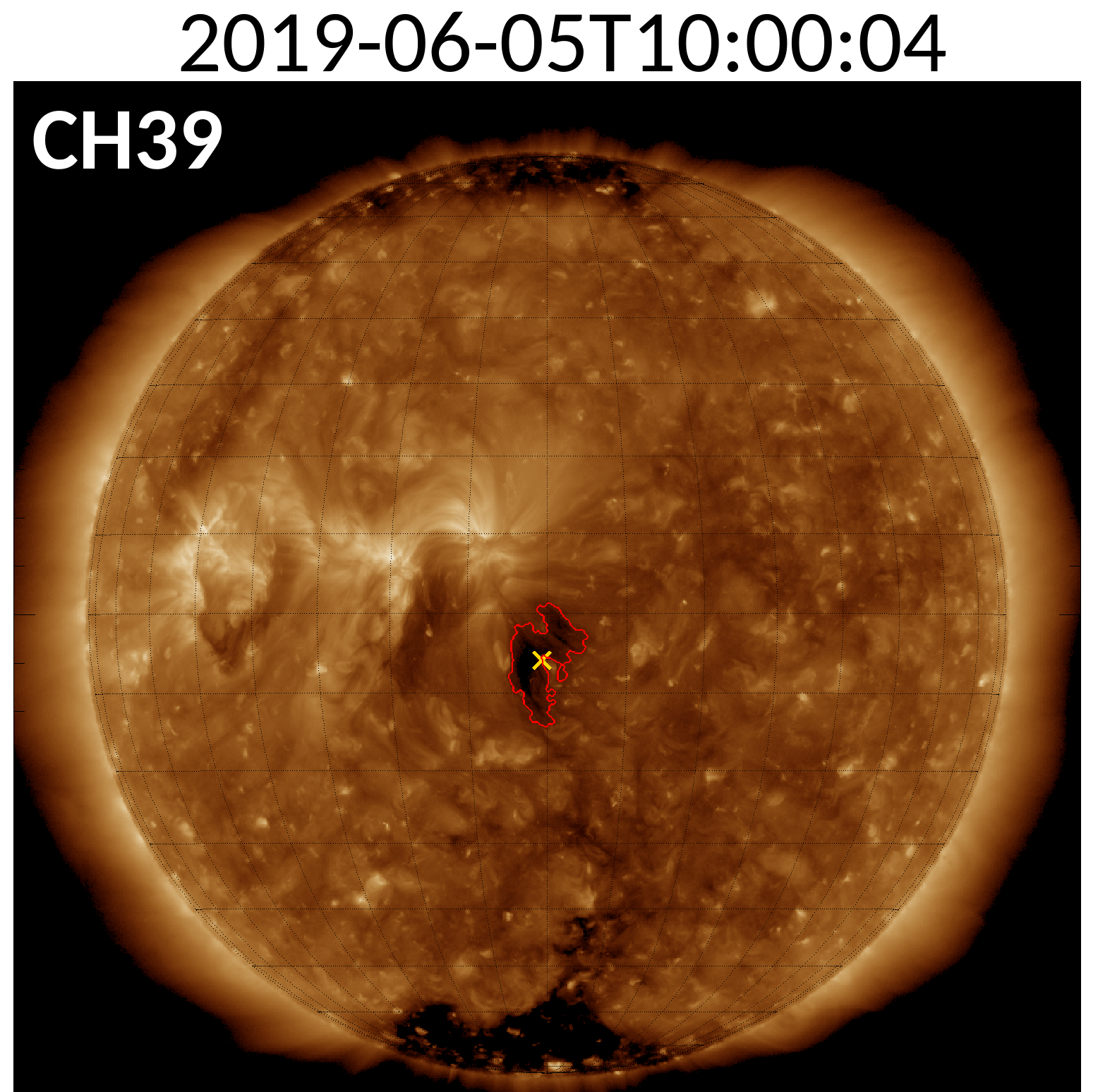}
\end{subfigure}%
 \hspace*{0.1\fill}
 \begin{subfigure}[]{0.24\linewidth}
\includegraphics[width=\linewidth]{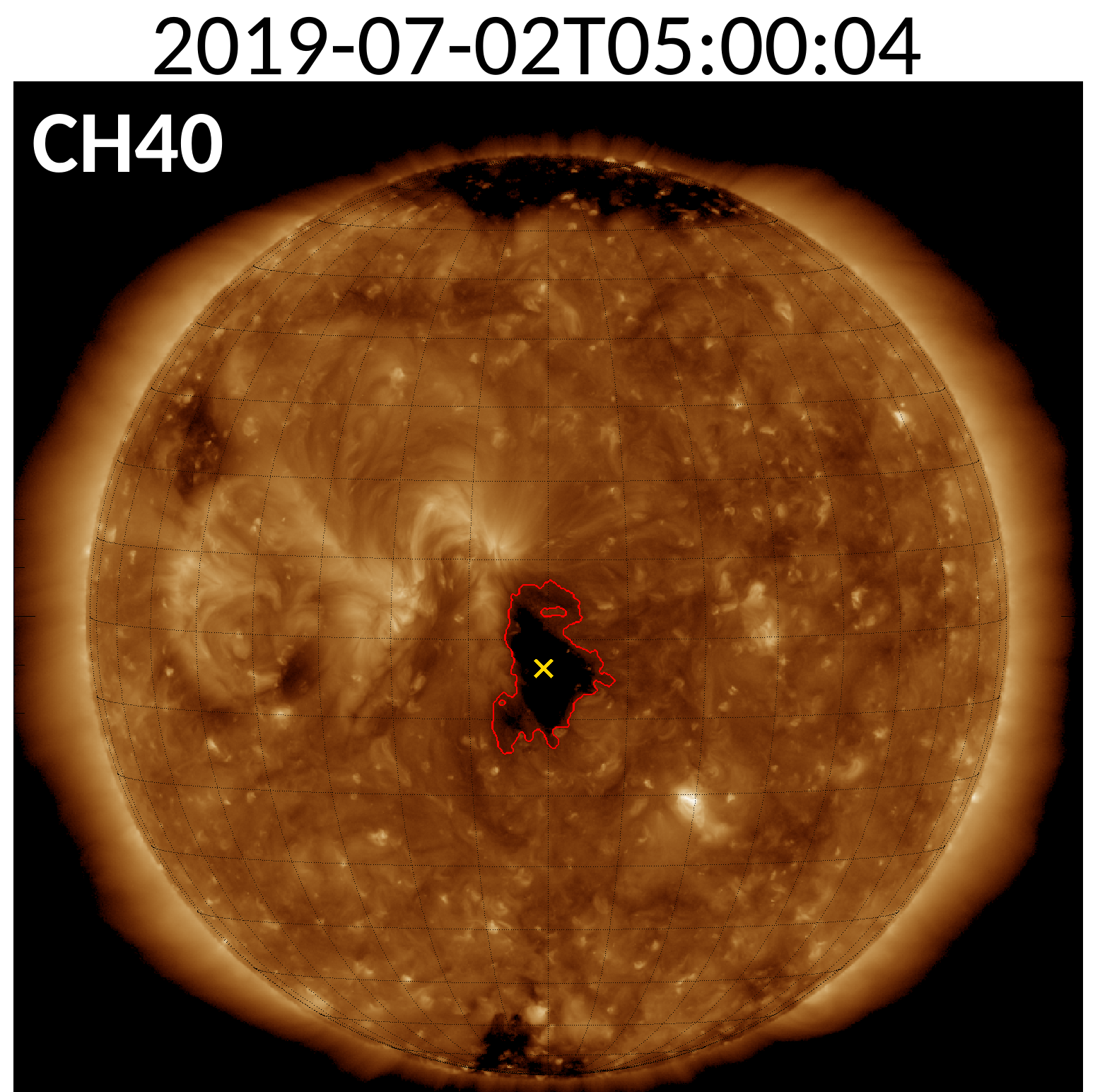}
\end{subfigure}%

\caption{Same as Fig.~\ref{Fig:EUV_images}, but for CHs 21-40. }
\label{Fig:EUV_images2}
\end{figure*}

\begin{figure*}[h!]

\begin{subfigure}[]{0.24\linewidth}
\includegraphics[width=\linewidth]{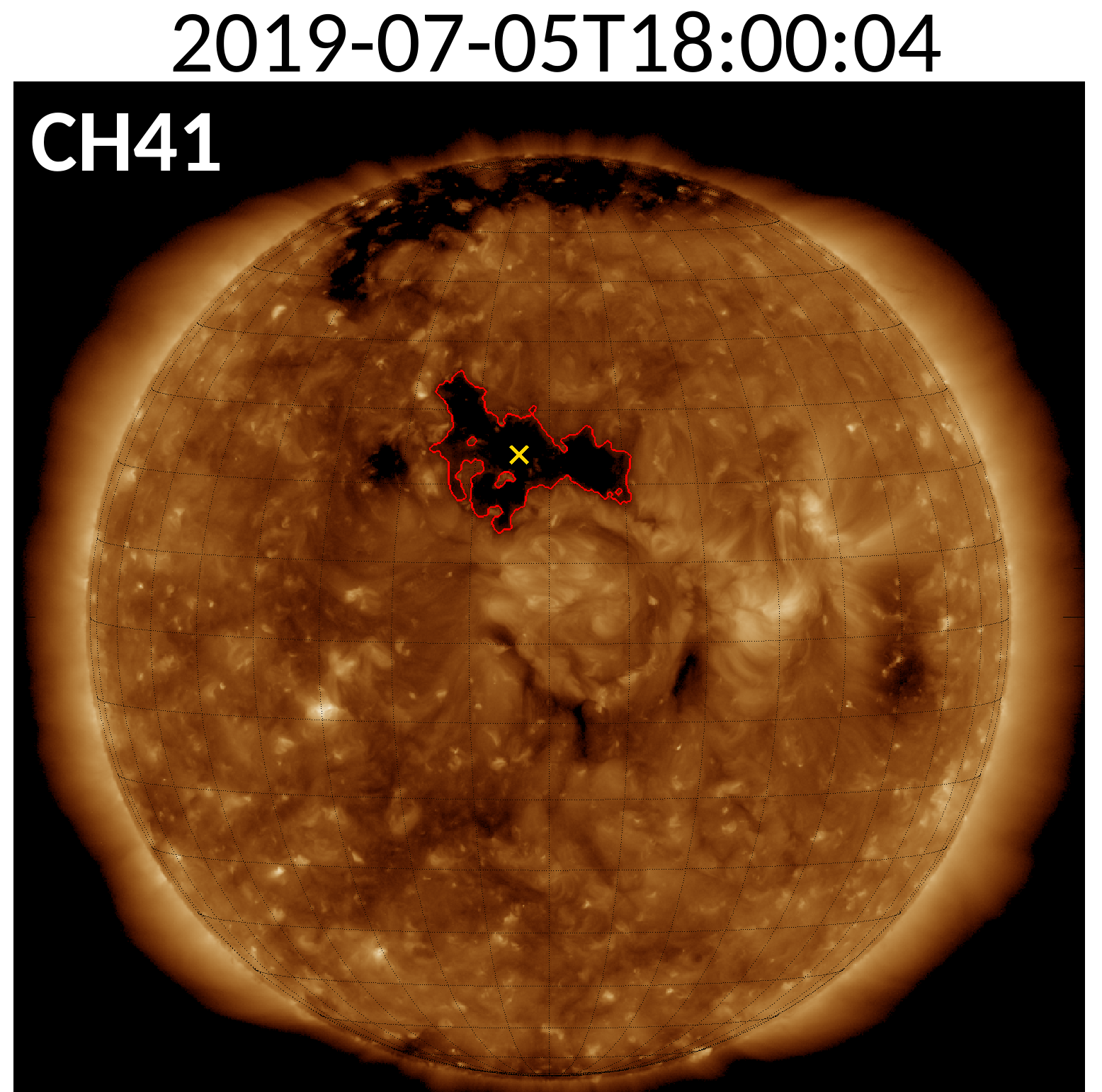}
\end{subfigure}
 \hspace*{0.1\fill}
\begin{subfigure}[]{0.24\linewidth}
\includegraphics[width=\linewidth]{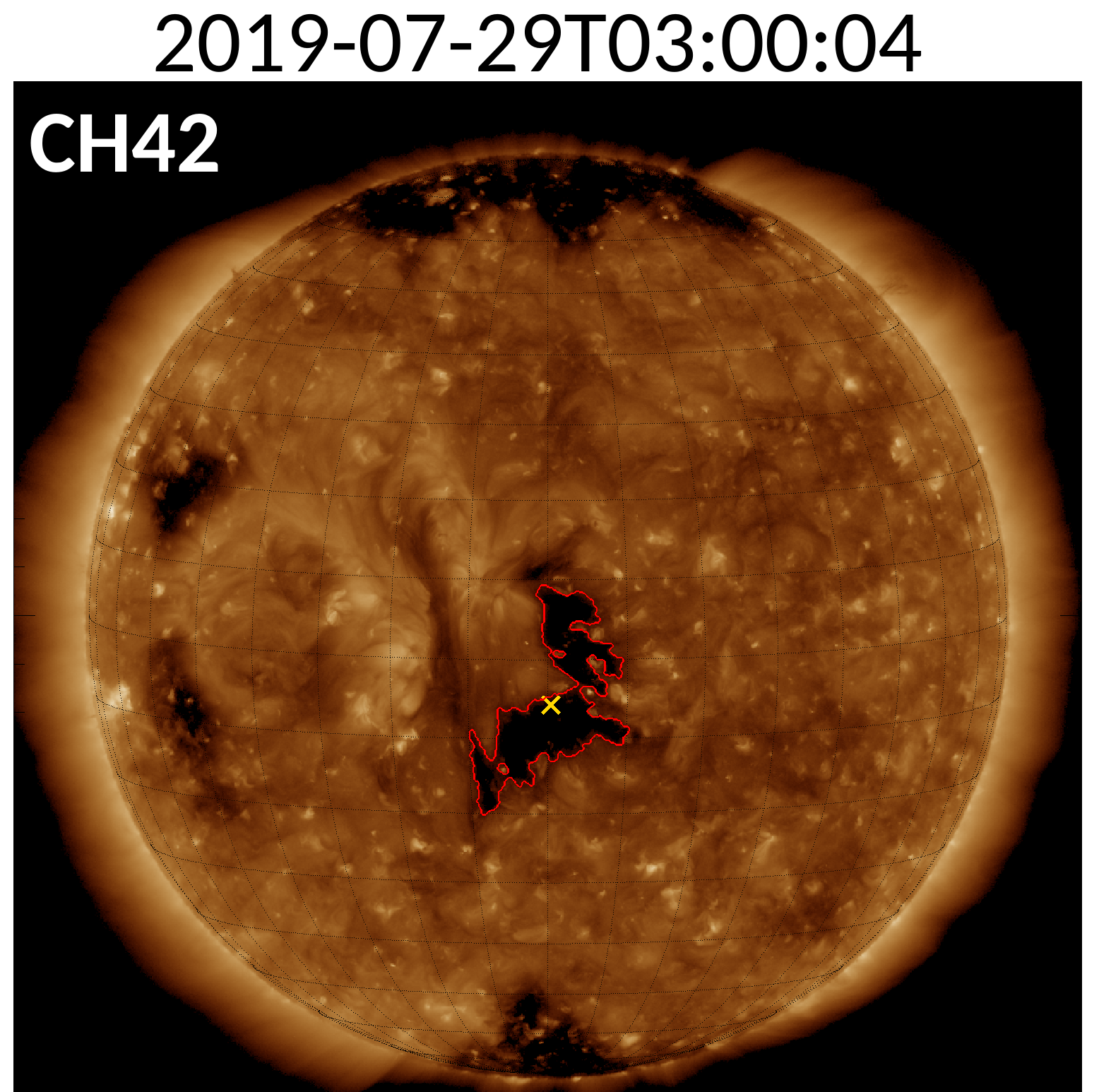}
\end{subfigure}%
 \hspace*{0.1\fill}
\begin{subfigure}[]{0.24\linewidth}
\includegraphics[width=\linewidth]{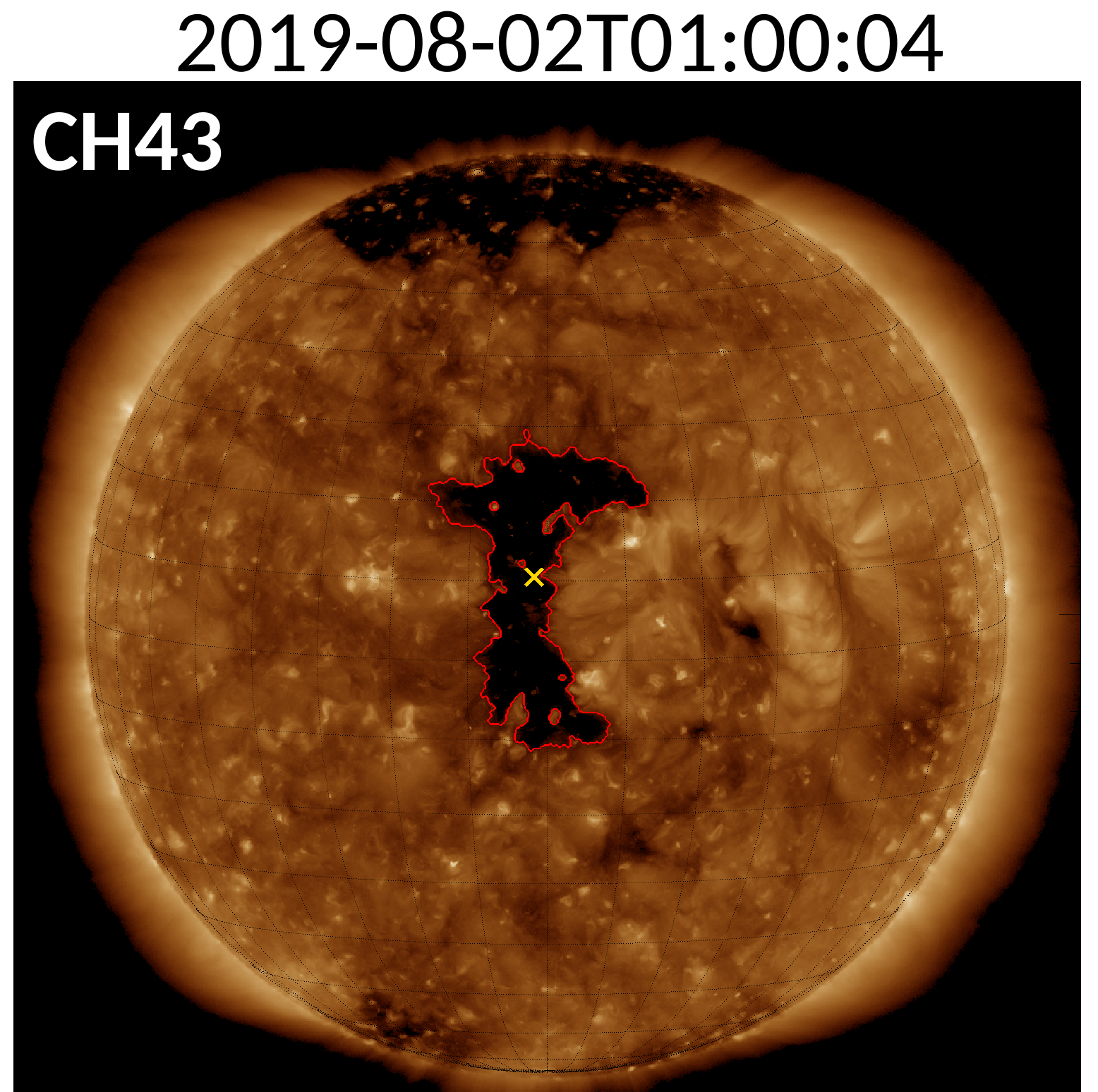}
\end{subfigure}%
 \hspace*{0.1\fill}
\begin{subfigure}[]{0.24\linewidth}
\includegraphics[width=\linewidth]{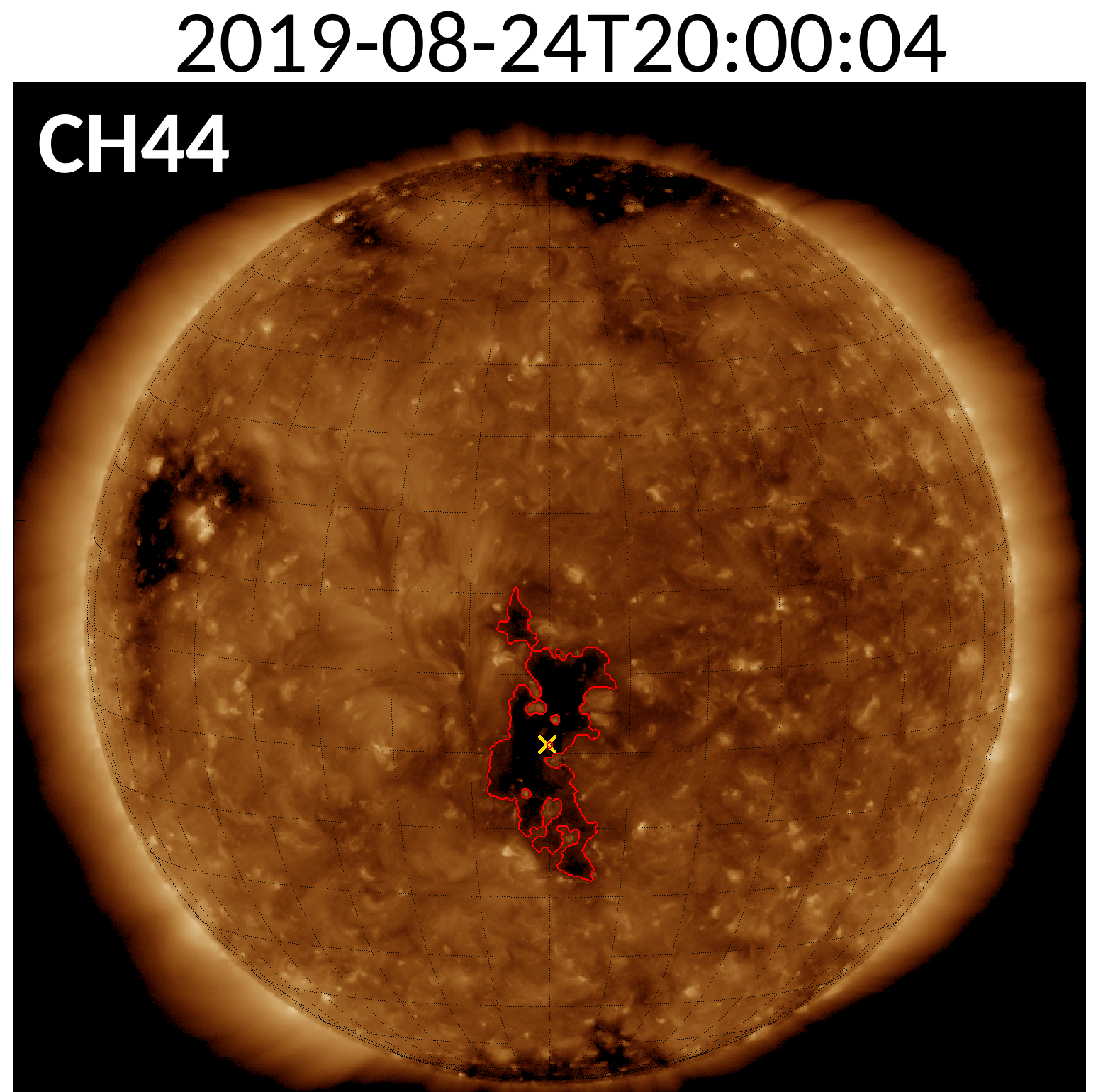}
\end{subfigure}%

\begin{subfigure}[]{0.24\linewidth}
\includegraphics[width=\linewidth]{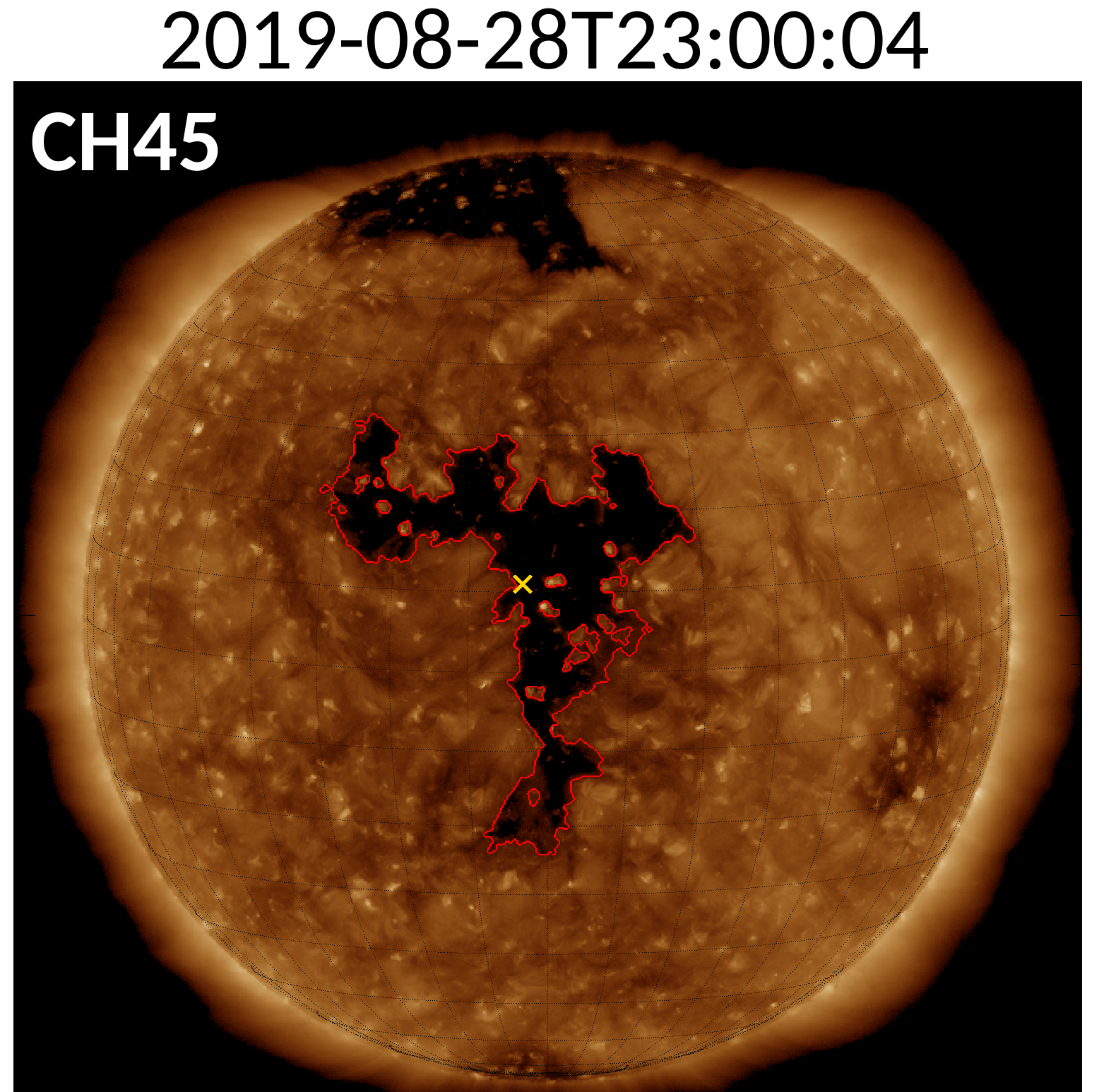}
\end{subfigure}%

\caption{Same as Fig.~\ref{Fig:EUV_images}, but but for CHs 41-45.}
\label{Fig:EUV_images3}
\end{figure*}

The EUV images of the 45 CHs used in this study during their crossing from the solar central meridian are presented in Fig.~\ref{Fig:EUV_images}, Fig.~\ref{Fig:EUV_images2}, and Fig.~\ref{Fig:EUV_images3}. The extraction of each CH was performed using CATCH \citep{Heinemann2019} and the boundaries of each such extraction are colored in red. The point represented by a yellow cross in each CH is their geomagnetic CoM. 

\section{Colatitude correction}
\label{appendix:colatitude_corr}

\begin{figure}[h!]
\centering
\begin{subfigure}[]{0.98\linewidth}
\includegraphics[width=\linewidth]{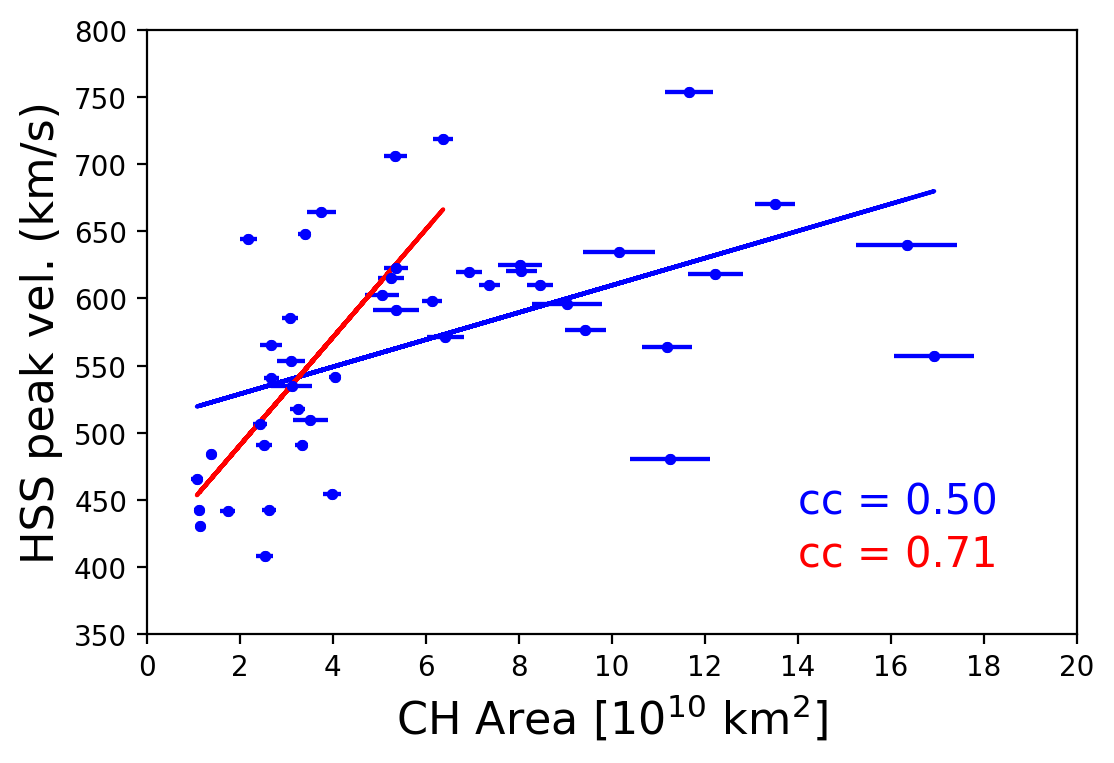}
\caption{}
\end{subfigure}%

\begin{subfigure}[]{0.98\linewidth}
\includegraphics[width=\linewidth]{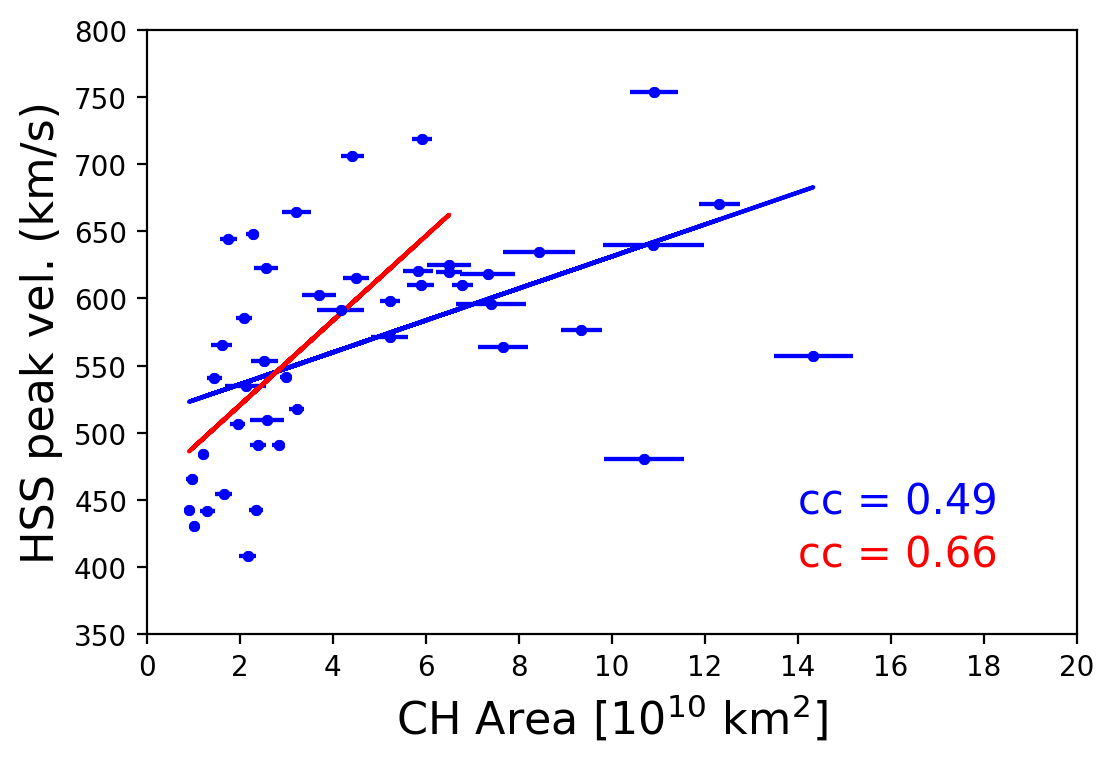}
\caption{}
\end{subfigure}

\caption{HSS peak velocity as a function of the CH area. Panel a,c: Linear regression line fitted for the whole sample. Panel b,d: Linear regression line fitted up until the point that the highest cc was achieved. The top and bottom plots show the bulk speed versus CH area relationship without and with applying the \citet{Hofmeister2018} correction, respectively.}
\label{Fig:CHarea_vs_speed_WithandWithout_colat_corr}
\end{figure}

\begin{figure}[h!]
\centering
\begin{subfigure}[]{0.98\linewidth}
\includegraphics[width=\linewidth]{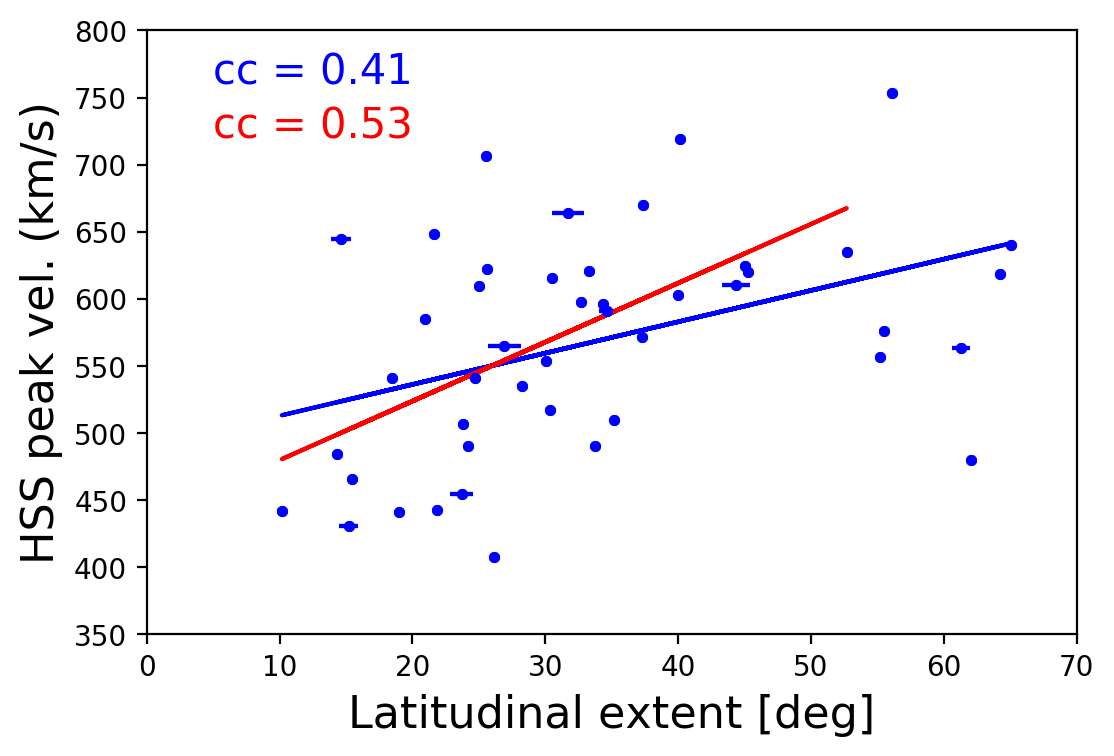}
\caption{}
\end{subfigure}%

\begin{subfigure}[]{0.98\linewidth}
\includegraphics[width=\linewidth]{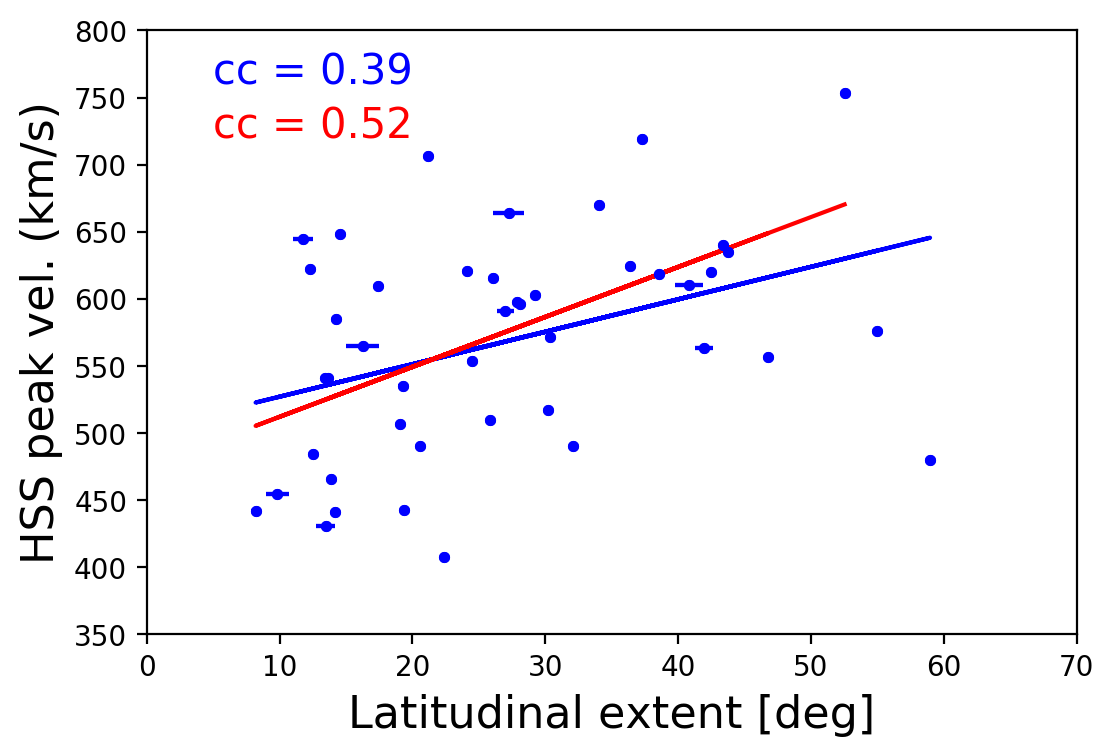}
\caption{}
\end{subfigure}%

\caption{HSS peak velocity as a function of the CH latitudinal extent. Panel a,c: Linear regression line fitted for the whole sample. Panel b,d: Linear regression line fitted up until the point that the highest cc was achieved. The top and bottom plots show the bulk speed versus latitudinal CH extent relationship without and with applying the \citet{Hofmeister2018} correction, respectively.}
\label{Fig:CHlat_vs_speed_WithandWithout_colat_corr}
\end{figure}

According to \citet{Hofmeister2018}, a specific correction is used to provide the geoeffective parts of the areas and latitudinal extents for each structure based on the heliospheric latitudinal angle between the position of the CH and the position of the measuring satellite. The aim is to understand if, for our sample, this so-called colatitude correction leads to better correlations with the maximum HSS velocity in situ, or not. %Co-latitude here refers to the latitudinal angle between the position of the CH and the position of the measuring satellite. 
In Fig.~\ref{Fig:CHarea_vs_speed_WithandWithout_colat_corr} we present a comparison between the CH area and the HSS peak speed at Earth, with and without the colatitude correction \citep{Hofmeister2018}, complementary to Fig.~\ref{Fig:CHarea_vs_speed} in section 3.1. Figure~\ref{Fig:CHlat_vs_speed_WithandWithout_colat_corr} has a similar setup, only we are inspecting how the HSS peak velocity depends on the latitudinal extent of CHs (complementary to Fig.~\ref{Fig:long_lat_extent_vs_speed}b). 

\mbox{}  % empty space for first column
\newpage % start next column

From Figs.~\ref{Fig:CHarea_vs_speed_WithandWithout_colat_corr} and \ref{Fig:CHlat_vs_speed_WithandWithout_colat_corr}, we see that the difference in the Pearson's cc is very small. The same is valid when we fit the whole sample or only the left part of the distribution. Nevertheless, the correlation is slightly better when we did not apply the correction.

\section{Bootstrapping results for CHs of different complexities}
\label{appendix:Bootstrapping}

\begin{figure*}[h!]
\centering
\begin{subfigure}[]{0.31\linewidth}
\includegraphics[width=\linewidth]{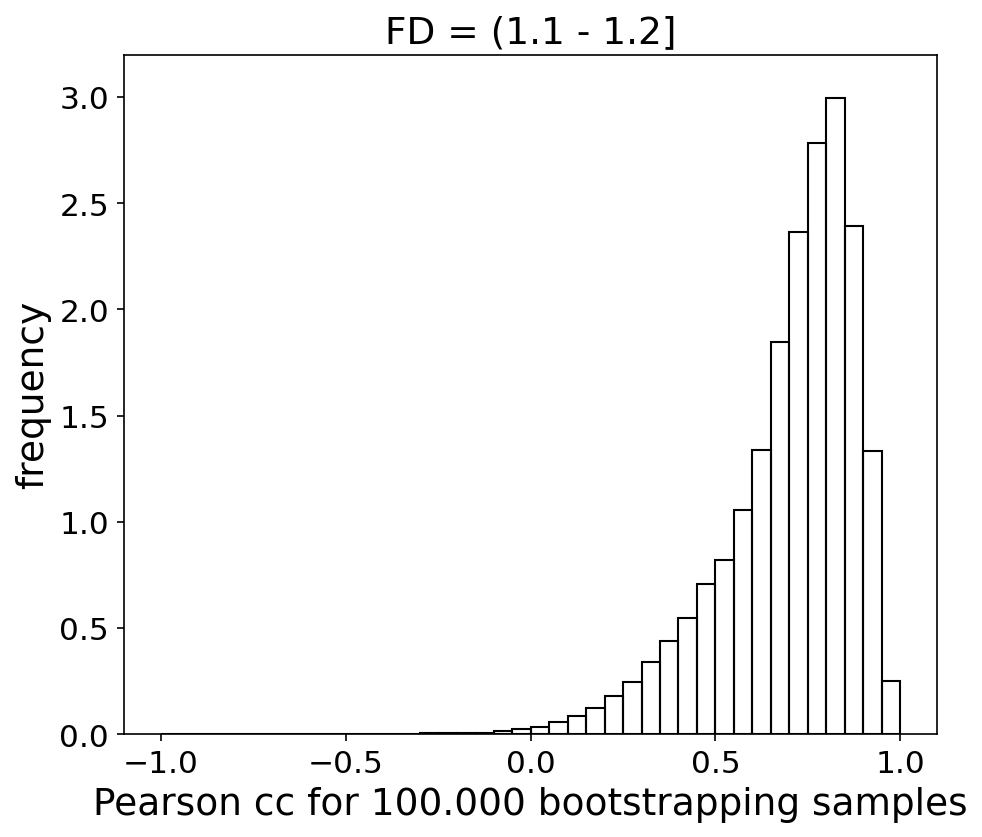}
\caption{}
\end{subfigure}
\hspace*{\fill}
\begin{subfigure}[]{0.31\linewidth}
\includegraphics[width=\linewidth]{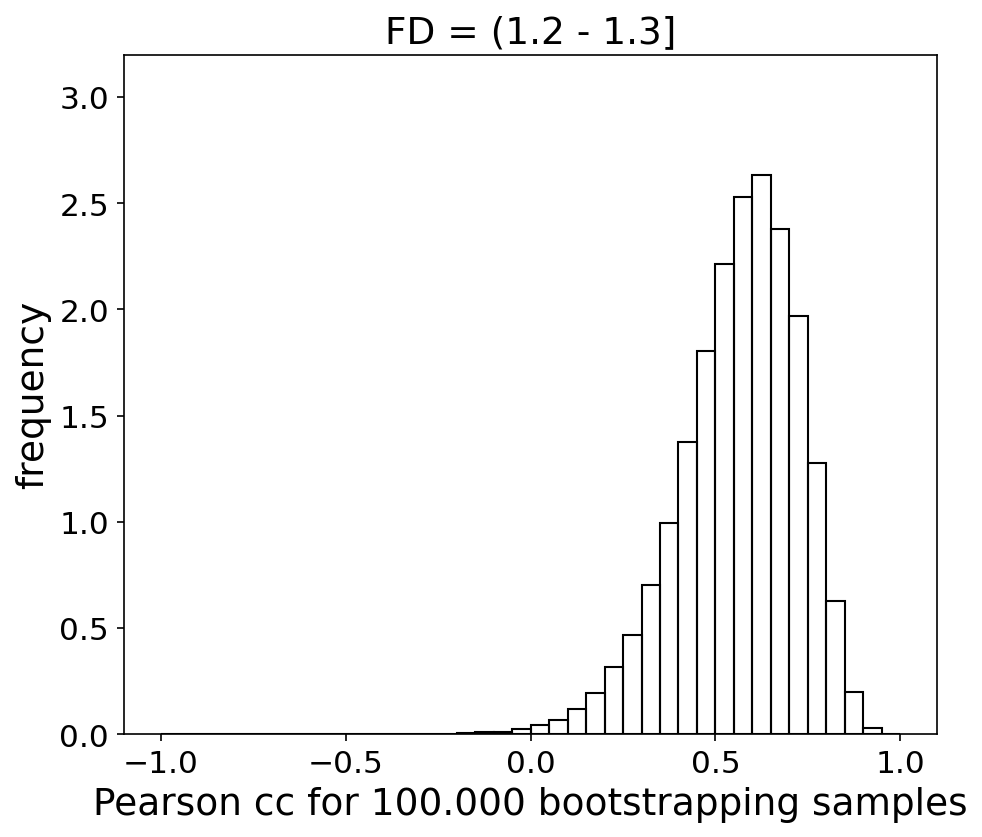}
\caption{}
\end{subfigure}%
\hspace*{\fill}
\begin{subfigure}[]{0.31\linewidth}
\includegraphics[width=\linewidth]{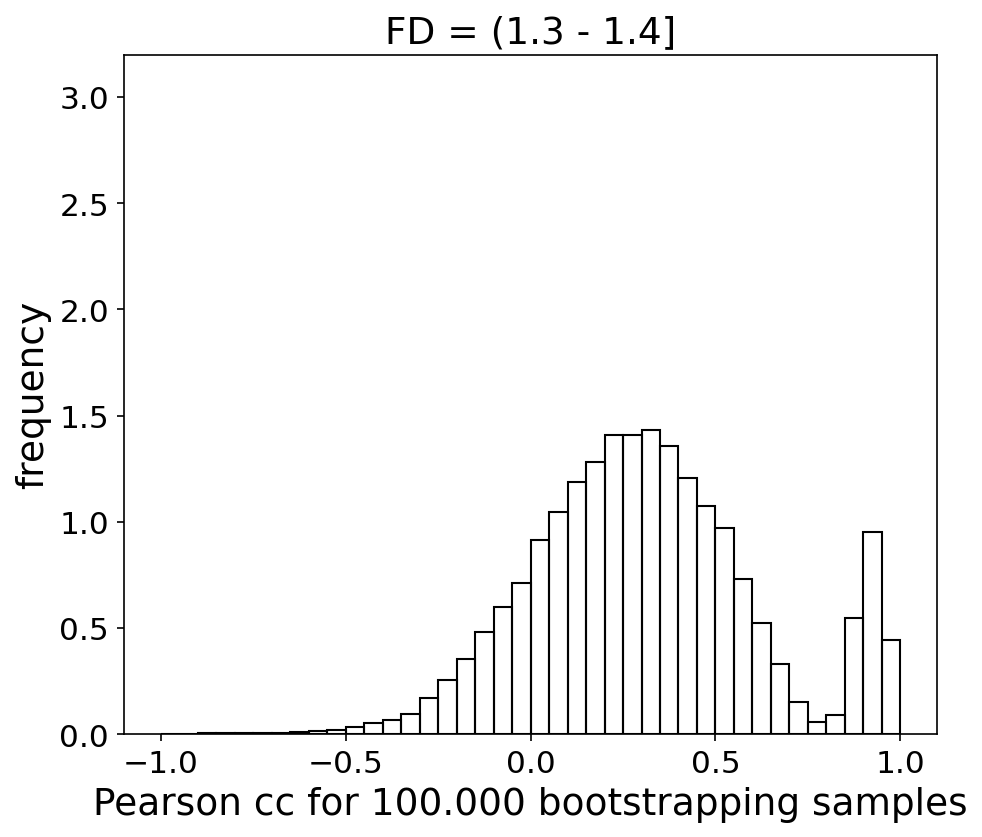}
\caption{}
\end{subfigure}%

\caption{Pearson's cc distributions after the bootstrapping of 100.000 resampled data sets for the CH groups of different FD complexities.}

\label{Fig:FD_appendix}
\end{figure*}

\begin{figure*}[h!]
\centering
\begin{subfigure}[]{0.31\linewidth}
\includegraphics[width=\linewidth]{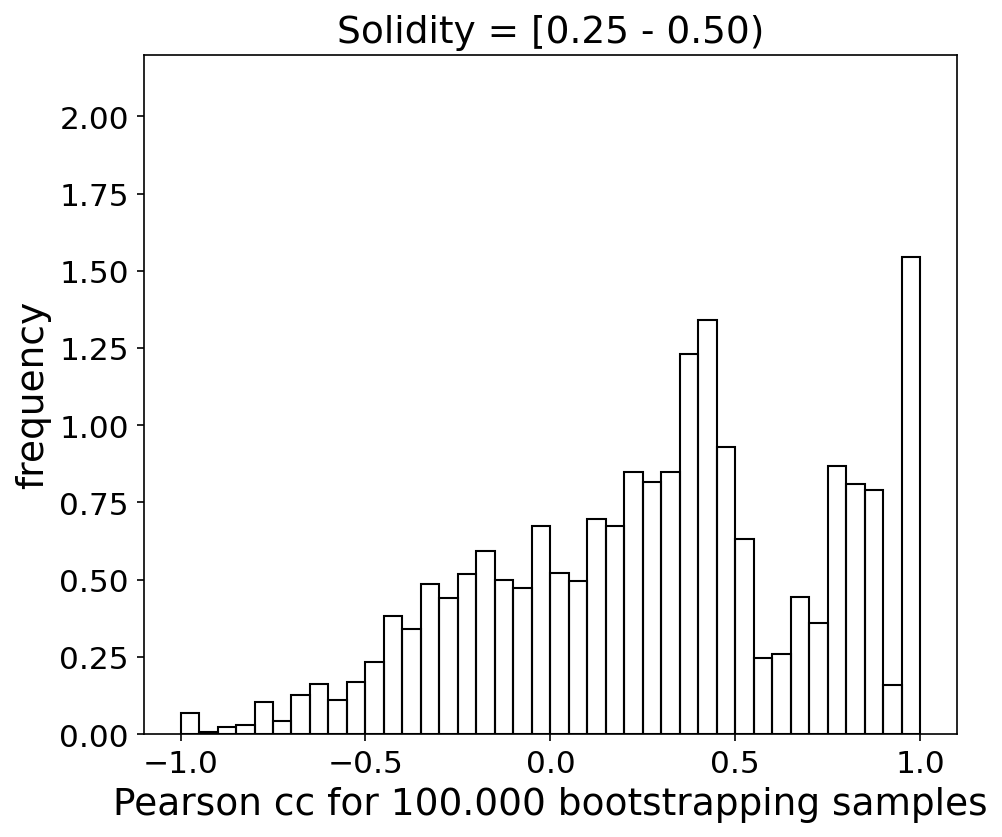}
\caption{}
\end{subfigure}
\hspace*{\fill}
\begin{subfigure}[]{0.31\linewidth}
\includegraphics[width=\linewidth]{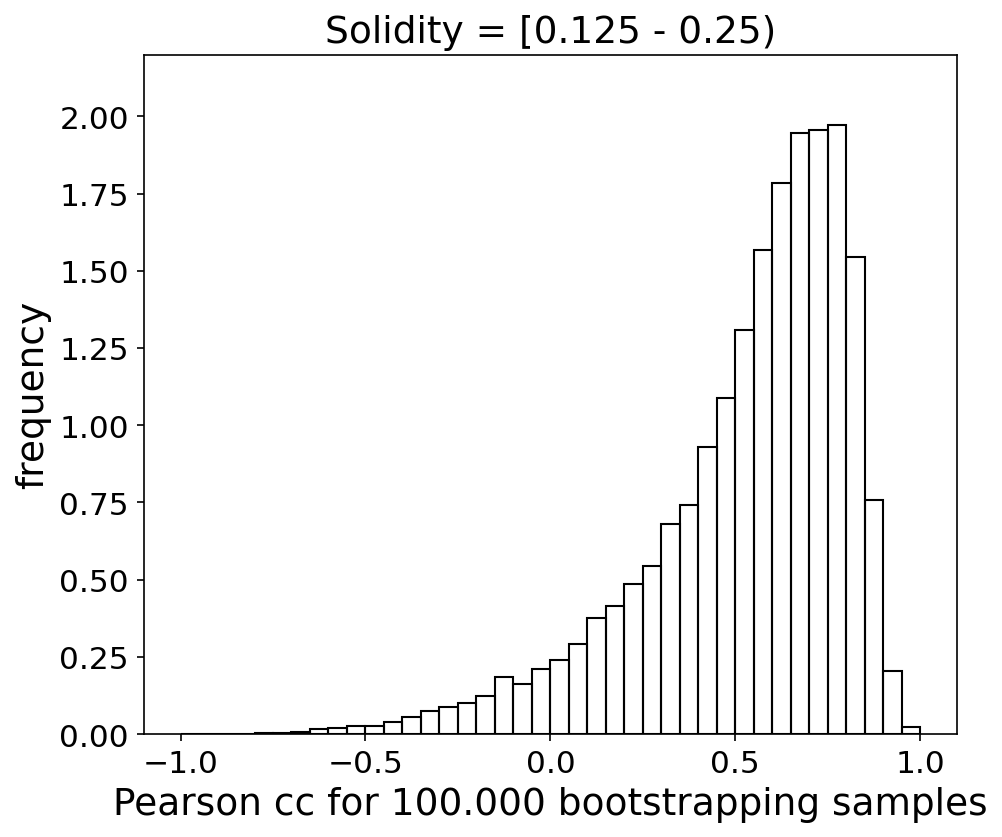}
\caption{}
\end{subfigure}%
\hspace*{\fill}
\begin{subfigure}[]{0.31\linewidth}
\includegraphics[width=\linewidth]{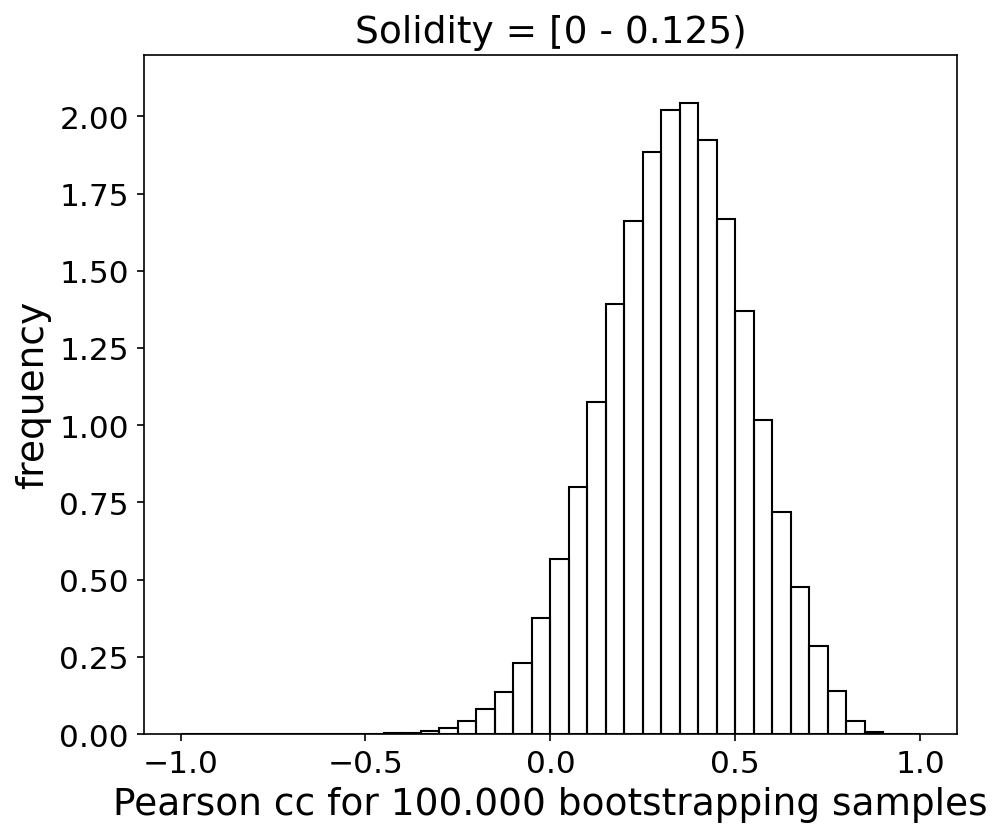}
\caption{}
\end{subfigure}%

\caption{Pearson's cc distributions after the bootstrapping of 100.000 resampled data sets for the CH groups of a different Solidity. }

\label{Fig:Solidity_Appendix}
\end{figure*}

In this section, we present the Pearson's cc distributions after the bootstrapping application of 100.000 resampled data sets for the CH groups of different complexities based on both the FD (Fig.~\ref{Fig:FD_appendix}) and Solidity (Fig.~\ref{Fig:Solidity_Appendix}). Figure \ref{Fig:FD_appendix}c shows the double-peaked distribution produced for the CHs with FD between 1.3 and 1.4 (the most complex CHs). The second peak toward very high cc values originates from the two data points below the linear regression line of Fig.~\ref{Fig:Speed_area_FDingroups}c (circled in green), when they are not taken into account during the bootstrapping resampling. The two points were not removed from the sample because they correspond to very patchy and big CHs, as identified based on the criterion set for that subset. Figure~\ref{Fig:Solidity_Appendix}a is also very ambiguous and leads to a median of low Pearson's cc values because of the very big scattering of the points of Fig.~\ref{Fig:Speed_area_RATIOgroups}a. 

\end{appendix}
\end{document}